\documentclass[vecphys]{svmult}

\usepackage{makeidx}         
\usepackage{graphicx}        
\usepackage{multicol}        
\usepackage[bottom]{footmisc}

\makeindex             

\begin{document}

\title*{Magnetic fields, strings and cosmology}
\author{Massimo Giovannini\inst{1}}
\institute{  Centro ``Enrico Fermi",  Via Panisperna 89/A, 00184 Rome, Italy\\
Department of Physics, Theory Division, CERN, 1211 Geneva 23, Switzerland
\texttt{massimo.giovannini@cern.ch}}
%
%
\maketitle

\begin{center}
To appear in the book \\
{\em String theory and fundamental interactions}\\
published on the occasion of the celebration \\ 
of the 65th birthday of Gabriele Veneziano,\\
eds. M. Gasperini and J. Maharana \\
(Lecture Notes in Physics, Springer Berlin/Heidelberg, 2007),\\
  www.springerlink.com/content/1616-6361.
\end{center}

\section{Half a century of large-scale magnetic fields}
\label{sec1}

\subsection{A premise}

The content of the present contribution is devoted to large-scale 
magnetic fields whose 
origin, evolution and implications constitute today a rather 
intriguing triple point in the phase diagram of physical theories.
Indeed, sticking to the existing literature (and refraining from dramatic 
statements on the historical evolution of theoretical physics) it 
appears that the subject of large-scale magnetization  
thrives and prosper at the crossroad of astrophysics, cosmology and theoretical high-energy physics. 

Following the kind invitation of Jnan Maharana and Maurizio Gasperini,
I am delighted to contribute to this set of lectures whose guideline is 
dictated by the inspiring efforts of Gabriele Veneziano in understanding 
the fundamental forces of Nature. 
My voice joins the choir of gratitude proceeding from the whole physics community 
for the novel and intriguing  results obtained  by Gabriele through the various stages of his manifold activity. 
I finally ought to convey my personal thankfulness for the teachings, 
advices and generous clues received during the last fifteen years. 

\subsection{Length scales} 
The typical magnetic field strengths, in the Universe, 
range from few $\mu {\mathrm G}$ (in the case of galaxies and clusters), 
to few G (in the case of planets, like the earth or Jupiter) and up to $10^{12} {\mathrm G}$ 
in neutron stars. Magnetic fields 
are not only observed in planets and stars but also in the interstellar medium, in the intergalactic medium and, last but not least, in the intra-cluster medium. 

Magnetic fields whose correlation length is larger 
than the astronomical unit ( $1\,{\rm AU} 
= 1.49 \times 10^{13} {\rm cm}$) will be named  {\em large-scale magnetic fields}. 
In fact, magnetic fields  with approximate correlation scale comparable with the earth-sun distance are 
not observed (on the contrary, both the magnetic field of the sun and the one of the earth 
have a clearly distinguishable localized structure). Moreover,  in magnetohydrodynamics (MHD), 
the magnetic diffusivity scale (i.e. the scale below which magnetic fields are diffused because 
of the finite value of the conductivity) turns out to be, amusingly 
enough, of the order of the AU.

\subsection{The early history}

In the forties large-scale magnetic field had no empirical evidence.
For instance, there was no evidence of magnetic 
fields associated with the galaxy as a whole with a rough correlation 
scale of \footnote{Recall that $1\, \mathrm{kpc} = 3.085\times 10^{21} \mathrm{cm} $. Moreover, 
$1 \mathrm{Mpc} = 10^{3} \,\mathrm{kpc}$. The present size of the Hubble radius is $H_{0}^{-1} = 
1.2 \times 10^{28} \mathrm{cm} \equiv 4.1 \times 10^{3} \, \mathrm{Mpc}$ for $h=0.73$.}
$30 \mathrm{kpc}$. More specifically, the theoretical situation can be summarized as follows.
The seminal contributions of H. Alfv\'en \cite{alv1} convinced the community that 
magnetic fields can have a very large life-time in a highly conducting plasma.
 Later on, in the seventies, Alfv\'en will be awarded by the Nobel prize 
 ``for fundamental work and discoveries in magnetohydrodynamics with fruitful 
 applications in different parts of plasma physics".

Using the discoveries  of Alfv\'en,  Fermi \cite{fermi} postulated, in 1949,
the existence of a large-scale magnetic field permeating 
the galaxy with approximate intensity of $\mu$ G and, hence, in equilibrium with the cosmic rays 
\footnote{In this contribution magnetic fields will be expressed in Gauss. In the SI units $ 1 {\rm T} = 10^{4} {\rm G}$.
For practical reasons, 
in cosmic ray physics and in cosmology it is also useful to 
express the magnetic field in $\mathrm{GeV}^2$ (in units $\hbar=c=1$).
Recalling that the Bohr magneton is about $ 5.7\times 10^{-11} 
{\rm MeV}/{\rm T}$ the conversion factor will then be $1 {\rm G} = 1.95 \times 
10^{-20} {\rm GeV}^2 $. The use of Gauss (G) instead of Tesla (T) is justified 
by the existing astrophysical literature where magnetic fields are typically 
expressed in Gauss.}

 Alfv\'en \cite{alv2} did not react positively to the proposal of Fermi, insisting, in a somehow 
 opposite perspective, that cosmic rays are in equilibrium with stars and 
 disregarding completely the possibility 
of a galactic magnetic field.
 Today we do know that this may be the case 
for low-energy cosmic rays but certainly not for the most energetic ones 
around, and beyond, the knee in the cosmic ray spectrum.

At the historical level it is amusing to notice that the mentioned 
controversy can be fully understood from the issue $75$ of Physical Review
where it is possible to consult the paper of Fermi \cite{fermi}, the paper of Alfv\'en 
\cite{alv2} and even a paper by  R. D. Richtmyer and  E. Teller \cite{alv3}
supporting the views and doubts of Alfv\'en.

In 1949 Hiltner \cite{hiltner} and, independently, 
Hall \cite{hall} observed polarization of starlight which was later on interpreted 
by Davis and Greenstein \cite{davis} as an effect of galactic magnetic field 
aligning the dust grains. 

According to the presented chain of events it is legitimate to conclude that
\begin{itemize}
\item{} the discoveries of Alfv\'en were essential in the Fermi proposal 
who was pondering on the origin of cosmic 
rays in 1938 before leaving Italy \footnote{The author is indebted with Prof. G. Cocconi who was so kind to share his personal recollections of the scientific 
discussions with E. Fermi.} because of the infamous fascist legislation;
\item{} the idea that cosmic rays are in equilibrium with the galactic 
magnetic fields (and hence that the galaxy possess a magnetic field) 
was essential in the correct interpretation of the first, fragile, optical 
evidence of  galactic magnetization.
\end{itemize}
The origin of the galactic magnetization, according to \cite{fermi}, had to be somehow primordial. 
It should be noticed, for sake of completeness, 
 that the observations of Hiltner \cite{hiltner} and Hall \cite{hall}
took place from November 1948 to January 1949. The paper of Fermi \cite{fermi} was submitted in January 1949 but 
it contains no reference to the work of Hiltner and Hall. This indicates the Fermi was probably not aware 
of these optical measurements.

The idea that large-scale magnetization should 
somehow be the remnant of the initial conditions of the gravitational  collapse 
of the protogalaxy idea was further pursued by Fermi in collaboration 
with S. Chandrasekar \cite{fermi2,fermi3} who tried, rather ambitiously, to connect the magnetic field of the galaxy to its angular momentum.

\subsection{The middle ages}
 
In the fifties various observations on polarization
of Crab nebula suggested that the Milky Way is not the only magnetized 
structure in the sky. The 
effective new twist in the observations of large-scale magnetic fields 
was the development (through the fifties and sixties) of radio-astronomical 
techniques. From these measurements, the first unambiguous 
evidence of radio-polarization from the Milky Way (MW)
was obtained (see \cite{wiel} and references therein for an account of these 
developments).  

It was also soon realized that the radio-Zeeman effect (counterpart of the optical Zeeman splitting employed to determine the magnetic field of the sun)
could offer accurate determination of (locally very strong) magnetic fields 
in the galaxy. The observation of Lyne and Smith \cite{lyne} 
that pulsars could be used to determine the column density 
of electrons along the line of sight opened 
the possibility of using not only synchrotron 
emission as a diagnostic of the presence of a large-scale magnetic field, but also Faraday rotation. For a masterly written introduction 
to pulsar physics the reader may consult the book of Lyne and Smith 
\cite{lynebook}.

In the seventies all the basic experimental tools 
for the analysis of galactic and extra-galactic magnetic fields 
were ready. Around this epoch also 
extensive reviews on the experimental endeavors 
started appearing and a very nice account 
could be found, for instance, in the review of Heiles \cite{heiles}.

It became gradually evident in the early eighties, that measurements 
of large-scale magnetic fields in the MW and in the external galaxies 
are two complementary aspects of the same problem. While MW studies 
can provide valuable informations  concerning the {\em local} 
structure of the galactic magnetic field, 
the observation of external galaxies provides 
the only viable tool for the reconstruction
of the {\em global } features of the 
galactic magnetic fields. 

Since the early seventies, some relevant attention 
has been paid not only to the magnetic fields of the 
galaxies but also to the magnetic fields of the {\em clusters}.
A cluster is a gravitationally bound system of galaxies. 
The {\em local group} (i.e. {\em our} cluster containing the MW, Andromeda
together with other fifty galaxies) is an {\em irregular} cluster 
in the sense that it contains fewer galaxies than typical clusters 
in the Universe. Other clusters (like Coma, Virgo) are more typical 
and are then called {\em regular} or Abell clusters. As an order 
of magnitude estimate, Abell clusters can contain $10^{3}$ galaxies.

\subsection{New twists}

In the nineties magnetic fields have been measured in single Abell
clusters but around the turn of the century these estimates became 
more reliable thanks to improved experimental techniques.
In order to estimate magnetic fields in clusters, 
an independent knowledge of the electron density along 
the line of sight is needed. Recently Faraday rotation measurements
obtained by radio telescopes (like VLA \footnote{The Very Large Array Telescope, 
consists of 27 parabolic antennas spread over a surface of 20 ${\rm km}^2$ 
in Socorro (New Mexico)}) have been combined with independent measurements 
of the electron density in the intra-cluster medium. This was made possible
by the maps of the x-ray sky obtained with satellites measurements 
(in particular  ROSAT \footnote{The  ROegten SATellite (flying 
from June 1991 to  February 1999) provided maps of the x-ray sky in the 
range $0.1$--$2.5$ keV. A catalog of x-ray bright Abell  clusters was compiled.}).
This improvement in the experimental capabilities seems to have partially 
settled the issue confirming the measurements of the early nineties and implying 
that also clusters are endowed with a magnetic field of $\mu $G strength 
which is {\em not associated with individual galaxies} \cite{gov,fer}. 

While entering the new millennium the capabilities of the observers 
are really confronted with a new challenge: the possibility that also 
superclusters are endowed with their own magnetic field. 
Superclusters are (loosely) gravitationally bound systems of clusters. An 
example is the local supercluster formed by the local group 
and by the VIRGO cluster.  Recently a large new sample of Faraday 
rotation measures of polarized extragalactic sources has been 
compared with galaxy counts in Hercules and Perseus-Pisces (two 
nearby superclusters) \cite{kro2}. First attempts to detect
magnetic fileds associated with superclusters have been reported \cite{kro3}.
A cautious and conservative approach suggests 
that these fragile evidences must be corroborated with more 
conclusive observations (especially in the light of the, sometimes dubious, 
independent determination of the electron density \footnote{In \cite{kro} 
it was cleverly argued that informations on the plasma 
densities from direct observations can be gleaned from detailed multifrequency observations of few giant radio-galaxies (GRG) having 
dimensions up to $4$ Mpc. The estimates based on this 
observation suggest column densities of electrons between $10^{-6}$ and 
$10^{-5}$ $\mathrm{cm}^{-3}$.}). However it is 
not excluded that as the nineties gave us a firmer evidence of cluster 
magnetism, the new millennium may give us more solid understanding 
of supercluster magnetism. 
In the present historical introduction various experimental techniques 
have been swiftly mentioned. A more extensive introductory 
description of these techniques can be found in \cite{f1}.

\subsection{Hopes for the future}

The hope for the near future is connected with the possibility of a 
next generation radio-telescope. Along this line the SKA (Square 
Kilometer Array) has been proposed \cite{fer} (see also \cite{SKA}).
While the technical features of the instrument cannot be thoroughly 
discussed in the present contribution, it suffices to notice that the collecting 
area of the instrument, as the name suggest, will be of $10^{6}\, {\rm m}^2$.
The specifications for the SKA require an angular resolution of $0.1$
arcsec at $1.4$ GHz, a frequency capability of $0.1$--$25$ GHz, and a field of view of at 
least $1\,{\rm deg}^2$ at $1.4$ GHz \cite{SKA}.  The number of independent beams is expected to be larger than $4$ and the number of instantaneous 
pencil beams will be roughly 100 with a maximum 
primary beam separation of about $100$ $\mathrm{deg}$ at low frequencies 
(becoming $1$ $\mathrm{deg}$ at high frequencies, i.e. of the order 
of $1$ GHz).
These specifications will probably allow full sky surveys of Faraday Rotation.

The frequency range of SKA is rather suggestive if we compare it with 
 the one of the Planck experiment \cite{planck}. Planck 
 will operate in $9$ frequency channels from $30$ to, approximately, 
 $900$ GHz. While the three low-frequency channels (from $30$ to 
 $70$ GHz) are not sensitive to polarization the six
  high-frequency channels (between $100$ and $857$ GHZ) will be definitely sensitive to CMB polarization. Now, it should be appreciated that the Faraday 
  rotation signal {\em decreases} with the frequency $\nu$ as $\nu^{-2}$. Therefore, for lower frequencies the Faraday Rotation signal will be 
  larger than in the six high-frequency channels. 
Consequently it is legitimate to hope for a fruitful interplay between 
the next generation of SKA-like radio-telescopes and CMB satellites. 
Indeed, as suggested above, the upper branch of the frequency capability 
of SKA almost overlaps with the lower frequency of Planck so that 
possible effects of large-scale magnetic fields on CMB polarization
could be, with some luck, addressed with the combined action 
of both instruments. 
In fact, the same mechanism leading to the 
Faraday rotation in the radio leads to a Faraday  rotation 
of the CMB {\em provided} the CMB is linearly polarized. These 
considerations suggest, as emphasized in a recent topical 
review, that CMB anisotropies are germane to several 
aspects of large-scale magnetization \cite{maxcqg}. 
The considerations reported so far suggest that during 
the next decade the destiny of radio-astronomy and CMB 
physics will probably be linked together and not 
only for reasons of convenience. 

\subsection{Few burning questions}

In this general and panoramic view of the history of the subject 
we started from the relatively old controversy opposing 
E. Fermi to H. Alfv\'en with the still uncertain but foreseeable 
future developments. 
While the nature of the future developments 
is inextricably connected with the advent of new instrumental 
capabilities, it is legitimate to remark that, in more than fifty years,
magnetic fields have been detected over scales that are progressively 
larger.  From the historical development of the subject a series of 
questions arises naturally:
\begin{itemize}
\item{} what is the origin of large-scale magnetic fields? 
\item{} are magnetic fields primordial as assumed by Fermi more than fifty years ago? 
\item{} even assuming that large-scale magnetic fields are primordial, is there a theory for their generation? 
\item{} is there a way to understand if large-scale magnetic fields are 
really primordial?
\end{itemize}
In what follows we will not give definite answers to these 
important questions but we shall be content of outlining 
possible avenues of new developments.

The plan of the present lecture will be the following.
In Sect.~\ref{sec2} the main theoretical problems 
connected with the origin of large-scale magnetic fields will 
be discussed. In Sect.~ \ref{sec3} the attention will be focused 
on the problem of large-scale magnetic field generation 
in the framework of string cosmological model, a subject where 
the pre-big bang model, in its various incarnations, plays a 
crucial r\^ole. But, finally, large-scale magnetic fields are 
really primordial? Were they really present prior to matter-radiation 
equality? A modest approach to these important questions suggests 
to study the physics of magnetized CMB anisotropies which will be introduced, in its 
essential lines,  in Sect.~\ref{sec4}.
The concluding remarks are collected in Sect.~ \ref{sec5}.

\section{Magnetogenesis}
\label{sec2}
While in the previous Section the approach has been purely 
historical, the experimental analysis of large-scale magnetic fields 
prompts a collection of interesting theoretical problems. They can be summarized by the following 
chain of evidences (see also \cite{f1}):
\begin{itemize}
\item{} In spiral galaxies magnetic fields follow the orientation of the spiral 
arms, where matter is clustered because of differential rotation. While 
there may be an asymmetry in the intensities of the magnetic field in the 
northern and southern emisphere (like it happens in the case of the Milky Way) the typical strength is in the range of the $\mu$ G. 
\item{} Locally magnetic fields may even be in the $\mathrm{mG}$ range and, in this case, 
they may be detected through Zeeman splitting techniques.
\item{} In spiral galaxies the magnetic field is predominantly toroidal 
with a poloidal component present around the nucleus of the galaxy 
and extending for, roughly, $100$ pc.
\item{} The correlation scale of the magnetic field in spirals is of the 
order of $30$ kpc.
\item{} In elliptical galaxies magnetic fields have been measured 
at the $\mu$ G level but the correlation scale is shorter than in the case 
of spirals: this is due to the different evolutionary history of elliptical galaxies
and to their lack of differential rotation;
\item{} Abell clusters of galaxies exhibit magnetic fields present in the so-called 
intra-cluster medium: these fields, always at the $\mu$ G level, are not associated with individual galaxies;
\item{} superclusters {\em might} also be magnetized even if, at the moment, conclusions are premature, as partially explained in Section \ref{sec1} (see also \cite{kro3} and \cite{f1}).
\end{itemize}

The statements collected above rest on various detection techniques
ranging from Faraday rotation, to synchrotron emission, to Zeeman 
splitting of clouds of molecules with an unpaired electron spin. 
The experimental evidence swiftly summarized above 
seems to suggest that different and distant objects have magnetic fields 
of comparable strength. The second suggestion seems 
also to be that the strength of the magnetic fields is, in the first (simplistic)
approximation, independent on the physical scale.

These empirical coincidences 
reminds a bit of one of the motivations of the standard hot big-bang model, 
namely the observation that the light elements are equally abundant 
in rather different parts of our Universe.  
The approximate equality of the abundances implies
that, unlike the heavier elements, the light elements 
have primordial origin. The four light isotopes 
${\rm D}$, $^{3}{\rm He}$, $^{4}{\rm He}$ and $^{7} {\rm Li}$ 
are mainly produced at a specific stage of the hot big bang model 
named nucleosynthesis occurring below the a typical temperature of $0.8$ 
MeV when neutrinos decouple from the plasma and the neutron abundance 
evolves via free neutron decay \cite{bernstein}. The abundances calculated in the 
simplest big-bang nucleosythesis model agree fairly well with the 
astronomical observations.

In similar terms it is plausible to argue that large-scale magnetic fields have comparable strengths at 
large scales because the initial conditions for their evolutions 
were the same, for instance at the time of the gravitational collapse 
of the protogalaxy. The way the initial conditions for the evolution
of large-scale magnetic fields are set  is generically named {\em magnetogenesis} \cite{f1}.
 
There is another comparison which might be useful. Back in the seventies
the so-called Harrison-Zeldovich spectrum was postulated. 
Later, with the developments 
of inflationary cosmology the origin of a flat spectrum of curvature and density 
profiles has been justified on the basis of a period of quasi-de Sitter expansion named {\em inflation}.
It is plausible that in some inflationary models not only the fluctuations 
of the geometry are amplified but also the fluctuations of the gauge fields. This 
happens if, for instance, gauge couplings are effectively dynamical. As the Harrison-Zeldovich 
spectrum can be used as initial condition for the subsequent Newtonian evolution, the primordial 
spectrum of the gauge fields can be used as initial condition for the subsequent  MHD evolution which may lead, eventually, to the observed large-scale magnetic fields. 
The plan of the present section is the following. In Subsect. \ref{subsect21} some 
general ideas of plasma physics will be summarized with particular attention to those 
tools that will be more relevant for the purposes of this lecture. 
In Subsect. \ref{subsect22} the concept of dynamo amplification will be introduced 
in a simplified perspective. In  Subsect. \ref{subsect23} it will be argued that the dynamo amplification, in one 
of its potential incarnations, necessitates  some {\em initial conditions} or as we say in 
the jargon, some {\em seed field}. In Subsect. \ref{subsect24} a panoramic view of astrophysical 
seeds will be presented with the aim of stressing the common aspects of, sometimes diverse, 
physical mechanisms. Subsect. \ref{subsect25} and \ref{subsect26} the two basic 
approaches to cosmological magnetogenesis will be illustrated. In the first case 
(see Subsect. \ref{subsect25}) magnetic fields are produced inside 
the Hubble radius at a given stage in the life of the Universe. In the second case 
(see Subsect. \ref{subsect26}) vacuum fluctuations of the hypercharge field 
are amplified during an inflationary stage of expansion.
Subsection \ref{subsect27} deals with the major problem of inflationary 
magnetogenesis, namely  conformal (Weyl) invariance whose breaking 
will be one of the themes of string cosmological mechanisms for the 
generation of large-scale magnetic fields.

\subsection{Magnetized plasmas}
\label{subsect21} 
Large-scale magnetic fields evolve in a plasma, i.e. a system 
often illustrated as the {\em fourth state of matter}. As we can walk 
in the phase diagram of a given chemical element by going from the solid to the liquid and to the gaseous state  
with a series of diverse phase transitions, a plasma can be obtained by ionizing a gas.  A typical example of weakly coupled plasma is therefore an ionized gas. Examples 
of strongly coupled plasmas can be found also in solid state physics. An essential 
physical scale that has to be introduced in the description of plasma properties 
is the so-called Debye length that will be discussed in the following paragraph.
 
Different 
descriptions of a plasma exist and they range from effective 
fluid models of charged particles \cite{boyd,krall,chen,biskamp} to kinetic approaches like the ones pioneered by Vlasov \cite{vlasov} and Landau \cite{landau}.
From a physical point of view, a plasma is a system of charged 
particles which is globally neutral for typical length-scales larger 
than the Debye length $\lambda_{\rm D}$:
\begin{equation}
\lambda_{\rm D} = \sqrt{\frac{T_0}{8 \pi n_0 e^2}},
\label{debye}
\end{equation}
where $T_{0}$ is the kinetic temperature and $n_{0}$ the mean charge 
density of the electron-ion system, i.e. $n_{\rm e} \simeq n_{\rm i} = n_{0}$.
For a test particle  the Coulomb potential 
will then have the usual Coulomb form but it will be suppressed, at large distances by a Yukawa term, i.e.  $e^{- r/\lambda_{\rm D}}$.
In the interstellar medium there are three kinds of regions which 
are conventionally defined:
\begin{itemize}
\item{} ${\rm H}_{2}$ regions, where the Hydrogen
is predominantly in molecular form (also denoted by  HII);
\item{}  ${\rm H}^{0}$ regions (where Hydrogen is 
in atomic form);
\item{}  and ${\rm H}^{+}$ regions, where Hydrogen is ionized, (also denoted by HI).
\end{itemize} 
In the ${\rm H}^{+}$ regions the typical temperature $T_0$ is of the order 
of $10$--$20$ eV while for $n_{0}$ let us take, for instance, 
 $n_0 \sim 3 \times 10^{-2} {\rm cm}^{-3}$.
Then $\lambda_{\rm D} \sim  30\, {\rm km}$.

For $r \gg 
\lambda_{\rm D}$ the Coulomb potential is screened by the global effect of 
the other particles in the plasma. Suppose now that particles 
exchange momentum through two-body interactions. Their cross 
section will be of the order of $\alpha_{\rm em}^2/T_0^2$ and the mean free 
path will be $\ell_{\rm m f p} \sim T_0^2/(\alpha_{\rm em}^2 n_0)$, i.e. recalling 
Eq. (\ref{debye}) 
$\lambda_{\rm D} \ll \ell_{\rm m f p}$. This means that the
plasma is a weakly collisional system which is, in general,  not
 in local thermodynamical 
equilibrium and this is the reason why we introduced $T_{0}$ as the kinetic (rather than thermodynamic) temperature. 

The last observation can be made even more explicit by defining another 
important scale, namely the plasma frequency which, in the system 
under discussion, is given by 
\begin{equation}
\omega_{\rm pe} = \sqrt{\frac{ 4 \pi n_0 e^2}{m_{\rm e}}} \simeq 2\,\, \biggl( \frac{n_{0}}{10^{3}\,\, {\rm cm }^{-3}} \biggr)^{1/2} \,\, {\rm MHz},,
\label{pl}
\end{equation}
where $m_{\rm e}$ is the electron mass. Notice that, in the interstellar 
medium (i.e. for $n_{0} \simeq 10^{-2} \,\, {\rm cm}^{-3}$) Eq. (\ref{pl}) 
gives a plasma frequency in the GHz range. This observation is important, 
for instance, in the treatment of Faraday rotation since the plasma 
frequency is typically much larger than the Larmor frequency i.e. 
\begin{equation}
\omega_{\rm Be} =\frac{e B_{0}}{m_{\rm e} } \simeq 
18.08 \biggl(\frac{B_0}{ 10^{-3}\,\,\, {\rm G}}\biggr)\,\,\,{\rm kHz},
\end{equation}
implying, for $B_{0} \simeq \mu {\rm G}$, $\omega_{\rm Be} \simeq
20 {\rm Hz}$. The same hierarchy holds also when the (free) electron density 
is much larger than in the interstellar medium, and, for instance, at 
the last scattering between electrons and photons for a redshift $z_{\mathrm{dec}} \simeq 1100$ 
(see Sect. \ref{sec4}).

The plasma frequency is 
the oscillation frequency of the electrons when they are displaced 
from their equilibrium configuration in a background of approximately 
fixed ions. Recalling that $v_{\rm ther}\simeq \sqrt{T_0/m_{\rm e}}$ is the thermal 
velocity of the charge carriers, the collision frequency $\omega_{\rm c} \simeq v_{\rm ther}/\ell_{\rm mfp}$ 
is always much smaller than $\omega_{\rm pe} \simeq v_{\rm ther}/\lambda_{\rm D}$.
Thus, in the idealized system described so far, the following 
hierarchy of scales holds:
\begin{equation}
\lambda_{\rm D} \ll \ell_{\rm mfp},~~~~~~\omega_{\rm c} \ll \omega_{\rm pe}, 
\end{equation} 
which means that before doing one collision the system undergoes 
many oscillations, or, in other words, that the mean free path 
is not the shortest scale in the problem.
Usually one defines also the {\em plasma parameter} ${\cal N} = n_0^{-1} \lambda_{\rm D}^{-3}$, 
i.e. the number of particles in the Debye sphere. In the approximation of weakly coupled plasma  
${\cal N}\ll 1$ which also imply that the mean kinetic energy of the 
particles is larger than the mean inter-particle potential.

The spectrum of plasma excitations is a rather vast subject and it
will not strictly necessary for the following considerations (for further 
details see\cite{boyd,krall,chen}). It 
is sufficient to remark that we can envisage, broadly 
speaking, two regimes that are physically different:
\begin{itemize}
\item{} typical length-scales much {\em larger} than $\lambda_{\rm D}$
and typical frequencies much {\em smaller} than $\omega_{\rm pe}$;
\item{}  typical length-scales smaller (or comparable) with $\lambda_{\rm D}$
and typical frequencies much {\em larger} than $\omega_{\rm pe}$.
\end{itemize}
In the first situation reported above it can be shown that a single fluid description suffices. 
The single fluid 
description is justified, in particular, for the analysis  of the dynamo 
instability which occurs for dynamical times of the order of the age of the 
galaxy and length-scales larger than the kpc. 
In the opposite regime, i.e. $\omega \geq \omega_{\rm pe}$ and 
$L \geq \lambda_{\rm D}$ the single fluid approach breaks down
and a multi-fluid description is mandatory. This is, for instance, the 
branch of the spectrum of plasma excitation where the displacement 
current (and the related electromagnetic propagation) cannot be neglected. A more reliable description is provided, in this regime, by the Vlasov-Landau 
(i.e. kinetic) approach \cite{vlasov, landau} (see also \cite{krall}).

Consider, therefore, a two-fluid system of electrons and protons.
This system will be described by the continuity equations of the 
density of particles, i.e.
\begin{equation}
\frac{\partial n_{\rm e}}{\partial t} + \vec{\nabla}\cdot( n_{\rm e} \vec{v}_{\rm e})=0, \qquad \frac{\partial n_{\rm p}}{\partial t} + \vec{\nabla}\cdot( n_{\rm p} \vec{v}_{\rm p})=0,
\label{density}
\end{equation}
and by the momentum conservation equations
\begin{eqnarray}
&& m_{\rm e} n_{\rm e} \biggl[ \frac{\partial}{\partial t} + \vec{v}_{\rm e} \cdot 
\vec{\nabla}\biggr] \vec{v}_{\rm e} = - e n_{\rm e} \biggl[ \vec{E} + \vec{v}_{\rm e} \times \vec{B}\biggr] - \vec{\nabla}p_{\rm e} - {\mathcal C}_{\rm ep},
\label{mome}\\
&& m_{\rm p} n_{\rm p} \biggl[ \frac{\partial}{\partial t} + \vec{v}_{\rm p} \cdot 
\vec{\nabla}\biggr] \vec{v}_{\rm p} = e n_{\rm p} \biggl[ \vec{E} + \vec{v}_{\rm p} \times \vec{B}\biggr] - \vec{\nabla}p_{\rm p} - {\mathcal C}_{\rm pe}.
\label{momp}
\end{eqnarray}
Equations (\ref{density}), (\ref{mome}) and (\ref{momp}) must 
be supplemented by Maxwell equations reading, in this case 
\begin{eqnarray}
&&\vec{\nabla}\cdot \vec{E} = 4\pi e (n_{\rm p} - n_{\rm e}),
\label{Mx1}\\
&& \vec{\nabla}\cdot \vec{B} =0,
\label{Mx2}\\
&& \vec{\nabla} \times \vec{E} + \frac{\partial \vec{B}}{\partial t} =0,
\label{Mx3}\\
&& \vec{\nabla} \times \vec{B} = \frac{\partial \vec{E}}{\partial t} + 
4\pi e ( n_{\rm p} \vec{v}_{\rm p} - n_{\rm e} \vec{v}_{\rm e}).
\label{Mx4}
\end{eqnarray}
The two fluid system of equations is rather useful to discuss 
various phenomena like the propagation of electromagnetic excitations 
at finite charge density both in the presence and in the absence 
of a background magnetic field \cite{boyd,krall,chen}. 
The previous observation implies that a two-fluid treatment is 
mandatory for the description of Faraday rotation of the Cosmic 
Microwave Background (CMB) polarization. This subject 
will not be specifically discussed in the present lecture 
(see, for further details,  \cite{maxbir} and references therein).

Instead of treating the two fluids as separated,  the plasma may be considered 
as a single fluid defined by an appropriate set of {\em global} variables: 
\begin{eqnarray}
&& \vec{J} = e (n_{\rm p} \vec{v}_{\rm p} - n_{\rm e} \vec{v}_{\rm e}),
\label{totalJ}\\
&& \rho_{\rm q} = e ( n_{\rm p} - n_{\rm e}),
\label{totrhoq}\\
&& \rho_{\rm m} = (m_{\rm e} n_{\rm e} + m_{\rm p} n_{\rm p}),
\label{totrhom}\\
&& \vec{v} = \frac{m_{\rm e}\,n_{\rm e} \vec{v}_{\rm e} + n_{\rm p} m_{\rm p} v_{\rm p}}{m_{\rm e} n_{\rm e} + m_{\rm p} n_{\rm p}},
\label{bulkv}
\end{eqnarray}
where $\vec{J}$ is the global current and $\rho_{\rm q}$ is the global charge 
density; $\rho_{\rm m}$ is the total mass density and $\vec{v}$ is the 
so-called bulk velocity of the plasma. From the definition of the bulk velocity 
it is clear that $\vec{v}$ is the centre-of-mass velocity of the electron-ion 
system. The interesting case is the one where the plasma is 
globally neutral, i.e. $n_{\rm e} \simeq n_{\rm p} = n_{0}$, implying, from 
Maxwell and continuity equations the following equations
\begin{equation}
\vec{\nabla}\cdot \vec{E} =0, \qquad \vec{\nabla}\cdot \vec{J} =0, \qquad 
\vec{\nabla}\cdot \vec{B}=0.
\label{solenoidal}
\end{equation} 
The equations reported in Eq. (\ref{solenoidal}) are the first characterization 
of MHD equations, i.e. a system where the total current as well as the 
electric and magnetic fields are all solenoidal. 
The remaining equations allow to obtain the relevant set of conditions 
describing the long wavelength modes of the magnetic field i.e.
\begin{eqnarray}
&& \vec{\nabla} \times \vec{B} = 4\pi \vec{J},
\label{mhd1}\\
&& \vec{\nabla} \times \vec{E} = - \frac{\partial \vec{B}}{\partial t}.
\label{mhd2}
\end{eqnarray}
In Eq. (\ref{mhd1}), the contribution of the displacement 
current has been neglected for consistency with the solenoidal nature 
of the total current (see Eq. (\ref{solenoidal})).  
Two other relevant equations can be obtained by summing and subtracting 
the momentum conservation equations, i.e. Eqs. (\ref{mome}) 
and (\ref{momp}). The result of this procedure is 
\begin{eqnarray}
&& \rho_{\rm m} \biggl[ \frac{\partial \vec{v}}{\partial t} + 
\vec{v} \cdot\vec{\nabla} \vec{v} \biggr]= \vec{J} \times \vec{B} - \vec{\nabla} P
\label{MOM}\\
&& \vec{E} + \vec{v}\times \vec{B} = \frac{\vec{J}}{\sigma} + 
\frac{1}{e n_{\rm q} }( \vec{J} \times \vec{B} - \vec{\nabla} p_{\mathrm{e}}),
\label{OHM}
\end{eqnarray}
where $n_{\mathrm{q}} \simeq n_{0} \simeq n_{\mathrm{e}}$ and 
 $P= p_{\rm e} + p_{\rm p}$.
Equation (\ref{MOM}) is derived from the sum of Eqs. (\ref{mome}) 
and (\ref{momp}) and  in (\ref{MOM}) 
$\vec{J} \times \vec{B}$ is the Lorentz force term which is quadratic in the magnetic field. In fact using Eq. (\ref{mhd1})  
\begin{equation}
\vec{J} \times \vec{B} = \frac{1}{4\pi} 
(\vec{\nabla}\times \vec{B}) \times \vec{B}.
\label{LF}
\end{equation}
Note that to derive Eq. 
(\ref{OHM}) the limit $m_{\rm e}/m_{\rm p} \to 0$ must be taken, at some point. 
There are some caveats related to this procedure since viscous 
and collisional effects may be relevant \cite{krall}. Equation (\ref{OHM}) 
is sometimes called one-fluid generalized Ohm law. 
In Eq. (\ref{OHM}) the term 
$\vec{J}\times \vec{B}$ is nothing but the {\em Hall current} 
and $\vec{\nabla} p_{\mathrm{e}}$ is often called
thermoelectric term. Finally the term $\vec{J}/\sigma$ is the resistivity term and $\sigma$ is 
the conductivity of the one-fluid description. In Eq. (\ref{OHM})  the pressure has been taken to be isotropic. 
Neglecting, the Hall and thermoelectric terms (that may play, however, a r\^ole 
in the Biermann battery mechanism for magnetic field generation) the 
Ohm law takes the form 
\begin{equation}
\vec{J} = \sigma (\vec{E} + \vec{v} \times \vec{B}).
\label{ohm2}
\end{equation}
Using Eq. (\ref{ohm2}) together with Eq. (\ref{mhd1}) it is easy to show that 
the Ohmic electric field is given by 
\begin{equation}
\vec{E} = \frac{\vec{\nabla}\times \vec{B}}{4 \pi \sigma} - \vec{v} \times 
\vec{B}.
\label{ohmel}
\end{equation}
Using then Eq. (\ref{ohmel}) into Eq. (\ref{mhd2}) and exploiting 
known vector identities we can get the 
canonical form of the magnetic diffusivity equation
\begin{equation}
\frac{\partial \vec{B}}{\partial t} = \vec{\nabla}\times (\vec{v} \times \vec{B})
+\frac{1}{4\pi\sigma} \nabla^2 \vec{B},
\label{magndiff}
\end{equation}
which is the equation to be used to discuss the general features 
of the dynamo instability.

MHD can be studied into two different  (but complementary) limits
\begin{itemize}
\item{} the ideal (or superconducting) limit where the conductivity 
is set to infinity (i.e. the $\sigma\to \infty$ limit);
\item{} the real (or resistive) limit where the conductivity is finite.
\end{itemize}

The plasma description following from MHD can be also phrased 
in terms of the conservation of two interesting quantities, i.e. 
the magnetic flux and the magnetic helicity \cite{biskamp,maxknot}:
\begin{eqnarray}
&& \frac{d}{d t} \biggl(\int_{\Sigma} \vec{B} \cdot d\vec{\Sigma} \biggr) = - \frac{1}{4\pi \sigma }\int \vec{\nabla} \times\vec{\nabla}\times \vec{B} \cdot d\vec{\Sigma},
\label{fluxcons}\\
&& \frac{d}{dt} \biggl(\int_{V} d^{3} x \vec{A} \cdot \vec{B} \biggr) = - \frac{1}{4\pi \sigma} \int_{V} d^{3} x \vec{B} \cdot \vec{\nabla} \times \vec{B}.
\label{helcons}
\end{eqnarray}
In Eq. (\ref{fluxcons}), $\Sigma$ is an arbitrary closed surface that 
moves with the plasma. In the ideal MHD limit the magnetic flux is 
exactly conserved and the the flux is sometimes said to be frozen into the 
plasma element. In the same limit also the magnetic helicity is 
conserved.  In the resistive limit the magnetic flux and helicity are dissipated 
with a rate proportional to $1/\sigma$ which is small provided the conductivity 
is sufficiently high. The term appearing at the right hand side off Eq. (\ref{helcons}) is called magnetic gyrotropy.

The conservation of the magnetic helicity is a statement on the conservation 
of the {\em topological}  properties of the 
magnetic flux lines. If the magnetic field is completely 
stochastic, the magnetic flux lines will be closed loops 
evolving independently in the plasma and the helicity 
will vanish. There could be, however, 
more complicated topological situations
where a single magnetic loop is twisted (like some 
kind of M\"obius stripe) or the case where 
the magnetic loops are connected like the rings of a chain.
In both cases the magnetic helicity will not be zero 
since it measures, essentially, the number of links and twists 
in the magnetic flux lines. 
The conservation of the magnetic flux and of the magnetic 
helicity is a consequence of the fact that, in ideal 
MHD, the Ohmic electric field is always orthogonal 
both to the bulk velocity field and to the magnetic 
field. In the resistive MHD approximation this is 
no longer true \cite{biskamp}.

\subsection{Dynamos}
\label{subsect22}
The dynamo theory has been developed 
starting from the early fifties through
the eighties and various extensive presentations exist in the literature
 \cite{parker,zeldovich,ruzmaikin}.  
Generally speaking a {\em dynamo} is a process where the kinetic energy 
of the plasma is transferred to magnetic energy.  
There are different sorts of dynamos. Some of the 
dynamos that are currently addressed in the existing literature
are large-scale dynamos, small-scale dynamos, nonlinear dynamos,
$\alpha$-dynamos...

It would be difficult, in the present lecture, even to review such a vast literature and, therefore, 
it is more appropriate to refer to some review articles where 
the modern developments in dynamo theory and in mean 
field electrodynamics are reported \cite{kulsrud1,brandenburg}.
As a qualitative example of the dynamo action it is practical do 
discuss the magnetic diffusivity equation obtained, from general 
considerations, in Eq. (\ref{magndiff}). 

Equation (\ref{magndiff}) simply stipulates that the first time derivative 
of the magnetic fields intensity results from the balance 
of two (physically different)  contributions. The first term at the right hand side of Eq. (\ref{magndiff})
is the the {\em dynamo} term and it contains the bulk velocity 
of the plasma $\vec{v}$. If this term dominates the magnetic field 
may be amplified thanks to the differential rotation of the plasma. 
The dynamo term provides then the coupling allowing the transfer 
of the {\em kinetic} energy into {\em magnetic} energy. 
The second term at the right hand side  of Eq. (\ref{magndiff}) is the 
{\em  magnetic diffusivity} whose 
effect is to damp the magnetic field intensity. 
Defining then as $L$ the typical scale of spatial variation of the 
magnetic field intensity,   
the typical time scale of resistive phenomena turns out to be
\begin{equation}
t_{\sigma} \simeq 4\pi \sigma L^2 .
\label{difftime}
\end{equation}
In a non-relativistic plasma 
the conductivity $\sigma$ goes typically as $T^{3/2}$  \cite{boyd,krall}. 
In the case of planets, like 
the earth, one can wonder why a sizable magnetic field can still be present. One of 
the theories is that the dynamo term regenerates continuously 
the magnetic field which is dissipated by the diffusivity term \cite{parker}. 
In the case of the galactic disk the 
value of the conductivity \footnote{It is common use in the astrophysical applications 
to work directly with $\eta = (4\pi \sigma)^{-1}$. In the case of the galactic disks 
$\eta = 10^{26} {\rm cm}^{2}~{\rm Hz}$. }
is given by $\sigma \simeq 7\times 10^{-7} {\rm Hz}$. Thus, for  $L \simeq {\rm kpc}$ 
$t_{\sigma} \simeq 10^{9} (L/{\rm kpc})^2 {\rm sec}$. 

Equation (\ref{difftime}) can also give the typical 
resistive length scale once the time-scale 
of the system is specified. Suppose that the time-scale of the system is 
given by $t_{U} \sim H_{0}^{-1} \sim 10^{18} {\rm sec}$ where $H_0$ is the 
present order of magnitude of the Hubble parameter. Then  
\begin{equation}
L_{\sigma} = \sqrt{\frac{t_{U}}{\sigma}},
\label{diffscale}
\end{equation} 
leading to $L_{\sigma } \sim {\rm AU}$. The scale (\ref{diffscale}) gives then the 
upper limit on the diffusion scale for a magnetic field whose lifetime is 
comparable with the age of the Universe at the present epoch. Magnetic fields with typical 
correlation scale larger than $L_{\sigma}$ are not affected by resistivity. On the other 
hand, magnetic fields with typical correlation scale $ L< L_{\sigma}$ are diffused. The
value $L_{\sigma} \sim {\rm AU}$ is consistent with the phenomenological 
evidence that there are no magnetic fields coherent over scales smaller than $10^{-5}$ pc.

The dynamo term may  be responsible for the origin of the magnetic field of the galaxy. 
The galaxy has a typical rotation period of 
$3 \times 10^{8}$ yrs and comparing this figure with the typical age of the galaxy, ${\cal O}(10^{10} {\rm yrs})$, 
it can be appreciated that the galaxy performed about $30$ rotations since the time 
of the protogalactic collapse. 

The effectiveness of the dynamo action depends on the physical 
properties of the bulk velocity field.
In particular, a necessary requirement to have a potentially successful dynamo action is that
the velocity field is non-mirror-symmetric or that, in other words, 
$\langle \vec{v} \cdot \vec{\nabla}\times \vec{v} \rangle \neq 0$.
Let us see how this statement can be made reasonable in the framework of 
Eq. (\ref{magndiff}). 
From Eq.  (\ref{magndiff}) the usual structure of the dynamo term may be derived by carefully averaging
over the velocity filed according to the procedure of \cite{vains,matt}.
By assuming that the motion of the  fluid is random and with zero mean
velocity the average is taken over the ensemble of the possible
velocity fields.
In more physical terms this averaging procedure of Eq. (\ref{magndiff}) is
equivalent to average over scales and times exceeding the
characteristic correlation scale and time $\tau_{0}$ of the velocity
field. This procedure assumes that the correlation scale of the
magnetic field is much bigger than the correlation scale of the
velocity field which is required to be divergence-less
(${\vec{\nabla}}\cdot \vec{v}=0$).
In this approximation the magnetic diffusivity equation can be written
as:
\begin{equation}
\frac{\partial\vec{B}}{\partial t} =
\alpha(\vec{\nabla}\times\vec{B}) +
\frac{1}{4\pi\sigma}\nabla^2\vec{B} ,
\label{dynamored}
\end{equation}
where 
\begin{equation}
\alpha 
= -\frac{\tau_{0}}{3}\langle\vec{v}\cdot\vec{\nabla}
\times\vec{v}\rangle,
\label{alphafirst}
\end{equation}
 is the so-called $\alpha$-term
in the absence of vorticity. In Eqs. (\ref{dynamored})--(\ref{alphafirst}) $\vec{B}$ is
the magnetic field averaged 
over times longer that $\tau_{0}$ which is the typical correlation
time of the velocity field. 

The fact that the velocity field must be 
globally non-mirror
symmetric \cite{zeldovich} suggests, already at this qualitative level, 
the deep connection between dynamo action and fully developed turbulence. 
In fact, if the system would 
be, globally, invariant under parity transformations, then, the $\alpha$ term would 
simply be vanishing. This observation may also be  related to the turbulent features 
of cosmic systems. In cosmic turbulence the systems are 
usually rotating and, moreover, they possess a gradient in the 
matter density (think, for instance, to the case of the galaxy). It is then 
plausible that parity is broken at the level of the galaxy since terms 
like $ \vec{\nabla} \rho_{\rm m} \cdot \vec{\nabla} \times \vec{v}$ 
are not vanishing \cite{zeldovich}.

The dynamo term, as it appears in Eq. (\ref{dynamored}),
 has a simple electrodynamical meaning,
namely, it can be interpreted as a mean ohmic current directed along
the magnetic field :
\begin{equation}
\vec{J} = - \alpha \vec{B}.
\label{vort}
\end{equation}
Equation stipulates that an ensemble of screw-like vortices with
zero mean helicity is able to generate loops in the magnetic flux
tubes in a plane orthogonal to the one of the original field.
As a simple (and known) application of Eq. (\ref{dynamored}), it is appropriate 
to consider the case where the magnetic field 
profile is given by a sort of Chern-Simons wave
\begin{equation}
B_{x}(z,t) = f(t) \sin{k z}, ~~~~B_{y}=f(t) \cos{k z},~~~~ B_{z}(k,t) =0.  
\label{knot1}
\end{equation}
For this profile the magnetic gyrotropy is non-vanishing, i.e. 
$\vec{B}\cdot\vec{\nabla}\times\vec{B} = k f^2(t)$. From Eq. (\ref{dynamored}), using Eq. (\ref{knot1}) 
$f(t)$ obeys the following equation
\begin{equation}
\frac{d f}{dt} = \biggl( k \alpha - \frac{k^2}{4\pi\sigma}\biggr) f
\label{f1}
\end{equation}
admits exponentially growing solutions for sufficiently large scales, i.e. $k < 4\pi |\alpha| \sigma$.
Notice that in this naive example the $\alpha$ term is assumed to be constant. However, as the amplification proceeds, 
$\alpha$ may develop a dependence upon $|\vec{B}|^2$, i.e. $\alpha \to \alpha_0 ( 1 - \xi |\vec{B}|^2) \alpha_0 [ 1 - \xi f^2(t)]$. In the case 
of Eq. (\ref{f1}) this modification will introduce non-linear terms  whose effect will be to stop the growth 
of the magnetic field. 
This regime is often called {\em saturation of the dynamo} and the non-linear equations 
appearing in this context are sometimes 
called Landau equations \cite{zeldovich} in analogy with the Landau equations appearing in hydrodynamical 
turbulence.

In spite of the fact that in the previous example  the velocity field has been averaged, its evolution 
obeys  the Navier-Stokes equation which we have already written but without the diffusion term
\begin{equation}
\rho_{\rm m}\biggl[ \frac{\partial \vec{v}}{\partial t} + (\vec{v}\cdot \vec{\nabla}) \vec{v} - \nu \nabla^2 \vec{v}\biggr] = - \vec{\nabla}P + 
\vec{J}\times \vec{B},
\label{NS1}
\end{equation}
where $\nu$ is the thermal viscosity coefficient. There are idealized cases where the Lorentz force term can be 
neglected. This 
is the so-called  force free approximation. Defining the kinetic helicity as 
$ \vec{\Omega} = \vec{\nabla} \times \vec{v}$, 
the magnetic diffusivity and Navier-Stokes equations can be written in a rather simple and symmetric form 
\begin{eqnarray}
&& \frac{\partial \vec{B}}{\partial t} = \vec{\nabla} \times (\vec{v} \times \vec{B}) + \frac{1}{4\pi \sigma} \nabla^2 \vec{B},
\nonumber\\
&& \frac{\partial \vec{\Omega}}{\partial t} = \vec{\nabla} \times (\vec{v} \times \vec{\Omega}) + \nu \nabla^2 \vec{\Omega}.
\label{symm}
\end{eqnarray}

In MHD various dimensionless ratios can be defined. The most frequently used are 
the magnetic Reynolds number, the kinetic Reynolds number and the Prandtl number:
\begin{eqnarray}
&& {\rm R}_{\rm m} = v L_{B} \sigma,
\label{magnrey}\\
&& {\rm R}= \frac{v L_{v}}{\nu},
\label{rey}\\
&& {\rm Pr} = \frac{ {\rm R}_{\rm m}}{{\rm R}} = \nu\sigma \biggl(\frac{L_{B}}{L_{v}}\biggr),
\label{Pr}
\end{eqnarray}
where $L_{B}$ and $L_{v}$ are the typical scales of variation of the magnetic and velocity fields. 
If ${\rm R}_{\rm m} \gg 1$ the system is said to be {\em magnetically} turbulent. If ${\rm R} \gg 1 $ 
the system is said to be {\em kinetically} turbulent. In realistic situations the plasma is both 
kinetically and magnetically turbulent and, therefore, the ratio of the two Reynolds numbers 
will tell which is the dominant source of turbulence. There have been, in recent years, various 
studies on the development of magnetized turbulence (see, for instance, \cite{biskamp}) whose 
features differ slightly from the ones of hydrodynamic turbulence. While the details of this discussion 
will be left aside, it is relevant to mention that, in the early Universe, turbulence may develop. 
In this situation a typical phenomenon, called inverse cascade, can take place. A direct 
cascade is a process where energy is transferred from large to small scales. 
Even more interesting, for the purposes of the present lecture, is the opposite 
process, namely the inverse cascade where the energy transfer goes from small to large length-scales.
One can also generalize the 
the concept of energy cascade to the cascade of any conserved quantity in the plasma, like, for instance, 
the helicity. Thus, in general terms, the transfer process of a conserved 
quantity is a cascade. 

The concept of cascade (either direct or inverse) is related with the concept 
of turbulence, i.e. the class of phenomena taking place in fluids and plasmas
at high Reynolds numbers.
It is very difficult to reach, with terrestrial plasmas, the physical situation
where the magnetic  and the kinetic Reynolds numbers are both large 
but, in such a way that their ratio is also large i.e. 
\begin{equation}
{\rm R}_{\rm m} \gg 1, ~~~~~{\rm R}\gg1, ~~~~~~ {\rm Pr} = \frac{{\rm R}_{\rm m}}{{\rm R}} \gg 1.
\label{ratio}
\end{equation}
The physical regime expressed through Eqs. (\ref{ratio}) rather common 
in the early Universe. Thus, MHD turbulence is probably one of the key aspects 
of magnetized plasma dynamics at very high temperatures and densities.
Consider, for instance, the plasma at the electroweak 
epoch when the temperature was of the order of $100$ GeV. One can compute the Reynolds 
numbers and the Prandtl number from their definitions given in Eqs. (\ref{magnrey})--(\ref{Pr}).
In particular, 
\begin{equation}
{\rm R}_{\rm m} \sim 10^{17}, ~~~~~~~~~~{\rm  R} = 10^{11}, ~~~~~~{\rm  Pr} \simeq 10^{6},
\label{reyew}
\end{equation}
which can be obtained from  Eqs. (\ref{magnrey})--(\ref{Pr}) using as fiducial 
parameters $ v \simeq 0.1$, $\sigma T/\alpha$, $\nu\simeq ( \alpha T)^{-1}$ and $L \simeq 0.01 ~H_{\rm ew}^{-1} 
\simeq 0.03~{\rm cm}$ for $T \simeq 100 ~{\rm GeV}$.  

If an inverse energy cascade takes place, many (energetic) magnetic domains coalesce giving rise 
to a magnetic domain of larger size but of smaller energy. This phenomenon can be 
viewed, in more quantitative terms, as an effective increase of the correlation scale of the magnetic field.
This consideration plays a crucial r\^ole for the viability of mechanisms where the magnetic field
is produced in the early Universe inside the Hubble radius (see Subsect. \ref{subsect25}).

\subsection{Initial conditions for dynamos}
\label{subsect23}
According to the qualitative description of the dynamo instability presented in the previous 
subsection, the origin of large-scale  magnetic fields in spiral galaxies can be reduced to the 
three keywords: {\em seeding}, {\em amplification} and 
{\em ordering}. The first stage, i.e. the seeding, is the most controversial one and will be briefly reviewed in the following sections 
of the present review. In more quantitative terms the amplification and the ordering may be summarized as follows:
\begin{itemize}
\item{} during the $30$ rotations performed by the galaxy 
since the protogalactic collapse, the magnetic field should be amplified 
by about $30$ e-folds;
\item{}  if the large scale magnetic field of the 
galaxy is, today, ${\cal O}(\mu {\rm G})$ the magnetic field 
at the onset of galactic rotation might have been even $30$ e-folds smaller, i.e. 
${\cal O}(10^{-19} {\rm G})$ over a typical scale of $30$--$100$ kpc.;
\item{} assuming perfect flux freezing during the gravitational 
collapse of the protogalaxy (i.e. $\sigma \to \infty$) the 
magnetic field at the onset of gravitational collapse should 
be  ${\cal O}(10^{-23})$ G over a typical scale of 1 Mpc.
\end{itemize}
This picture is  oversimplified and each of the three steps mentioned above 
can be questioned. In what follows the main sources of debate, emerged 
in the last ten years, will be briefly discussed. 

There is a simple way to relate the value of the magnetic fields 
right after gravitational collapse to the value of the magnetic field 
right before gravitational collapse. Since the gravitational collapse 
occurs at high conductivity the magnetic flux and the magnetic helicity
are both conserved (see, in particular, Eq. (\ref{fluxcons})). 
Right before the formation of the galaxy a patch 
of matter of roughly $1$ Mpc collapses by gravitational 
instability. Right {\em before} the collapse the mean energy density  
of the patch, stored in matter, 
 is of the order of the critical density of the Universe. 
Right {\em after} collapse the mean matter density of the protogalaxy
is, approximately, six orders of magnitude larger than the critical density.

Since the physical size of the patch decreases from $1$ Mpc to 
$30$ kpc the magnetic field increases, because of flux conservation, 
of a factor $(\rho_{\rm a}/\rho_{\rm b})^{2/3} \sim 10^{4}$ 
where $\rho_{\rm a}$ and $\rho_{\rm b}$ are, respectively the energy densities 
right after and right before gravitational collapse. The 
correct initial condition in order to turn on the dynamo instability
would be $|\vec{B}| \sim 10^{-23}$ Gauss over a scale of $1$ Mpc, right before 
gravitational collapse. 

The estimates presented in the last paragraph are
 based on the (rather questionable) assumption that the amplification 
occurs over thirty e-folds while the magnetic flux is 
completely frozen in. In the real situation, the 
achievable amplification is much smaller. Typically a good 
seed would not be $10^{-19}$ G after collapse (as we assumed for 
the simplicity of the discussion) but rather \cite{kulsrud1}
\begin{equation}
|\vec{B}| \geq  10^{-13} {\rm G}. 
\label{dynreq}
\end{equation}

The galactic rotation period is of the order of $3\times 10^{8}$ yrs. This 
scale should be compared with the typical age of the galaxy.
 All along this  rather large dynamical time-scale the effort has been directed, from the fifties,
to the justification that a substantial portion of the kinetic energy of the system 
(provided by the differential rotation) may be converted into magnetic 
energy amplifying, in this way, the seed field up to the observed value 
of the magnetic field, for instance in galaxies and in clusters. 
In recent years a lot of progress has been made both in the context 
of the small and large-scale dynamos \cite{lazarian,brandenburg} 
(see also \cite{bs1,bs2,bs3}). This progress was also driven 
by the higher resolution of the numerical simulations and by the 
improvement in the understanding of the largest magnetized 
system that is rather close to us, i.e. the sun \cite{brandenburg}. 
More complete 
accounts of this progress can be found in the second 
paper of Ref. \cite{lazarian} and, 
more comprehensively, in Ref. \cite{brandenburg}. 
Apart from the aspects involving solar physics and numerical 
analysis, better physical understanding of the r\^ole of the 
magnetic helicity in the dynamo action has been reached. This 
point is crucially connected with the two conservation laws 
arising in MHD, i.e. the magnetic flux and magnetic helicity 
conservations whose relevance has been already emphasized, 
respectively, in Eqs. (\ref{fluxcons}) and (\ref{helcons}).
Even if the rich interplay between small and large scale dynamos 
is rather important, let us focus on the problem of large-scale 
dynamo action that is, at least superficially, more central for the 
considerations developed in the present lecture. 

Already at a qualitative level it is clear that there is a clash between 
the absence of mirror-symmetry of the plasma, the quasi-exponential 
amplification of the seed and the conservation of magnetic flux and 
helicity in the high (or more precisely infinite) conductivity limit. 
The easiest clash to understand, intuitively, is the flux 
conservation versus the exponential amplification: both flux freezing 
and exponential amplification have to take place in the 
{\em same} superconductive (i.e. $\sigma^{-1} \to 0$) limit. 
The clash between helicity conservation and dynamo action 
can be also understood in general terms: the dynamo action implies 
a topology change of the configuration
since  the magnetic flux lines cross each other constantly \cite{lazarian}.

One of the recent progress in this framework  is a more consistent 
formulation of the large-scale dynamo problem
\cite{lazarian,brandenburg}: 
 large scale dynamos produces small scale helical fields that quench 
 (i.e. prematurely saturate) the $\alpha$ effect.  In other words, the 
 conservation of the magnetic helicity can be seen, according to the 
 recent view, as a fundamental constraint on the dynamo action.
 In connection with the last point, it should be mentioned that, 
 in the past, 
 a rather different argument was suggested \cite{kulsrud2}: 
 it was argued that the dynamo action not only leads to the amplification of the large-scale field but also of the random field component. The random field would then 
 suppress strongly the dynamo action. 
 According to the considerations 
 based on the conservation of the magnetic helicity this argument seems to be incorrect since the increase of the random component would also entail and increase of the rate of the topology change, i.e. 
 a magnetic helicity non-conservation.

The possible applications of dynamo mechanism to  clusters is still
under debate and it seems more problematic.  
The typical scale of the gravitational collapse of a cluster 
is larger (roughly by one order of magnitude) than the scale of gravitational
collapse of the protogalaxy. Furthermore, the mean mass density 
within the Abell radius ( $\simeq 1.5 h^{-1} $ Mpc) is roughly 
$10^{3}$ larger than the critical density. Consequently, clusters 
rotate much less than galaxies. Recall that clusters are 
formed from peaks in the density field. The present overdensity 
of clusters is of the order of $10^{3}$. Thus, in order to get 
the intra-cluster magnetic field, one could think that 
magnetic flux is exactly conserved and, then, from an intergalactic 
magnetic field $|\vec{B}| > 10^{-9}$ G  an intra cluster magnetic field
$|\vec{B}| > 10^{-7}$ G can be generated. This simple estimate 
shows why it is rather important to improve the accuracy of magnetic 
field measurements in the intra-cluster medium: the 
change of a single order of magnitude in the estimated magnetic field 
may imply rather different conclusions for its origin.

\subsection{Astrophysical mechanisms}
\label{subsect24}
Many (if not all) the astrophysical mechanisms proposed so far
are related to what is called, in the jargon, a {\em battery}. 
In short, the idea is the following. 
The explicit form of the generalized Ohmic electric field 
in the presence of thermoelectric corrections can be written 
as in Eq. (\ref{OHM}) where we set $n_{\rm q} = n_{\rm e}$ to stick to the 
usual conventions\footnote{For simplicity, we shall neglect 
the Hall contribution arising in the generalized Ohm law. The Hall 
contribution would produce, in Eq. (\ref{thermo}) a term $\vec{J}\times \vec{B}/
n_{\rm e} e$ that is of higher order in the magnetic field and that is 
proportional to the Lorentz force. The Hall term will play no r\^ole 
in the subsequent considerations. However, it should be borne in mind 
that the Hall contribution may be rather interesting in connection 
with the presence of strong magnetic fields like the ones 
of neutron stars (i.e. $10^{13}$ G). This occurrence is even more 
interesting since in the outer regions of neutron stars strong density 
gradients are expected.}
\begin{equation}
\vec{E} = - \vec{v} \times \vec{B} + 
\frac{\vec{\nabla}\times \vec{B}}{4\pi \sigma} - 
\frac{\vec{\nabla}P_{\rm e}}{ e n_{\rm e}}.
\label{thermo}
\end{equation}
By comparing Eq. (\ref{ohmel}) with Eq. (\ref{thermo}), it is clear 
that the additional term at the right hand side, receives contribution
from a temperature gradient. In fact, restoring for a moment the Boltzmann
constant $k_{B}$ we have that since  $P_{\rm e} = k_{B} \, n_{\rm e} \, T_{\rm e}$, the additional term depends upon the gradients of the temperature, hence 
the name thermoelectric. It is interesting to see under which conditions 
the curl of the electric field receives contribution from the thermoelectric 
effect. Taking the curl of both sides of Eq. (\ref{thermo}) 
we obtain 
\begin{equation}
\vec{\nabla}\times \vec{E} = \frac{1}{4\pi \sigma} \nabla^2 \vec{B} + 
\vec{\nabla}(\vec{v} \times \vec{B}) - \frac{\vec{\nabla}n_{\rm e} \times 
\vec{\nabla} P_{\rm e}}{e n_{\rm e}^2}  = - \frac{\partial \vec{B}}{\partial t},
\label{diffusivity}
\end{equation}
where the second equality is a consequence of Maxwell's equations.
From Eq. (\ref{diffusivity}) it is clear that the evolution of the magnetic field 
inherits a source term iff  the gradients in the pressure and electron 
density are not parallel. If $\vec{\nabla} P_{\rm e} \parallel \vec{\nabla} n_{\rm e}$ a fully valid solution of Eq. (\ref{diffusivity}) is $\vec{B}=0$. In the opposite 
case a seed magnetic field is naturally provided by the thermoelectric  term.
The usual (and rather general) observation that one can make in connection with the geometrical properties of the thermoelectric term
is that cosmic ionization fronts may play an important r\^ole. 
For instance, when quasars emit ultraviolet photons, cosmic 
ionization fronts are produced. Then the intergalactic medium 
may be ionized. It should also be recalled, however, that the temperature 
gradients are usually normal to the ionization front. In spite of this, it is 
also plausible to think that density gradients can arise in arbitrary 
directions due to the stochastic nature of density fluctuations.

In one way or in another, astrophysical mechanisms for the generation 
of magnetic fields use an incarnation of the thermoelectric effect 
\cite{rees1} (see also \cite{subr1,zweibel}). 
In the sixties and seventies, for instance, it was rather popular to think 
that the correct ``geometrical" properties of the thermoelectric term
may be provided by a large-scale vorticity. As it will also be discussed 
later, this assumption seems to be, at least naively, in contradiction
with the formulation of inflationary models whose prediction would actually 
be that the large-scale vector modes are completely washed-out by 
the expansion of the Universe. Indeed, all along the eighties and nineties 
the idea of primordial vorticity received just a minor attention. 

The attention then focused on the possibility that objects of rather 
small size may provide intense seeds. After all we do know that 
these objects may exist. For instance the Crab nebula has 
a typical size of a roughly 1 pc and a magnetic field that is a 
fraction of the m G. These seeds will then combine and diffuse 
leading, ultimately, to a weaker seed but with large correlation scale.
This aspect, may be, physically, a bit controversial since we do
observe magnetic fields in galaxies and clusters that are ordered over very 
large length scales. It would then seem necessary  that the seed fields 
produced in a small object (or in several small  objects) undergo 
some type of dynamical self-organization whose final effect
is a seed coherent over length-scales 4 or 5 orders of magnitude 
larger than the correlation scale of the original battery.

An interesting idea could be that qualitatively different 
batteries lead to some type of conspiracy that may produce a strong 
large scale seed. In \cite{rees1} it has been suggested that Population III
stars may become magnetized thanks to a battery operating at 
stellar scale. Then if these stars would explode as supernovae (or if they 
would eject a magnetized stellar wind) the pre-galactic environment 
may be magnetized and the remnants of the process incorporated 
in the galactic disc. In a complementary perspective, a similar 
chain of events may take place over a different physical scale. A battery 
could arise, in fact in active galactic nuclei at high red-shift. Then 
the magnetic field could be ejected leading to intense fields in the lobes 
of ``young" radio-galaxies. These fields will be somehow inherited by 
the ``older" disc galaxies and the final seed field may be, according 
to \cite{rees1} as large as $10^{-9}$ G at the pre-galactic stage.

 In summary we can therefore say that:
 \begin{itemize}
 \item{} both the primordial and the astrophysical hypothesis 
 for the origin of the seeds demand an efficient (large-scale) 
 dynamo action;
 \item{} due to the constraints arising from the conservation 
 of magnetic helicity and magnetic flux the values of the 
 required seed fields may turn out to be larger than previously 
 thought at least in the case when the amplification is only driven 
 by a large-scale dynamo action \footnote{The situation may change if 
 the magnetic fields originate from the combined action  of small
 and large scale dynamos like in the case of the two-step 
 process described in \cite{rees1}.};
 \item{} magnetic flux conservation during gravitational collapse 
 of the protogalaxy 
 may increase, by compressional amplification, the initial seed 
 of even 4 orders of magnitude;
 \item{} compressional amplification, as well as large-scale dynamo, 
 are much less effective in clusters: therefore, the magnetic field
 of clusters is probably connected to the specific way the dynamo saturates, and, in this sense, harder to predict from a specific value of the initial 
 seed.
 \end{itemize}

\subsection{Magnetogenesis: inside the Hubble radius}
\label{subsect25}
One of the weaknesses of the astrophysical hypothesis 
is connected with the smallness of the correlation scale 
of the obtained magnetic fields. This type of impasse
led the community to consider the option that the initial 
conditions for the MHD evolution are dictated not by astrophysics 
but rather by cosmology. The first ones to think about cosmology 
as a possible source of large-scale magnetization were 
Zeldovich \cite{zel1,zel2}, and Harrison \cite{harrison1,harrison2,harrison3}.

The emphasis of these two authors was clearly different. While Zeldovich 
thought about a magnetic field which is {\em uniform} (i.e. homogeneous 
and oriented, for instance, along a specific Cartesian direction) 
Harrison somehow anticipated the more modern view by considering 
the possibility of an {\em inhomogeneous} magnetic field.
In the scenario of Zeldovich the uniform magnetic field would induce a slight 
anisotropy in the expansion rate along which the magnetic field is aligned. 
So, for instance, by considering a constant (and uniform) magnetic field pointing along the $\hat{x}$ 
Cartesian axis, the induced geometry compatible with such a configuration 
will fall into the Bianchi-I class
\begin{equation}
ds^2 = dt^2 - a^2(t) d x^2 - b^2(t)[ dy^2 + dz^2].
\label{bianchim}
\end{equation}
By solving Einstein equations in this background geometry it turns out that, during a radiation 
dominated epoch, the expansion rates along the $\hat{x}$ and the $\hat{y}-\hat{z}$ plane 
change and their difference is proportional to the magnetic energy density \cite{zel1,zel2}.
This observation is not only relevant for magnetogenesis but also for Cosmic Microwave
Background (CMB) anisotropies since the difference in the expansion rate turns out to be proportional 
to the temperature anisotropy. While we will get back to this point later, in Section \ref{sec4}, 
as far as magnetization is concerned we can just remark that the idea of Zeldovich was that 
a uniform magnetic field would modify the initial condition of the standard hot big bang model 
where the Universe would start its evolution already in  a radiation-dominated phase.

The model of Harrison \cite{harrison1,harrison2, harrison3} is, in a sense, more 
{\em dynamical}.  Following earlier work of Biermann \cite{biermann},
Harrison thought that inhomogeneous MHD equations could be used to gennerate 
large-scale magnetic fields {\em provided} the velocity field was turbulent enough.
The Biermann battery was simply a battery (as the ones described above in this session) but
operating prior to decoupling of matter and radiation. The idea of Harrison 
was instead that vorticity was already present so that the effective MHD equations 
will take the form 
\begin{equation}
\frac{\partial}{\partial \tau} ( a^2 \vec{\Omega} + \frac{e}{m_{\rm p}} \vec{B}) = \frac{e}{4\pi \sigma m_{\rm p}} \nabla^2\vec{B},
\label{harrison2}
\end{equation}
where, as previously defined, $\vec{\Omega} = \vec{\nabla} \times \vec{v}$ and $m_{\rm p}$ is the ion mass.  Equation 
(\ref{harrison2}) is written in a conformally flat Friedmann-Robertson-Walker metric of the form 
\begin{equation}
ds^2 = G_{\mu\nu} dx^{\mu} dx^{\nu} = a^2(\tau) [ d\tau^2 - d\vec{x}^2], 
\label{FRW}
\end{equation}
where $\tau$ is the conformal time coordinate and 
where, in the  conformally flat case, $G_{\mu\nu} = a^2(\tau)\eta_{\mu\nu}$ , $\eta_{\mu\nu}$ being the four-dimensional 
Minkowski metric.
If we now postulate that some vorticity was present prior to decoupling, then Eq. (\ref{harrison2}) 
can be solved  and the magnetic field can be related to the initial vorticity as 
\begin{equation}
\vec{B} \sim - \frac{m_{\rm p}}{e} \vec{\omega}_{\rm i} \biggl(\frac{a_i}{a}\biggr)^2.
\label{harrison3}
\end{equation}  

If the estimate of the vorticity is made prior to equality (as originally 
suggested by Harrison\cite{harrison1}) of after decoupling as also suggested, a bit later, in Ref. \cite{mishustin},
the result can change even by two orders of magnitude. 
Prior to equality $|\vec{\Omega}(t) \simeq 0.1/t$ and, therefore,
$|\vec{B}_{\rm eq}| \sim 10^{-21}$G. If a similar estimate is made after decoupling the typical value of the generated magnetic field is of the order of $10^{-18}$ G. So, in this context, the 
problem of the origin of magnetic fields is circumvented by postulating an appropriate 
form of vorticity whose origin must be explained.

The Harrison mechanism is just one of the first examples of magnetic field generation 
{\em inside the Hubble radius}. In cosmology we define the Hubble radius as the inverse 
of the Hubble parameter, i.e. $r_{H} = H^{-1}(t)$. 
The first possibility we can think of implies that magnetic fields 
are produced, at a given epoch in the life of the Universe,
inside the Hubble radius, for instance by a phase transition 
or by any other phenomenon able to generate a charge separation and, 
ultimately, an electric current.  In this context, the 
correlation scale of the field is much smaller that the typical scale 
of the gravitational collapse of the proto-galaxy which is 
of the order of the ${\rm Mpc}$. 
In fact, if the Universe is decelerating and if 
the correlation scale evolves as 
the scale factor, the Hubble radius grows  much faster 
than the correlation scale. 
Of course, one might invoke the possibility that the correlation 
scale of the magnetic field evolves more rapidly than the scale 
factor. A well founded physical rationale for this occurrence 
is what is normally called inverse cascade, i.e. the possibility 
that  magnetic (as well as kinetic) energy density is transferred 
from small to large scales. This implies, in real space, that 
 (highly energetic) small scale magnetic domains may coalesce
 to form magnetic domains of smaller energy but over larger 
 scales.  
 In the best of all possible situations, i.e. when inverse cascade 
 is very effective, it seems rather hard 
 to justify a growth of the correlation scale that would eventually 
 end up into a ${\rm Mpc}$ scale at the onset of gravitational collapse.

\begin{figure}
\centering
\includegraphics[height=5cm]{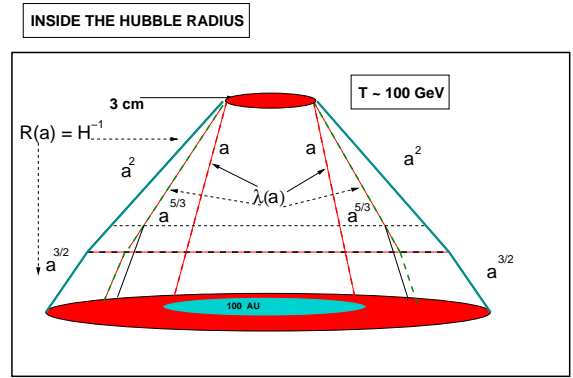}
\caption{Evolution of the correlation scale 
for magnetic fields produced inside the Hubble radius. The horizontal thick dashed line marks the end of the radiation-dominated phase and the onset of the matter-dominated phase. The horizontal thin dashed line marks 
the moment of $e^{+}$--$e^{-}$ annihilation (see also footnoote 2). 
The full (vertical) lines represent the evolution of the Hubble radius 
during the different stages of the life of the Universe. The dashed (vertical) lines 
illustrate the evolution of the correlation scale of the magnetic fields. 
In the absence of inverse cascade the evolution of the correlation scale is given 
by the (inner) vertical dashed lines. If inverse cascade takes place the evolution 
of the correlation scale is faster than the first power of the scale factor (for instance
$a^{5/3}$) but always slower than the Hubble radius.}
\label{FIG1}      
\end{figure}

In Fig. \ref{FIG1} we report a schematic illustration of the evolution 
of the Hubble radius $R_{H}$ and of the correlation scale of 
the magnetic field as a function of the scale factor.  In Fig. \ref{FIG1} the 
horizontal dashed line simply marks the end of the radiation-dominated 
phase and the onset of the matter dominated phase: while above the dashed line 
the Hubble radius evolves as $a^2$ (where $a$ is the scale factor), below the dashed 
line the Hubble radius evolves as $a^{3/2}$. 

We consider, for 
simplicity, a magnetic field whose typical correlation scale is as 
large as the Hubble radius at the electro-weak epoch when the 
temperature of the plasma was of the order of $100\,\, {\rm GeV}$.
This is roughly the regime contemplated by the considerations 
presented around Eq. (\ref{reyew}).
If the correlation scale evolves as the scale factor, the Hubble radius 
at the electroweak epoch (roughly $3$ cm) projects today over a scale 
of the order of the astronomical unit. If inverse cascades are invoked, the 
correlation scale may grow, depending on the specific features 
of the cascade, up to $100$ A.U.  or even up to $100$ pc. In both
 cases the final scale is too small if compared with the typical 
scale of the gravitational collapse of the proto-galaxy. In Fig. \ref{FIG1} 
a particular model for the evolution of the correlation scale $\lambda(a)$ has 
been reported \footnote{Notice, as it will be discussed later, that the inverse 
cascade lasts, in principle, only down to the time of $e^{+}-e^{-}$ 
annihilation (see also thin dashed horizontal line in Fig. \ref{FIG1}) since for temperatures smaller than $T_{e^{+}-e^{-}}$ the 
Reynolds number drops below 1. This is the result of the sudden drop in the number of charged particles that leads to a rather long mean free path for 
the photons. }.

\subsection{Inflationary magnetogenesis}
\label{subsect26}
If magnetogenesis takes place inside the Hubble radius the main problem is therefore the correlation scale 
of the obtained seed field. The cure for this problem is  to look for a mechanism producing 
magnetic fields that are coherent over large-scales (i.e. Mpc and, in principle, even larger). 
This possibility may arise in the context of inflationary models. 
Inflationary models may be conventional (i.e. based on a quasi-de Sitter stage of expansion)
or unconventional (i.e. not based on a quasi-de Sitter stage of expansion). Unconventional inflationary models are, for instance, pre-big bang 
models that will be discussed in more depth in Section \ref{sec3}. 

The rationale for the previous statement is that, in inflationary models, the zero-point 
(vacuum) fluctuations of fields of various spin are amplified. Typically fluctuations 
of spin 0 and spin 2 fields. The spin 1 fields enjoy however of a property, called Weyl invariance, 
that seems to forbid the amplification of these fields. While Weyl invariance and its possible 
breaking will be the specific subject of the following subsection, it is useful for the moment 
to look at the kinematical properties by assuming that, indeed, also spin 1 field 
can be amplified.

\begin{figure}
\centering
\includegraphics[height=5cm]{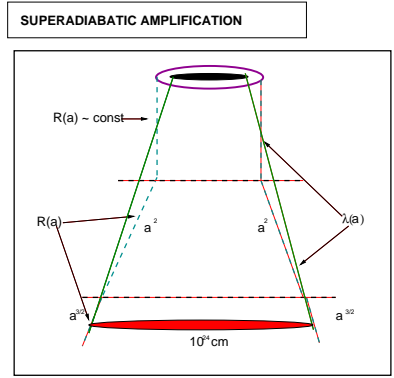}
\caption[a]{Evolution of the correlation scale if magnetic fields would be produced 
 by superadiabatic amplification during a conventional inflationary phase. The dashed 
 vertical lines denote, in the present figure, the evolution of the Hubble 
 radius while the full line denotes the evolution 
 of the correlation scale (typically selected to smaller than the Hubble radius during inflation).}
\label{FIG2} 
\end{figure}

Since during inflation the Hubble 
 radius is roughly constant (see Fig. \ref{FIG2}), the correlation scale evolves much faster than the Hubble radius itself and, therefore, large scale 
 magnetic domains can naturally be obtained. Notice that, in Fig. \ref{FIG2}
 the (vertical) dashed lines illustrate the evolution of the Hubble radius 
 (that is roughly constant during inflation) while the full line denotes 
 the evolution of the correlation scale. Furthermore, the horizontal (dashed) lines 
 mark, from top to bottom, the end of the inflationary phase and the onset of the 
 matter-dominated phase.
 This phenomenon
 can be understood as  the gauge counterpart of the super-adiabatic 
 amplification of the scalar and tensor modes of the geometry.
The main problem, in such a framework, is to 
get large amplitudes  for scale of the order of the ${\rm Mpc}$ at the 
onset of gravitational collapse. Models where the gauge couplings 
are effectively dynamical (breaking, consequently, the Weyl invariance 
of the evolution equations of Abelian gauge modes) 
may provide rather intense magnetic fields.

The two extreme possibilities 
mentioned above may be sometimes combined.
 For instance, it can happen that 
magnetic fields are produced by super-adiabatic amplification of vacuum 
fluctuations during an inflationary stage of expansion. After 
exiting the horizon, the  gauge modes  will reenter at different moments 
all along the radiation and matter dominated epochs.
The spectrum of the primordial gauge fields after reentry 
 will not only be determined by the amplification mechanism but also on the plasma effects. As soon as the magnetic inhomogeneities reenter, some 
other physical process, taking place inside the Hubble radius, may be triggered by the presence of large scale magnetic fields. An example, in this 
context, is the production of topologically non-trivial configurations 
of the hypercharge field (hypermagnetic knots) from a stochastic 
background of hypercharge fields with vanishing helicity 
\cite{hmk1,hmk2,hmk3} (see also \cite{anomaly,bdv1,bdv2,bamba0,angel}).

\subsection{Breaking of conformal invariance}
\label{subsect27}
Consider the action for an Abelian gauge field in four-dimensional curved space-time
\begin{equation}
S_{\rm em} = - \frac{1}{4} \int d^{4} x \sqrt{-G} F_{\mu\nu} F^{\mu\nu}.
\label{action}
\end{equation}
Suppose, also, that the geometry is characterized by a conformally flat 
line element of Friedmann-Robertson-Walker type as the one 
introduced in Eq. (\ref{FRW}). 
The equations of motion derived from Eq. (\ref{action}) 
can be written as 
\begin{equation}
\partial_{\mu} \biggl( \sqrt{-G} F^{\mu\nu} \biggr)=0.
\label{eqomem}
\end{equation}
Using Eq. (\ref{FRW}) and recalling that $\sqrt{-G} = a^4(\tau)$, we will have 
\begin{equation}
\sqrt{-G} F^{\mu\nu} =
 a^4(\tau) \frac{\eta^{\mu\alpha}}{a^2(\tau)} \frac{\eta^{\nu\beta}}{a^2(\tau)}
 F_{\alpha\beta} = F^{\mu\nu}
\label{confinv}
\end{equation}
where the second equality follows from the explicit form of the metric. 
Equation (\ref{confinv}) shows that the evolution equations of Abelian gauge fields are the 
same in flat space-time and in a conformally flat FRW space-time. This property is  
correctly called Weyl invariance or, more ambiguously, conformal invariance.
Weyl invariance is realized also in the case of chiral (massless) 
fermions always in the case of conformally flat space-times.

One of the reasons of the success of inflationary models in making predictions is deeply 
related with the lack of conformal invariance of the evolution equations 
of the fluctuations of the geometry. In particular it can be shown that the tensor modes
of the geometry (spin 2) as well as the scalar modes (spin 0) obey evolution equations that 
are {\em not} conformally invariant. This means that these modes of the geometry 
can be amplified and eventually affect, for instance, the temperature autocorrelations as well
as the polarization power spectra in the microwave sky.

To amplify large-scale magnetic fields, therefore, we would like to break 
conformal invariance. Before considering this possibility, let us discuss an even 
more conservative approach consisting in studying the evolution of Abelian 
gauge fields coupled to another field whose evolution is {\em not} Weyl invariant.
An elegant way to achieve this goal is to couple the action of the hypercharge 
field to the one of a complex scalar field (the Higgs field). The Abelian-Higgs 
model, therefore, leads to the following action
\begin{equation}
S = \int d^{4} x \sqrt{- G}\biggl[ G^{\mu\nu} ({\mathcal D}_{\mu})^{\ast} \phi{\mathcal D}_{\nu}\phi - m^2 \phi^{\ast} \phi  - \frac{1}{4} {\mathcal F}_{\mu\nu} {\mathcal F}^{\mu\nu}\biggr],
\label{abhiggs1}
\end{equation}
where ${\mathcal D}_{\mu} = \partial_{\mu} - i e {\mathcal A}_{\mu}$ and 
${\mathcal F}_{\mu\nu} = \partial_{[\mu} {\mathcal A}_{\nu]}$. 
Using Eq. (\ref{FRW}) into Eq. (\ref{abhiggs1}) and assuming that the complex scalar 
field (as well as the gauge fields) are not a source of the background geometry, the 
canonical action for the normal modes of the system can be written as 
\begin{equation}
S= \int d^{3}x d\tau \biggl[ \eta^{\mu\nu} (D_{\mu} \Phi)^{\ast} D_{\nu} \Phi + 
\biggl( \frac{a''}{a} - m^2 a^2 \biggr) \Phi^{\ast} \Phi - \frac{1}{4} F_{\alpha\beta} F^{\alpha\beta} \biggr],
\label{abhiggs2}
\end{equation}
where $\Phi = a \phi$; $D_{\mu} =\partial_{\mu} - ie A_{\mu}$ and $F_{\mu\nu} = \partial_{[\mu} A_{\nu]}$.
From Eq. (\ref{abhiggs2}) it is clear that also when the Higgs field is massless the 
coupling to the geometry breaks explicitly Weyl invariance. Therefore,
current density and charge density fluctuations will be induced. Then, by 
employing a Vlasov-Landau description similar the resulting magnetic field will be of the order of 
$B_{\rm dec} \sim 10^{-40} T_{\rm dec}^2$  \cite{gioshap} which is, by far, too small to seed 
any observable field even assuming, optimistically, perfect flux freezing 
and maximal efficiency for the dynamo action. The results of \cite{gioshap} disproved 
earlier claims (see \cite{mgs2} for a critical review) neglecting the r\^ole of the 
conductivity in the evolution of large-scale magnetic fields after inflation.

 The first attempts to analyze the Abelian-Higgs model in De Sitter space have been made 
by Turner and Widrow \cite{turnerwidrow} who just listed such a possibility as an open question. These two authors also analyzed different 
scenarios where conformal invariance for spin 1 fields could be broken 
in 4 space-time dimensions.   Their  first suggestion was that 
conformal invariance may be broken, at an effective level, 
through the coupling of photons to the geometry \cite{drummond}. Typically, the 
breaking of conformal invariance occurs through  products 
of gauge-field strengths and curvature tensors, i.e.
\begin{equation} 
\frac{1}{m^2}F_{\mu\nu}F_{\alpha\beta} R^{\mu\nu\alpha\beta},~~~~~~~\frac{1}{m^2} R_{\mu\nu} F^{\mu\beta} F^{\nu\alpha} g_{\alpha\beta},~~~~~~
\frac{1}{m^2} F_{\alpha\beta}F^{\alpha\beta} R
\label{curv}
\end{equation}
where $m$ is the appropriate mass scale; $R_{\mu\nu\alpha\beta}$ and  $R_{\mu\nu}$ are the Riemann and Ricci tensors 
and $R$ is the Ricci scalar. If the evolution of gauge fields is studied during 
phase of de Sitter (or quasi-de Sittter) 
expansion, then the amplification of the vacuum fluctuations induced by the 
couplings listed in Eq. (\ref{curv}) is minute. The price in order to get large amplification 
should be, according to \cite{turnerwidrow}, an explicit breaking of gauge-invariance 
by direct coupling of the vector potential to the Ricci tensor or to the Ricci scalar, i.e. 
\begin{equation}
R A_{\mu} A^{\mu},~~~~~~~~~R_{\mu\nu} A^{\mu} A^{\nu}.
\label{vector}
\end{equation}
In \cite{turnerwidrow} two other different models were proposed (but not scrutinized in detail) 
namely  scalar electrodynamics and the axionic coupling to the Abelian field strength.

Dolgov \cite{dolgov} considered the possible breaking of conformal invariance due 
to the trace anomaly. The idea is that the conformal invariance of 
gauge fields is broken by the triangle diagram where two photons in the external 
lines couple to the graviton through a loop of fermions. The local 
contribution to the effective action 
leads to the vertex $(\sqrt{-g})^{1+ \epsilon} F_{\alpha\beta}F^{\alpha\beta}$ where 
$\epsilon$ is a numerical coefficient depending upon the number of scalars and fermions present 
in the theory. The evolution 
equation for the gauge fields, can be written, in Fourier space, as 
\begin{equation}
{\cal A}_{k}'' + \frac{\epsilon}{8} {\cal H} {\cal A}_{k}' + k^2 {\cal A}_{k} =0,
\end{equation}
and it can be shown that only if $\epsilon >0$ the gauge fields are amplified.
Furthermore, only 
is $\epsilon \sim 8$ substantial amplification of gauge fields is possible.

In a series of papers \cite{carroll1,carroll2,carroll3} the possible 
effect of the axionic coupling to the amplification of gauge fields has been investigated.
The idea is here that conformal invariance is broken through 
the explicit coupling of a pseudo-scalar field to the gauge field (see Section 5), i.e. 
\begin{equation}
\sqrt{-g} c_{\psi\gamma}\alpha_{\rm em} \frac{\psi}{8\pi M} F_{\alpha\beta}\tilde{F}^{\alpha\beta},
\label{pseudo}
\end{equation}
where $\tilde{F}^{\alpha\beta}$ is the dual field strength and where 
$c_{\psi\gamma}$ is a numerical factor of order one. Consider now the case of  a standard 
pseudoscalar potential, for instance $m^2 \psi^2$, evolving in a de Sitter (or quasi-de Sitter space-time).
It can be shown, rather generically, that the vertex given in Eq. (\ref{pseudo}) leads to negligible 
amplification at large length-scales. The coupled system of evolution equations 
to be solved in order to get the amplified field is 
\begin{eqnarray}
&& \vec{B}'' - \nabla^2 \vec{B} - \frac{\alpha_{\rm em}}{2 \pi M} \psi' \vec{\nabla} \times \vec{B} =0,
\label{bpsi}\\
&& \psi'' + 2 {\cal H} \psi' + m^2 a^2 \psi =0,
\label{psib}
\end{eqnarray}
where $\vec{B} = a^2 \vec{B}$.
From Eq. (\ref{bpsi}),  there is a maximally amplified physical frequency 
\begin{equation}
\omega_{\rm max} \simeq \frac{\alpha_{{\rm em}}}{2 \pi M} \dot{\psi}_{\rm max} \simeq  \frac{\alpha_{{\rm em}}}{2 \pi } m
\end{equation}
where the second equality follows from $\psi \sim a^{-3/2} M \cos{m t}$ (i.e. $\dot{\psi}_{\rm max} \sim m M$).
The amplification for $\omega \sim \omega_{\rm max}$ is of the order of $\exp{[m \alpha_{\rm em}/(2 \pi H)]}$ 
where $H$ is the Hubble parameter during the de Sitter phase of expansion. From the above expressions one can 
argue that the modes which are substantially amplifed are the ones for which $\omega_{\rm max} \gg H$. The modes 
interesting for the large-scale magnetic fields are the ones which are in the opposite range, i.e.  $\omega_{\rm max} \ll H$.
Clearly, by lowering the curvature scale of the problem the produced seeds may be larger and the 
conclusions much less pessimistic \cite{carroll3}.

Another interesting idea pointed out by Ratra \cite{ratra} is that the electromagnetic field may be 
directly coupled to the inflaton field. In this case the coupling is specified 
through a parameter $\alpha$, i.e. $e^{\alpha \varphi} F_{\alpha\beta}F^{\alpha\beta}$ where $\varphi$ is the 
inflaton field in Planck units. In order to get 
sizable large-scale magnetic fields the effective gauge coupling must be larger than one during 
inflation (recall that $\varphi$ is large, in Planck units, at the onset of inflation).

In \cite{variation} it has been suggested that the evolution of the Abelian gauge 
coupling during inflation induce the growth of the two-point function of magnetic inhomogeneities.
This model is different from the one previously discussed \cite{ratra}. Here 
the dynamics of the gauge coupling is not related to the dynamics of the 
inflaton which is not coupled to the Abelian field strength. In particular, 
$r_{B}({\rm Mpc}) $ can be as large as $10^{-12}$. In \cite{variation}
the MHD equations have been generalized to the case of evolving gauge coupling.
Recently a scenario similar to \cite{variation} has been discussed in \cite{bamba}.

In the perspective of generating large scale magnetic fields Gasperini \cite{gravphot} 
 suggested to consider the possible mixing between the photon and the 
graviphoton field appearing in supergravity theories (see also, in a related context \cite{okun}). 
The graviphoton is 
the massive vector component of the gravitational supermultiplet and its 
interaction with the photon is specified by an interaction term of the 
type $ \lambda F_{\mu\nu} G^{\mu\nu}$ where $G_{\mu\nu}$ is 
the filed strength of the massive vector. Large-scale magnetic fields 
with $r_{B}({\rm Mpc})\geq 10^{-34}$ can be obtained 
if $\lambda \sim {\cal O}(1)$ and for a mass of the vector $m \sim 10^{2} {\rm TeV}$.

Bertolami and Mota \cite{bertolami} argue that if Lorentz invariance 
is spontaneously broken, then  photons acquire naturally a coupling 
to the geometry which is not gauge-invariant and which is similar to the 
coupling considered in \cite{turnerwidrow}.

\section{Why string cosmology?}
\label{sec3}
The moment has come to review my personal interaction with Gabriele 
Veneziano on the study of large-scale magnetic fields. While we had other 15 
joined papers with Gabriele (together with different combinations of authors) two of them \cite{GAB1,GAB2}
(both in collaboration with Maurizio Gasperini) are directly related to large-scale magnetic fields. Both papers reported in Refs. \cite{GAB1,GAB2}
appeared in 1995 while I was completing my Phd at the theory 
division of CERN. 

My scientific exchange with Gabriele Veneziano started at least four years 
earlier and the first person mentioning Gabriele to me was Sergio Fubini.
At that time Sergio was professor of Theoretical Physics at the University
of Turin and I had the great opportunity of discussing physics 
with him at least twice a month. Sergio was rather intrigued by the possibility 
of getting precise measurements on macroscopic quantum phenomena like 
superfluidity, superconductivity, quantization of the resistivity in the 
(quantum) Hall effect. I started working, under the supervision 
of Maurizio Gasperini, on the spectral properties of relic gravitons and we bumped into the concept of squeezed state \cite{gg}, a generalization of the concept of coherent state (see, for instance, \cite{stoler,luciano,yuen}).
Sergio got very interested and, I think, he was independently thinking about possible applications of squeezed states to superconductivity, a topic that became later on the subject of a paper \cite{mol}.
Sergio even suggested a review by Rodney Loudon \cite{loudon}, an author that I knew already beacuse of his inspiring book on quantum optics \cite{loudon2}. Ref. \cite{loudon} 
together with a physics report of B. L. Schumaker \cite{schumi} was very useful for my understanding of the subject. Nowadays a very complete 
and thorough presentation of the intriguing problems arising in quantum optics
can be found in the book of Leonard Mandel and Emil Wolf \cite{mandel}.

It is amusing to notice the following parallelism between quantum optics 
and the quantum treatment of gravitational fluctuations. While quantum 
optics deals with the coherence properties of systems of many photons, 
we deal, in cosmology, with the coherence properties of many gravitons 
(or phonons) excited during the time-evolution of the background fields.
The background fields act, effectively, as a "pump field". This terminology, now 
generally accepted, is exactly borrowed by quantum optics where the 
pump field is a laser. In the sixties and seventies the main 
problem of optics can be summarized by the following question:
why is {\em classical} optics so precise? Put it into different words, it is 
known that the interference of the amplitudes of the radiation field 
(the so-called Young interferometry)  can be successfully treated at a 
classical level. Quantum effects, in optics, arise not from the first -order 
interference effects (Young interferometry) but from 
the second-order interference effects, i.e. the so-called 
Hanbury-Brown-Twiss interferometry \cite{mandel} where the quantum nature 
of the radiation field is manifest since it leads, in the jargon 
introduced by Mandel \cite{mandel} to light which is 
either bunched or anti-bunched. 
A similar problem also arises in the treatment of cosmological perturbations 
when we ask the question of the classical limit of a quantum mechanically 
generated fluctuation (for instance relic gravitons). 

The interaction with Sergio led, few years later, 
to a talk that I presented at the physics 
department of the University of Torino. The title was {\em Correlation properties of many photons systems}. 
I mentioned my interaction with Sergio Fubini 
since it was Sergio who suggested that, eventually, I should talk to Gabriele 
about squeezed states. 

During the first few months of 1991, Gabriele submitted a seminal paper on the cosmological 
implications of the low-energy string effective action \cite{PBB1}.
This paper, together with another one written in collaboration 
with Maurizio Gasperini \cite{PBB2} represents 
the first formulation of pre-big bang models. A relatively recent introduction to pre-big 
bang models can be found in Ref. \cite{PBB3}. 

In \cite{GAB1,GAB2} it was argued that the string cosmological scenario 
provided by pre-big bang models \cite{PBB1,PBB2} 
would be ideal for the generation of large-scale magnetic fields. The rationale 
for this statement relies on two different observations:
\begin{itemize}
\item{} in the low-energy string effective action gauge fields are 
coupled to the dilaton whose expectation value, at the string 
energy scale, gives the unified value of the gauge and gravitational 
coupling;
\item{} from the mathematical analysis of the problem it is 
clear that to achieve a sizable amplification of large-scale 
magnetic fields it is necessary to have a pretty long phase 
where the gauge coupling is sharply growing in time \cite{GAB1}.
\end{itemize}

Let us therefore elaborate on the two mentioned points.
In the string frame the low-energy string effective action 
can be schematically written as \cite{LEST1,LEST2,LEST3}
\begin{eqnarray}
&& S_{\rm eff} = - \int d^{4} x \sqrt{- G} \biggl[ \frac{e^{- \varphi}}{2\lambda_{\rm s}^2} \biggl( R + G^{\alpha\beta} \partial_{\alpha} \varphi \partial_{\beta} \varphi - \frac{1}{12} H_{\mu\nu\alpha} H^{\mu\nu\alpha} 
\biggr ) 
\nonumber\\
&&+ \frac{e^{-\varphi}}{4} F_{\alpha\beta} F^{\alpha\beta}  
+ 
e^{-\varphi}\overline{\psi}\biggl(\frac{i}{2} 
\gamma^{\mathrm a} D_{\mathrm a} \psi + {\mathrm h.c.}\biggr) + 
{\mathcal R}^2+\, .......\biggr] + {\mathcal O}(g^2)+\,....
\label{LEST}
\end{eqnarray}
In Eq. (\ref{LEST}) the ellipses stand, respectively, for an expansion
in powers of $(\lambda_{\rm s}/L)^2$ and for an expansion
in powers of the gauge coupling constant $g^2 = e^{\varphi}$.
This action is written in the so-called string frame metric where 
the dilaton field $\varphi$ is coupled to the Einstein-Hilbert term.

Concerning the action (\ref{LEST}) few general comments are 
in order 
\begin{itemize}
\item{} the relation between the Planck and string scales
depends on time and, in particular, $ \ell_{\mathrm P}^2 = e^{\varphi} 
\lambda_{\mathrm s}^2$; the present ratio between the Planck and string scales gives the value, i.e. $g(\tau_{0}) = e^{\varphi_{0}/2} = 
\ell_{P} (\tau_0)/\lambda_{\rm s}$;
\item{} in four space-time dimensions the antisymmetric tensor 
field $H^{\mu\nu\alpha}$ can be written in terms of a pseudo-scalar 
field, i.e. 
\begin{equation}
H^{\mu\nu\alpha} = e^{\varphi}
 \frac{\epsilon^{\mu\nu\alpha\rho}}{\sqrt{-G}} \partial_{\rho}\sigma;
 \label{ax}
 \end{equation}
\end{itemize}
In critical superstring theory the dilaton field must have a potential 
that vanishes in the weak coupling limit (i.e. $\varphi \to -\infty$). Moreover, 
from the direct tests of Newton law at short distances it should 
also happen that the mass of the dilaton is such that $ m_{\varphi} > 10^{-4}$.
This requirement may be relaxed by envisaging non-perturbative 
mechanisms  where the dilaton is effectively decoupled from the matter 
fields and where a massless dilaton leads to observable violations 
of the equivalence principle.

From the structure of the action (\ref{LEST}),
Abelian gauge fields are amplified if the gauge coupling is dynamical.
Consider, in fact, the equations of motion for the hypercharge field 
strength 
\begin{equation}
\partial_{\mu}\biggl( e^{-\varphi} \sqrt{-G} F^{\mu\nu}\biggr)=0,
\label{emst}
\end{equation}
where $ F_{\mu\nu} = \partial_{[\mu} A_{\nu]}$. In the 
Coulomb gauge where $A_{0} =0$ and $\vec{\nabla}\cdot \vec{A}=0$
the equation for the rescaled vector potential ${\mathcal  A}_{\mu} = 
e^{\varphi/2} A_{\mu}$ becomes, for each independent polarization and in Fourier space, 
\begin{equation}
{\mathcal A}_{k}'' + \biggl[ k^2 - g \biggl(\frac{1}{g}\biggr)''\biggr]
 {\mathcal A}_{k}=0,
 \label{Ak}
\end{equation}
 where, as usual, the prime denotes a derivation with respect to the 
 conformal time coordinate. In Eq. (\ref{Ak}) $k$ denotes the comoving 
 wave-number From the structure of Eq. (\ref{Ak}) there exist two 
 different regimes.  For $k^2 \gg |g (g^{-1})''|$ the solution 
 off Eq. (\ref{Ak}) is oscillatory. In the opposite limit, i.e. 
 $k^2 \ll |g (g^{-1})''|$ the general solution can be written 
 as 
 \begin{equation}
 {\mathcal A}_{k}(\tau) = \frac{C_{1}(k)}{g(\tau)} + \frac{C_{2}(k)}{g(\tau)} 
 \int^{\tau} g^{2}(\tau') d\tau',
 \label{solampl}
 \end{equation}
 where $C_{1}(k)$ and $C_{2}(k)$ are two arbitrary constants. These 
two constants can be fixed by imposing quantum mechanical initial 
conditions for $\tau \to -\infty$.
Thus, depending on the evolution of $g(\tau)$ the Fourier 
amplitude ${\mathcal A}_{k}$ can be 
amplified.  

It can be shown \cite{GAB1,GAB2} 
that the the amplified magnetic energy density 
depends on the ratio  between the value of the gauge coupling 
at the reentry and at the exit of the typical scale of the gravitational 
collapse, i.e. 
\begin{equation}
r(k) =\frac{1}{\rho_{\gamma}} \frac{d \rho_{\mathrm{B}}}{d\ln{k}}  \simeq \frac{k^4}{ a^4 \rho_{\gamma}} \biggl(\frac{g_{\mathrm re}}{g_{\mathrm ex}}\biggr)^2. 
\label{amplB}
\end{equation}
The parameter $r(k)$ measures the relative weight of the magnetic energy density in units of the 
radiation background. To turn on the galactic dynamo in its simplest realization one 
should require that $r(k_{\mathrm G}) \geq 10^{-34}$ for a typical comoving wave-number
corresponding to the typical scale of the gravitational collapse 
of the protogalaxy. As explained before, this requirement seems to be 
too optimistic in the light of the most recent understanding of the dynamo theory.
The limit $r(k_{\rm G}) \geq 10^{-24}$ seems more reasonable. 

The fact that the gauge coupling must be sharply growing in order to produce large-scale 
magnetic fields fits extremely well with the pre-big bang dynamics where, indeed, the gauge 
coupling is expected to grow. The second requirement to obtain a phenomenologically 
viable mechanism for the amplification of large-scale gauge fields turned out to be the 
existence of a pretty long stringy phase. 

The "stringy" phase is simply the epoch where quadratic curvature corrections 
start being important and lead to an effective dynamics where 
the dilaton field is linearly growing in the cosmic time 
coordinate (see \cite{PBB3} and references therein). Towards the end of the stringy phase the dilaton freezes 
to its (constant) value and the Universe gets dominated 
by radiation. One possibility for achieving the transition to radiation 
is represented by the back-reaction effects of the produced particles  \cite{heating}.
In particular, the short wavelength modes play, in this context 
a crucial r\^ole. It is interesting that while the magnetic energy 
spectrum produced during the stringy phase is 
quasi-flat and the value of $r(k_{\mathrm G}) $ can be as 
large as $10^{-8}$ implying a protogalactic magnetic field of the order of 
$10^{-10}$ G. Under these conditions the dynamo mechanism would even be 
superfluous since the compressional amplification alone 
can amplify the seed field to its observed value.

The results reported above may be ``tested" in a framework where 
the pre-big bang dynamics is solvable. Consider, in particular, the situation where the evolution of the dilaton field as well as the one 
of the geometry is treated in the presence of a non-local dilaton 
potential \cite{G1,G2,meis0,meis1,gmv}. 

\begin{figure}
\centering
\includegraphics[height=5cm]{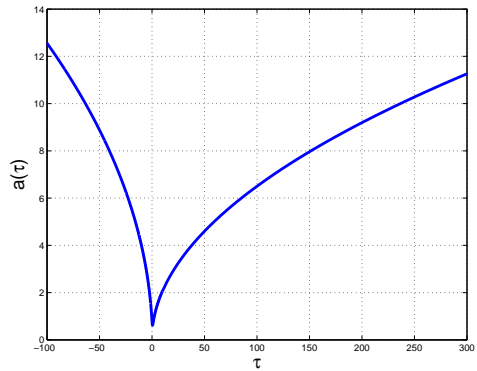}
\caption{The evolution of the scale factor in conformal time for a bouncing model regularized via 
non-local dilaton potential in the Einstein frame.}
\label{FB1}      
\end{figure}

In the Einstein frame description, the 
asymptotics of the (four-dimensional) pre-big bang dynamics  can be written as  \cite{heating} 
\begin{eqnarray}
&& a(\tau) \simeq a_{-} \sqrt{ -\frac{\tau}{2 \tau_0}}, ~~~~~~~~ a_{-} = e^{-\varphi_0/2} \sqrt{\frac{2(\sqrt{3} +1)}{\sqrt{3}}},
\nonumber\\
&& \varphi_{-} = \varphi_0 - \ln{2} - \sqrt{3} \ln{\biggl(\frac{\sqrt{3} +1}{\sqrt{3}}\biggr)} 
- \sqrt{3} \ln{\biggl(- \frac{\tau}{2 \tau_0}\biggr)},
\nonumber\\
&& {\cal H}_{-} = \frac{1}{2\tau}, ~~~~~~~~~~~~~~~\varphi_{-}' = - \frac{\sqrt{3}}{\tau},
\label{solminus}
\end{eqnarray}
 for $\tau \to -\infty$,  and 
\begin{eqnarray}
&& a(\tau) \simeq a_{+} \sqrt{ \frac{\tau}{2 \tau_0}}, ~~~~~~~~~~~~ a_{+} = e^{\varphi_0/2} \sqrt{\frac{2(\sqrt{3} -1)}{\sqrt{3}}}
\nonumber\\
&& \varphi_{+} = \varphi_0 - \ln{2} - \sqrt{3} \ln{\biggl(\frac{\sqrt{3} - 1}{\sqrt{3}}\biggr)}
+ \sqrt{3} \ln{\biggl( \frac{\tau}{2 \eta_0}\biggr)},
\nonumber\\
&&{\cal H}_{+} = \frac{1}{2\tau}, ~~~~~~~~~~~~~~~\varphi_{+}' =  \frac{\sqrt{3}}{\tau},
\label{solplus}
\end{eqnarray}
for $\tau \to +\infty$. In Eqs. (\ref{solminus}) and (\ref{solplus}), ${\mathcal H} = a'/a$ and, as usual, the prime denotes a derivation 
with respect to $\tau$.The branch of the solution denoted by minus 
describes, in the Einstein frame, an accelerated contraction, since the first derivative of the scale 
factor is negative while the second  is positive. The branch of the solution denoted 
with plus describes, in the Einstein frame, a decelerated expansion, since the first derivative of 
the scale factor is positive while the  derivative is negative. In both branches the 
dilaton grows and its derivative is always positive-definite (i.e. $\varphi'_{\pm} >0$ ) as required by the present approach to bouncing solutions.
The numerical solution corresponding to the asymptotics given in 
Eqs. (\ref{solminus}) and (\ref{solplus}) 
is reported in Fig. \ref{FB1}

In the 
Schr\"odinger description the vacuum state evolves, unitarily, to 
a multimode squeezed state, in full analogy with what happens in the case 
of relic gravitons \cite{squeezed1,squeezed2,squeezed3}. 
In the following the same process 
will be discussed within the Heisenberg representation.
The two physical polarizations of the photon can then 
be quantized according to the 
standard rules of quantization in the radiation gauge 
in curved space-times:  
\begin{equation}
\hat{\cal A}_{i}(\vec{x}, \tau) = \sum_{\alpha} \int \frac{d^3 k}{(2\pi)^{3/2}} \biggl[ \hat{a}_{k,\alpha} e^{\alpha}_{i}{\cal A}_{k}(\tau) e^{- i \vec{k}\cdot\vec{x}} +  \hat{a}_{k,\alpha}^{\dagger} e_{i}^{\alpha} {\cal A}_{k}(\tau)^{\star} e^{ i \vec{k}\cdot\vec{x}}\biggr],
\label{ahat}
\end{equation}
and 
\begin{equation}
\hat{\pi}_{i}(\vec{x}, \tau) = \sum_{\alpha} \int \frac{d^3 k}{(2\pi)^{3/2}} \biggl[ \hat{a}_{k,\alpha} e^{\alpha}_{i} \Pi_{k}(\tau) e^{- i \vec{k}\cdot\vec{x}} +  \hat{a}_{k,\alpha}^{\dagger} e_{i}^{\alpha} \Pi_{k}(\tau)^{\star} e^{ i \vec{k}\cdot\vec{x}}\biggr],
\label{pihat}
\end{equation}
where $e_{i}^{\alpha}(k)$ describe the polarizations of the photon and 
\begin{equation}
\Pi_{k}(\tau) = {\cal A}_{k}'(\tau), ~~~~~~~~~ [\hat{a}_{k,\alpha},\hat{a}_{p,\beta}^{\dagger} ] = \delta_{\alpha\beta} 
 \delta^{(3)}(\vec{k} - \vec{p}).
\label{defpi}
\end{equation}
The evolution equation for the mode functions will then be, in Fourier space,
\begin{equation}
{\cal A}_{k}'' + \biggl[ k^2 - g (g^{-1})''\biggr] {\cal A}_{k} =0,
\label{modef1}
\end{equation}
i.e. exactly the same equation obtained in (\ref{Ak}). The pump field
can  also be  expressed as:
\begin{equation}
g(g^{-1})''= \biggl( \frac{{\varphi '}^2}{4} - \frac{\varphi ''}{2} \biggr).
\end{equation}
 The maximally amplified modes are then the ones for which 
\begin{equation}
k_{\rm max}^2 \simeq |g (g^{-1})''|.
\end{equation}
The Fourier  modes appearing in Eq. (\ref{modef1})  have to be normalized while they are 
inside the horizon for large and negative $\tau$. In this limit the initial conditions provided by quantum mechanics are   
\begin{equation}
{\cal A}_{k}(\tau) = \frac{1}{\sqrt{2 k}} e^{- i k\tau},\qquad
\Pi_{k}(\tau) = - i \sqrt{\frac{k}{2}} e^{- i k\tau}. 
\label{NORM}
\end{equation}
\begin{figure}
\centering
\includegraphics[height=5cm]{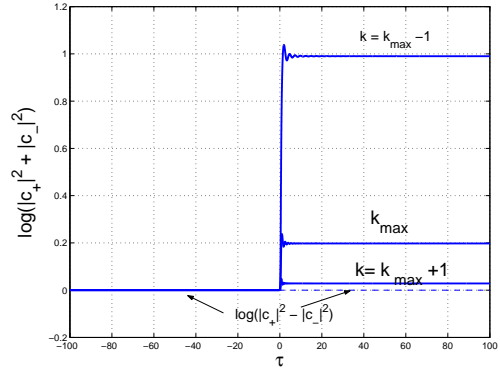}
\caption{The evolution of the mixing coefficients 
for $k \simeq k_{\mathrm{kmax}}$ in units of $\tau_{0}$.}
\label{FB2}      
\end{figure}
In the limit $\tau \to +\infty$ the positive and negative frequency modes will be mixed, 
so that the solution will be represented in the plane wave orthonormal basis as 
\begin{eqnarray}
{\cal A}_{k}(\tau) &=& \frac{1}{\sqrt{2 k}} \biggl[ c_{+}(k) e^{- i k\tau} + c_{-}(k) e^{ i k\tau}\biggr],
\nonumber\\
{\cal A}_{k}'(\tau) &=& - i \sqrt{\frac{k}{2}}  \biggl[ c_{+}(k) e^{- i k\tau} - c_{-}(k) e^{ i k\tau}\biggr].
\end{eqnarray}
where $c_{\pm}(k)$ are the (constant) mixing coefficients. The following two relations fully determine 
the square modulus of each of the two mixing coefficients in terms of the complex 
wave-functions obeying Eq. (\ref{modef1}): 
\begin{eqnarray}
&& |c_{+}(k)|^2 - |c_{-}(k)|^2 = i( {\cal A}_{k}^{\star} \Pi_{k} -{\cal A}_{k} \Pi_{k}^{\star}) ,
\label{CMIN}\\
&& |c_{+}(k)|^2 + |c_{-}(k)|^2 = \frac{1}{k^2}\biggl(|\Pi_{k}|^2 + k^2|{\cal A}_{k}|^2\biggr).
\label{CPL}
\end{eqnarray}
After having numerically computed the time evolution
of the properly normalized mode functions, Eqs. (\ref{CMIN}) and (\ref{CPL}) can be used to infer 
the value of the relevant mixing coefficient (i.e. $c_{-}(k)$).  Equation (\ref{CMIN}) is, in fact, 
the Wronskian of the solutions. If the second-order differential equation is written in the form 
(\ref{modef1}),  the Wronskian is always conserved throughout the time evolution of the system. 
Since, from Eq. (\ref{NORM}),  the Wronskian  is equal to $1$ initially, it will  be equal to $1$ all along the time evolution.
Thus, from Eq. (\ref{CMIN}) 
$|c_{+}(k)|^2 = |c_{-}(k)|^2 + 1$. The fact that the Wronskian must always be equal to $1$ is the measure 
of the precision of the algorithm. 

\begin{figure}
\centering
\includegraphics[height=5cm]{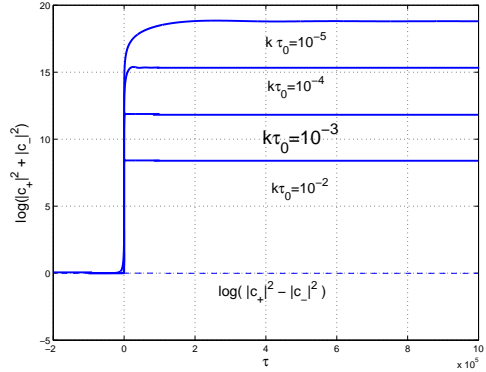}
\caption{The numerical estimate of the mixing coefficients in the case 
$k \tau_{0} \ll 1$.}
\label{FB3}      
\end{figure}

In Figs. \ref{FB2} and \ref{FB3} the numerical calculation of the spectrum is illustrated for different 
values of $k$. In Fig. \ref{FB3} the mixing coefficients 
are reported for modes $k\ll k_{\rm max}$. In Fig. \ref{FB2} the mixing coefficients are 
reported for modes around $k_{\rm max}$. Clearly, from Fig. \ref{FB3} a smaller 
$k$ leads to a larger mixing coefficient 
which means that the spectrum is rather blue. Furthermore by comparing the amplification of different modes 
it is easy to infer that the scaling law is $|c_{+}(k)|^2 + |c_{-}(k)|^2 \propto (k/k_{\rm max})^{- n_{g}}$ , with 
$n_{g} \sim 3.46$, which is in excellent agreement with the analytical determination of the mixing coefficients leading to $n_{g} = 2 \sqrt{3} \sim 3.46$[see  below, Eq. (\ref{cminanal})].

The second  piece information that  can be drawn from  Fig. \ref{FB2} concerns  $k_{\rm max}$,  whose specific value 
\begin{equation}
k_{\rm max} \simeq \frac{\sqrt{5}-0.5}{\tau_{0}}.
\label{kmax}
\end{equation}
can be determined numerically for different values of $\tau_0$.

For the value of $k_{\rm max}$  reported in Eq. (\ref{kmax}), 
the obtained mixing coefficient is $1$, i.e. $|c_{-}(k_{\rm max})| \simeq  1$.
According to Fig. \ref{FB2} as we move from $k_{\rm max}$ to larger $k$, $(|c_{+}(k)|^2 + |c_{-}(k)|^2) \simeq
(|c_{+}(k)|^2 - |c_{-}(k)|^2)$ implying that $|c_{-}(k)|  \sim 0$. 
Moreover, from the left plot of Fig. \ref{FB3} it can be appreciated that 
\begin{equation}
|c_{-}(k_{\rm max})|^2 =1, ~~~~~\log{(|c_{+}(k_{\rm max})|^2 +  |c_{-}(k_{\rm max})|^2)} = \log{3} \simeq 0.477.
\end{equation}
Thus the absolute normalization and slope of the relevant mixing coefficient can be 
numerically determined to be 
\begin{equation}
|c_{-}(k)|^2 = \biggl(\frac{k}{k_{\rm max}} \biggr)^{- 2\sqrt{3}}.
\label{numest1}
\end{equation}

 It can be concluded that Eq. (\ref{numest1}) is rather 
accurate as far as both the slope and the absolute normalization are concerned.
The numerical estimates presented so far can be also corroborated by the usual analytical treatment based on the matching of the 
solutions for the mode functions before and after the bounce. The evolution 
of the modes described by Eq. (\ref{modef1})  can be approximately determined 
from the exact asymptotic solutions given in Eqs. (\ref{solminus}) and (\ref{solplus}),  and  implying that 
$\varphi'_{\pm} \simeq \pm \sqrt{3}/\tau$. Thus the solutions of Eq. (\ref{modef1}) can be obtained in the two asymptotic regimes, i.e. for $\tau\leq -\tau_{1}$
\begin{equation}
{\cal A}_{k, -}(\tau) = \frac{\sqrt{-\pi\tau}}{2} e^{ i \frac{\pi}{2} ( \nu + 1/2)} H_{\nu}^{(1)}( - k\tau),
\label{exsol0}
\end{equation}
and for $\tau\geq \tau_{1}$ 
\begin{equation}
 {\cal A}_{k, +}(\eta) = \frac{\sqrt{\pi\tau}}{2} e^{ i \frac{\pi}{2} ( \mu + 1/2)}\biggl[ c_{-} 
H_{\mu}^{(1)}( k\tau) + c_{+} e^{- i \pi (\mu +1/2)} H_{\mu}^{(2)}(k\tau) \biggr] ,~\tau \geq -\tau_{1},
\label{exsol1}
\end{equation}
 where $H^{(1,2)}_{\alpha}(z)$ are Hankel functions of first and second kind  whose related indices are 
\begin{equation}
\nu= \frac{\sqrt{3} -1}{2},~~~~~~~~~~~\mu= \frac{\sqrt{3} +1}{2}.
\label{munu}
\end{equation}
The time scale $\tau_1$ defines the width of the bounce  and, typically, $\tau_{1} \sim \tau_0$.

The phases appearing in Eqs. (\ref{exsol0}) and (\ref{exsol1})
 are carefully chosen so that 
\begin{equation}
\lim_{\tau \to -\infty} {\cal A}_{k} = \frac{1}{\sqrt{2 k}} e^{- i k\tau}.
\end{equation}
Using then the appropriate matching conditions 
\begin{eqnarray}
&& {\cal A}_{k,-}(-\tau_1)= {\cal A}_{k,+}(\tau_1),
\nonumber\\
&& {\cal A}_{k,-}'(-\tau_1)= {\cal A}_{k,+}'(\tau_1),
\end{eqnarray}
and defining $x_1 = k\tau_1$, the obtained mixing coefficients are 
\begin{eqnarray}
&& c_{+}(k) = i \frac{\pi}{4} x_1 e^{i\pi( \nu +\mu + 1)/2} \biggl[ - \frac{\nu + \mu +1}{x_1} H_{\mu}^{(1)}(x_1) H_{\nu}^{(1)}(x_1) 
\nonumber\\
&&+
H_{\mu}^{(1)}(x_1) H_{\nu+ 1}^{(1)} (x_1) + H_{\mu+1}^{(1)}(x_1) H_{\nu}^{(1)}(x_1) \biggr],
\label{cpp}\\
&& c_{-}(k)=  i \frac{\pi}{4} x_1 e^{i\pi( \nu -\mu)/2} \biggl[ - \frac{\nu + \mu +1}{x_1} H_{\mu}^{(2)}(x_1) H_{\nu}^{(1)}(x_1) 
\nonumber\\
&& + H_{\mu}^{(2)}(x_1) H_{\nu+ 1}^{(1)}(x_1) + H_{\mu + 1}^{(2)}(x_1) H_{\nu}^{(1)}(x_1) \biggr],
\label{cpm}
\end{eqnarray}
satisfying the exact Wronskian normalization condition $|c_{+}(k)|^2 - |c_{-}(k)|^2 =1$.
In the small argument limit, i.e. $k\tau_1 \sim k\tau_0 \ll 1$ the leading term in Eq. (\ref{cpm}) leads to 
\begin{equation}
c_{-}(k) \simeq \frac{i~ 2^{\mu +\nu}}{4\pi} e^{i \pi(\nu - \mu)/2} x_1^{- \mu - \nu}  (\nu + \mu -1) \Gamma(\mu) \Gamma(\nu)
\label{cminanal}
\end{equation}
If we now insert the values given in Eq. (\ref{munu}) it turns out that $c_{-}(k) \simeq 0.41~ |k\tau_1|^{- \sqrt{3}}$. The 
spectral slope agrees with the numerical estimate, as already stressed. The absolute normalization cannot be determined
from Eq. (\ref{cminanal}),  where the small argument limit has already been taken. In order to determine the 
absolute normalization the specific value of $k_{\rm max}\tau_1$ has to be 
inserted in Eq. (\ref{cpm}). The result of this procedure, taking $\tau_1\sim \tau_0$ is $|c_{-}(k_{\rm max})|^2 = 0.14$,  which is 
roughly a factor of $10$ smaller than the interpolating formula given in Eq. (\ref{numest1}).


The observation that a dynamical gauge coupling implies 
a viable mechanism for the production of large-scale magnetic fields
can be interesting in general terms and, more specifically, in the context
of the pre-big bang models. In fact, in pre-big bang models, not only 
the fluctuations of the hypercharge field are amplified. In the minimal 
case we will have to deal with the fluctuations of the tensor \cite{gg,tens1} and scalar 
\cite{maxmuk} modes of the 
geometry and with the fluctuations of the antisymmetric tensor field \cite{sloth,ax1}.

\begin{figure}
\centering
\includegraphics[height=5cm]{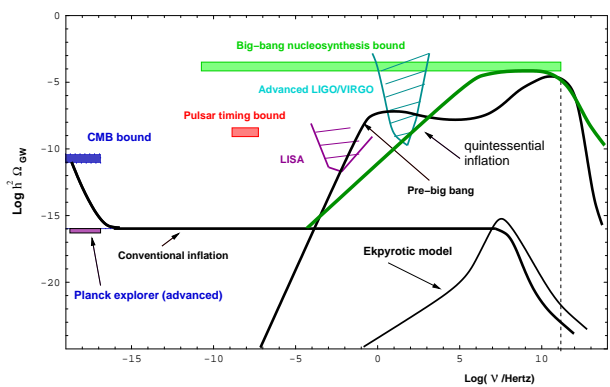}
\caption{The spectrum of relic gravitons from various 
cosmological models presented in terms 
of $h^2 \Omega_{\rm GW}$ .}
\label{F2a}      
\end{figure}

 The amplified tensor modes of the geometry 
lead to a stochastic background of gravitational waves (GW)
with  violet spectrum both 
in the GW  amplitude and energy density. In Fig. \ref{F2a} the  GW signal is parametrized 
in terms of the logarithm of
$\Omega_{\rm GW} = \rho_{\rm GW}/\rho_{\rm c}$, i.e. the fraction of 
critical energy density 
present (today) in GW. On the horizontal axis of Fig. \ref{F2a} the 
logarithm of the present (physical) 
frequency $\nu$ is reported.
In conventional inflationary models, for $\nu \geq 10^{-16} $ Hz, 
 $\Omega_{\rm GW}$, is constant (or slightly decreasing)  
as a function of the present 
frequency. In the case of string cosmological models $\Omega_{\rm GW}
 \propto \nu^3 \ln{\nu}$, which also 
implies a steeply increasing power spectrum.
This possibility spurred various experimental groups to analyse possible directs 
limits on the scenario arising from specific instruments such as  resonant 
mass detectors \cite{bars} 
and microwave cavities  \cite{picasso,cruise}. These attempts are justified since the
signal of pre-big bang models may be rather strong at high frequencies and, anyway, much stronger 
than the conventional inflationary prediction
 
The sensitivity of a pair of VIRGO detectors
to string cosmological gravitons has been  specifically analysed
 \cite{maxdan} with the 
conclusion that a VIRGO pair, in its upgraded stage, can certainly  probe 
wide regions of the parameter space of these models. If we  maximize the 
overlap between the two detectors \cite{maxdan} or 
if we reduce (selectively) the pendulum and pendulum's internal modes
contribution to the thermal noise of the instruments, the 
visible region (after one year of observation and with ${\rm SNR} = 1$)
of the parameter space will get even larger. Unfortunately, as in the 
case of the advanced LIGO detectors, the sensitivity to a flat $\Omega_{\rm GW}$ will be irrelevant for 
ordinary inflationary models also with the advanced VIRGO 
detector. It is worth mentioning 
that growing energy spectra of relic gravitons  can also arise 
in the context of quintessential inflationary models \cite{alex2,maxquint}.
In this case $\Omega_{\rm GW} \propto \nu \ln^2{\nu}$ (see \cite{maxquint} 
for a full discussion). 

The spectra of gravitational waves have features that are, in some sense, complementary 
to the ones of the large-scale magnetic fields. The parameter space leading to a possible 
signal of relic (pre-big bang) gravitons with wide-band interferometers 
has only a small overlap with the region of the parameter space leading 
to sizable large-scale magnetic fields. This conclusion can be evaded 
if the coupling of the dilaton to the hypercharge field is, in the 
action, of the type $e^{-\beta \varphi} F_{\mu\nu}F^{\mu\nu}$ \cite{nicotri} where 
the parameter $\beta$ has values $1$ and $1/2$, respectively, for heterotic and type I 
superstrings. In particular, in the case $\beta =1/2$, it is possible to find regions 
where both large-scale magnetic fields and relic gravitons are copiously 
produced.

Let us finally discuss the scalar fluctuations of the geometry. The spectrum of the 
scalar modes is determined by the spectrum of the Kalb-Ramond axion(s). If the axions 
would be neglected, the spectrum of the curvature fluctuations 
would be sharply increasing, or as we say in the jargon, the spectrum would 
be violet in full analogy with the spectrum of the tensor modes  of the geometry. This 
result \cite{maxmuk} has been recently analyzed  in the light of a recent 
controversy (see \cite{G1,G2}) and references therein). 

If the Kalb-Ramond axions are consistently included in the calculation, it is 
found that the large-scale spectrum of curvature perturbations 
becomes flat \cite{ax1} and essentially inherits the spectrum 
of the Kalb-Ramond axions.  If the axions decay (after a phase 
of coherent oscillations) the curvature perturbations will be adiabatic 
as in the case of conventional inflationary models but with some 
important quantitative differences \cite{ax1} since, in this case, the 
CMB normalization is explained in terms of the present value 
of the string curvature scale and in terms of the primordial 
slope of the axion spectrum.

\section{Primordial or not primordial, this is the question...}
\label{sec4}

While diverse theoretical models for the origin of large-scale 
magnetism can certainly be questioned on the basis 
of purely theoretical considerations, direct observations 
can tell us something more specific concerning the epoch of formation of
large-scale magnetic fields. It would be potentially useful  to give  some elements 
of response to the following burning question: are really magnetic fields primordial? 

The plan of the present section is the following. In Subsect. \ref{subs41} different meanings 
of the term {\em primordial} will be discussed. It will be argued that 
CMB physics can be used to constrain large-scale magnetic fields possibly 
present prior to matter-radiation equality. In Subsect. \ref{subs42} the scalar CMB anisotropies 
will be specifically discussed by deriving the appropriate set of evolution equations accounting for the presence of a fully inhomogeneous magnetic field.  In Subsect.  \ref{subs43} the
evolution of the different species composing the pre-decoupling plasma will be 
solved, in the tight-coupling approximation and in the 
presence of a fully inhomogeneous magnetic field. Finally Subsect. \ref{subs44} contains 
various numerical results and a strategy for parameter extraction.

\subsection{Pre-equality magnetic fields}
\label{subs41}

The term primordial seems to have 
slightly different meanings depending on the 
perspective of the various communities 
converging on the study of large-scale magnetic fields. 
Radio-astronomers have the hope 
that by scrutinizing the structure of magnetic fields in distant galaxies 
it would be possible, in the future, 
to understand if the observed magnetic fields 
are the consequence of a strong dynamo action or if their existence precedes 
the formation of galaxies. 

If the magnetic field does not flips its sign 
from one spiral arm to the other, then a strong dynamo action 
can be suspected \cite{beck1}. In the opposite case the magnetic field of galaxies should 
be {\em primordial} i.e. present already at the onset of gravitational collapse.
In this context,  primordial simply means 
protogalactic.  An excellent review on the evidence of magnetism in nearby galaxies can be found in \cite{beck2}. 
\begin{figure}
\centering
\includegraphics[height=5cm]{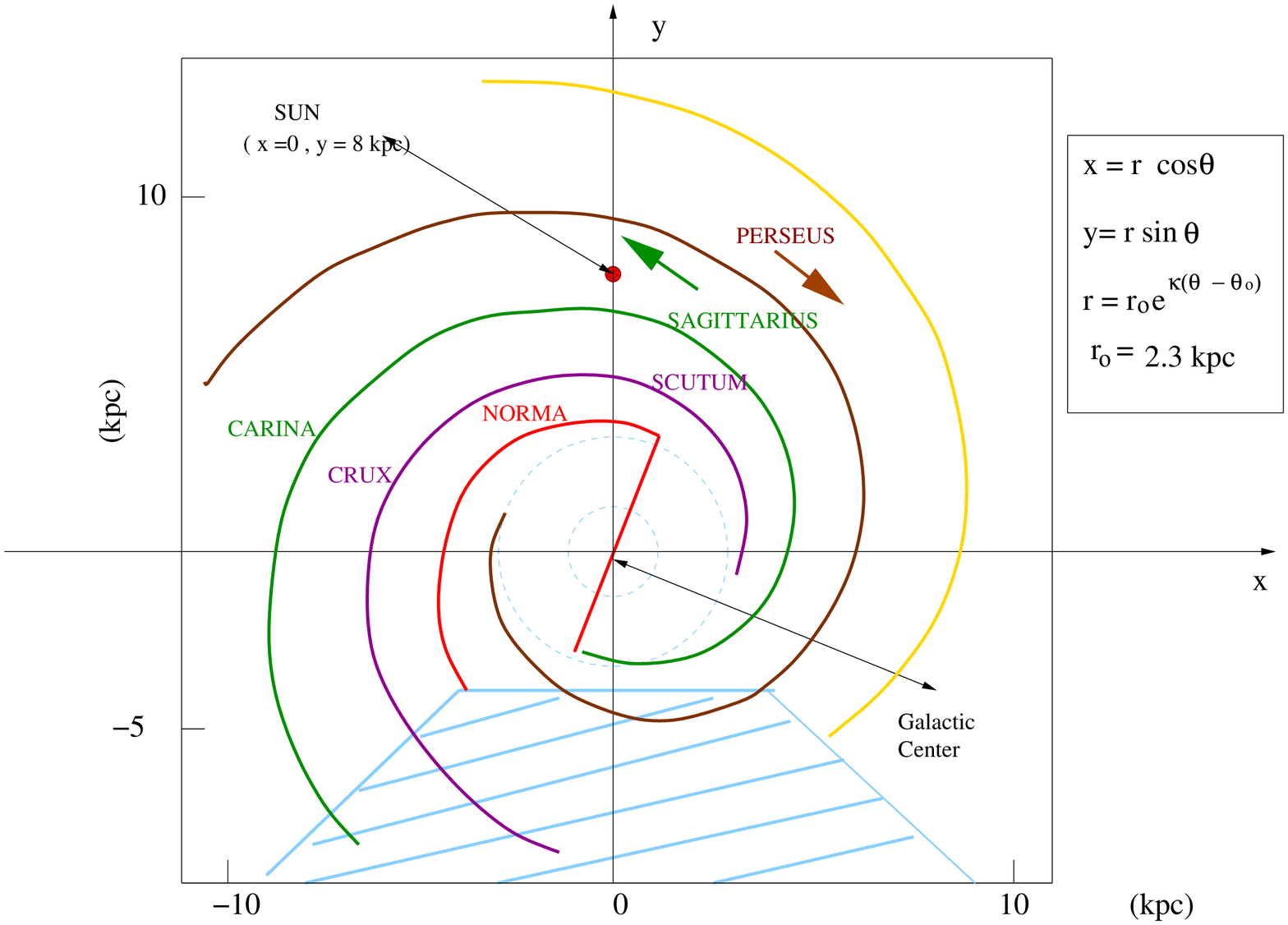}
\caption{The schematic map of the MW is illustrated. Following \cite{vallee} the origin of the two-dimensional 
coordinate system are in the Galactic center. The two large arrows indicate 
one of the possible (3 or 5) field  reversals observed so far.}
\label{F2b}      
\end{figure}
In Fig. \ref{F2b} a schematic view of the Milky Way is presented. The magnetic field follows the spiral 
arm. There have been claims, in the literature, of $3$ to $5$ field reversals. The arrows in Fig. 
\ref{F2b} indicate one of the possible field reversals. One reversal is certain beyond any doubt.
Another indication that would support 
the primordial nature of the magnetic field 
of galaxies would be, for instance, the evidence that not only spirals but also 
elliptical galaxies are magnetized (even if the magnetic field seems to have 
correlation scale shorter than in the case of spirals). Since elliptical 
galaxies have a much less efficient rotation, it seems difficult to postulate a strong dynamo action.
We will not pursue here the path of specific astrophysical signatures of 
a truly pre-galactic magnetic field and e refer the interested reader 
to \cite{beck1,beck2}. 

As a side remark, it should also be mentioned that magnetic fields may play a r\^ole in 
the analysis of rotation curves of spiral galaxies. This aspect has been investigated 
in great depth by E. Battaner, E. Florido and collaborators also in connection with possible effects of large-scale 
magnetic fields on structure formation \cite{ED1,ED2,ED3,ED4} (see also
\cite{ED5} and references therein). 

The large-scale magnetic fields produced via the parametric amplification of 
quantum fluctuations discussed earlier in the present lecture may also be defined primordial but, in this 
case, the term primordial has a much broader signification embracing 
the whole epoch that precedes the equality between matter and radiation 
taking place, approximately, at a redshift $z_{\mathrm eq} =3230$
for $h^2 \Omega_{\mathrm m0} =0.134$ and $h^2 \Omega_{\mathrm r0} = 4.15\times 10^{-5}$.
Consequently, large-scale magnetic fields may  affect, potentially,  
CMB anisotropies \cite{f1}. Through the years, various studies have been 
devoted to the effect of large-scale magnetic fields on the vector and tensor CMB anisotropies 
\cite{f3,f9} (see also \cite{f4} and references therein for some recent review articles). 

The implications of fully inhomogeneous magnetic fields on the scalar modes 
of the geometry remain comparatively less explored. 
By fully inhomogeneous we mean stochastically distributed fields 
that do not break the spatial isotropy of the background \cite{f4,f4a}.

CMB anisotropies are customarily described
in terms of a set of carefully chosen initial conditions 
for the evolution of the brightness perturbations of the radiation field.
One set of initial conditions corresponds to a purely adiabatic mode.
There are, however, more complicated situations where, on top of the 
adiabatic mode there is also one (or more) non-adiabatic mode(s).
A {\em mode}, in the present 
terminology, simply means a consistent solution of the governing 
equations of the metric and plasma fluctuations, i.e. 
a consistent solution of the perturbed Einstein equations and of the 
lower multipoles of the Boltzmann hierarchy. 

The simplest set of initial conditions for 
CMB anisotropies, implies, in a $\Lambda$CDM framework, that a nearly 
scale-invariant spectrum of adiabatic fluctuations is present after 
matter-radiation equality (but before decoupling) for typical wavelengths 
larger than the Hubble radius at the corresponding epoch \cite{f5}. 

It became relevant, through the years, to relax the assumption
of exact adiabaticity and to scrutinize the implications of a more general
mixture of adiabatic and non-adiabatic initial conditions (see \cite{f60,f61,f62} and references therein). 
In what follows it will be argued, along a similar perspective, 
that large-scale magnetic fields slightly modify the adiabatic paradigm 
so that their typical strengths may be constrained. To achieve such a goal, the first step is to 
solve the evolution equations of magnetized cosmological perturbations 
well before matter-radiation equality. The second step is to follow the solution through 
equality (and up to decoupling). On a more technical ground, 
the second step amounts to the calculation of the so-called transfer matrix \cite{maxprd06}
whose specific form is one of the the subjects of the present analysis.

\subsection{Basic Equations}
\label{subs42}
Consider then the system of cosmological perturbations 
of a flat Friedmann-Robertson-Walker (FRW) Universe, 
characterized by a conformal time scale factor $a(\tau)$ (see Eq. (\ref{FRW})),  
and consisting of a mixture of photons, baryons, CDM particles and massless neutrinos.
In the following the basic set of equations used in order to describe the magnetized curvature perturbations will be introduced and discussed. The perspective adopted here is closely related to the recent results obtained in Refs. \cite{maxprd206,maxcqg06} (see also \cite{f3a,f3b} for interesting 
developments).

 In the conformally Newtonian gauge \cite{f70,f71,f72,f73,f74}, the  
 {\em scalar} fluctuations of the metric tensor 
 $G_{\mu\nu} = a^2(\tau) \eta_{\mu\nu} $ 
 are parametrized in terms of the two longitudinal fluctuations i.e. 
 \begin{equation}
 \delta G_{00} = 2 a^2(\tau) \phi(\tau,\vec{x}),\qquad 
 \delta G_{ij} = 2 a^2(\tau) \psi(\tau,\vec{x}) \delta_{ij},
 \label{longfl}
 \end{equation}
 where $\delta_{ij}$ is the Kroeneker $\delta$. 
While the spatial curvature will be assumed 
to vanish, it is straightforward to 
 extend the present considerations to the case when the spatial curvature 
 is not negligible.
 
 In spite of the fact that the present discussion will be conducted within the conformally Newtonian 
 gauge, it can be shown 
that gauge-invariant descriptions of the problem are possible \cite{maxcqg06}. 
Moreover, specific non-adiabatic modes 
(like the ones related to the neutrino system) 
may be more usefully described in different gauges 
(like the synchronous gauge). The rationale 
for the last statement is that the neutrino 
isocurvature modes may be singular in the 
conformally Newtonian gauge. 
These issues will not be addressed here 
but have been discussed in the existing literature (see, for instance, 
 \cite{f73,f74} and references therein). Furthermore, for the 
benefit of the interested reader it is appropriate to mention 
that the relevant theoretical tools used in the present and in the following paragraphs follows the conventions of a recent review \cite{f74}.

\subsubsection{Hamiltonian and momentum constraints}
\label{ssubs421}
The Hamiltonian and momentum constraints, stemming from the $(00)$ and $(0i)$ 
 components of the perturbed Einstein equations are:
\begin{eqnarray}
&& \nabla^2 \psi - 3 {\cal H} ( {\cal H}\phi + \psi') = 4\pi G a^2 [ \delta \rho_{\mathrm t} + 
\delta\rho_{\mathrm B}],
\label{Ham1}\\
&&\nabla^2 ( {\mathcal   H} \phi + \psi') = - 4\pi G a^2 \biggl[ 
( p_{\rm t} + \rho_{\rm t}) \theta_{\rm t}  
+ \frac{ \vec{\nabla} \cdot ( \vec{E}\times \vec{B})}{4 \pi a^4}\biggr],
\label{Mom1}
\end{eqnarray}
where ${\cal H} = a'/a$ and the prime denotes a derivation with respect 
to the conformal time coordinate $\tau$.  In writing Eqs. (\ref{Ham1}) and (\ref{Mom1}) the following set 
of conventions has been adopted
\begin{eqnarray}
&& \delta \rho_{\mathrm t}(\tau,\vec{x}) = \delta \rho_{\gamma} (\tau,\vec{x})+ \delta \rho_{\nu} (\tau,\vec{x})+ 
\delta \rho_{\mathrm c}(\tau,\vec{x}) + \delta \rho_{\mathrm b}(\tau,\vec{x}),
\label{def1a}\\
&& \delta \rho_{\rm B}(\tau,\vec{x}) = \frac{B^2(\vec{x})}{8\pi a^4(\tau)},
\label{def1b}\\
&& (p_{\mathrm t} + \rho_{\mathrm t} ) \theta_{\mathrm t}(\tau,\vec{x})= (p_{\gamma} + \rho_{\gamma})\theta_{\gamma}(\tau,\vec{x})
+ (p_{\nu} + \rho_{\nu})\theta_{\nu} (\tau,\vec{x})
\nonumber\\
&& + (p_{\mathrm c} + \rho_{\mathrm c})\theta_{\mathrm c}(\tau,\vec{x})+ 
(p_{\mathrm b} + \rho_{\mathrm b})\theta_{\mathrm b}(\tau,\vec{x}).
\label{def1c}
\end{eqnarray}
Concerning Eqs. (\ref{def1a}), (\ref{def1b}) and (\ref{def1c}) the following comments are in order:
\begin{itemize}
\item{} in Eq. (\ref{def1a}) the total density fluctuation of the plasma, i.e. $\delta \rho_{\rm t}(\tau,\vec{x})$ 
receives contributions from all the 
species of the plasma; 
\item{} in  Eq. (\ref{def1b}) the fluctuation of the magnetic energy density 
$\delta \rho_{\mathrm B}(\tau,\vec{x})$ is quadratic in the magnetic field intensity; 
\item{}in  Eq. (\ref{def1c}) 
$\theta_{\mathrm t}(\tau,\vec{x}) = \partial_{i} v^{i}_{\mathrm t}$ is the divergence 
of the total peculiar velocity while $\theta_{\gamma}(\tau,\vec{x})$, $\theta_{\nu}(\tau,\vec{x})$, 
$\theta_{\mathrm c}(\tau,\vec{x})$ and 
$\theta_{\mathrm b}(\tau,\vec{x})$ are the divergences of the peculiar velocities of each individual species, i.e. 
photons, neutrinos, CDM particles and baryons.
\end{itemize}

The second term appearing at the right hand side of Eq. (\ref{Mom1}) is 
the divergence of the Poynting vector. In MHD the Ohmic electric field 
is subleading and, in particular, from the MHD expression of the Ohm law we will have
\begin{equation}
\vec{E} \times \vec{B} \simeq \frac{(\vec{\nabla}\times \vec{B})\times \vec{B}}{4\pi\sigma}.
\end{equation}
Since the Universe, prior to decoupling, is a very good conductor, the 
 ideal MHD limit can be safely adopted in the first approximation (see also \cite{bir}); thus 
for $\sigma \to \infty$ (i.e. infinite conductivity limit) the contribution of the Poynting vector 
vanishes. In any case, even if $\sigma $ would be finite but large, the second 
term at the right hand side of Eq. (\ref{Mom1}) 
would be suppressed in comparison with the contribution of  the divergence of 
the total velocity field.

The total (unperturbed) energy density and pressure of the mixture, i.e.  
\begin{eqnarray}
\rho_{\rm t} &=&  \rho_{\gamma} + \rho_{\nu} + \rho_{\mathrm c} +  \rho_{\mathrm b} + \rho_{\Lambda},
\nonumber\\
p_{\mathrm t} &=&  p_{\gamma} + p_{\nu} + p_{\mathrm c} +  p_{\mathrm b} + p_{\Lambda}.
\label{def2}
\end{eqnarray}
 determine the evolution of the background
geometry according to Friedmann equations: 
\begin{eqnarray}
&&{\cal H}^2 = \frac{8\pi G}{3} a^2 \rho_{\rm t}, 
\label{FL1}\\
&&{\cal H}^2 - {\cal H}' = 4\pi G a^2 (\rho_{\rm t} + p_{\rm t}),
\label{FL2}\\
&&\rho_{\rm t}' + 3 {\cal H} (\rho_{\rm t} + p_{\rm t})=0.
\label{FL3}
\end{eqnarray}
Notice that in Eq. (\ref{def2}) the contribution of the cosmological constant has been included.
If the dark energy is parametrized in terms of a cosmological constant (i.e. $p_{\Lambda} = - \rho_{\Lambda}$), 
then, $\delta \rho_{\Lambda} '=0$. Furthermore, the contribution of $\rho_{\Lambda}$ to the 
background evolution is negligible prior to decoupling. Slightly different situations (not contemplated 
by the present analysis)
may arise if the dark energy is parametrized in terms of one (or more) scalar 
degrees of freedom with suitable potentials.

\subsubsection{Dynamical equation and anisotropic stress(es)}
\label{ssubs422}
The spatial components of the perturbed Einstein equations, imply, instead 
\begin{eqnarray}
&& \biggl[\psi'' + {\mathcal H}(\phi' + 2 \psi') + ({\mathcal H}^2 + 2 {\mathcal H}') \phi + 
\frac{1}{2} \nabla^2 (\phi - \psi) \biggr] \delta_{i}^{j}
\nonumber\\
&& - 
\frac{1}{2} \partial_{i}\partial^{j} (\phi - \psi) = 4\pi G a^2 \biggl[ (\delta p_{\mathrm t} + 
\delta p_{\mathrm B}) \delta_{i}^{j} - \Pi_{i}^{j} - \tilde{\Pi}_{i}^{j} \biggr].
\label{psieq}
\end{eqnarray}
Equation (\ref{psieq}) contains, as source terms, not only the total fluctuation 
of the pressure of the palsma, i.e. $\delta p_{\mathrm t}$, but also 
\begin{eqnarray}
&& \delta p_{\mathrm B}(\tau,\vec{x}) = \frac{B^2(\vec{x})}{24\pi a^4(\tau)} = 
\frac{\delta \rho_{\mathrm B}(\tau,\vec{x})}{3}.
\label{magnp}\\
&& \tilde{\Pi}_{i}^{j}(\tau,\vec{x}) = \frac{1}{4\pi a^4} \biggl( B_{i}B^{j} - \frac{1}{3} B^2 \delta_{i}^{j}\biggr).
\label{magnanis}
\end{eqnarray}
Moreover, in Eq. (\ref{psieq}), $\Pi_{i}^{j}(\tau,\vec{x})$ is  the anisotropic stress of the fluid. As it will be mentioned 
in a moment (and later on heavily used) the main source of anisotropic stress of the fluid is provided 
by neutrinos which free-stream from temperature smaller than the MeV. Notice that both the anisotropic 
stress of the fluid, i.e. $\Pi_{i}^{j}(\tau,\vec{x})$  and the magnetic anisotropic stress,
 i.e. $\tilde{\Pi}_{i}^{j}(\tau,\vec{x})$,  are, by definition, traceless.

Using this last observation, Eq. (\ref{psieq}) can be separated into two independent equations.
Taking the trace of Eq. (\ref{psieq})  we do get 
\begin{equation}
 \psi'' + {\cal H} ( \phi' + 2 \psi') + ( 2 {\cal H}' + {\cal H}^2) \phi + 
\frac{1}{3} \nabla^2(\phi - \psi) = 4\pi G a^2 (\delta p_{\rm t} + \delta p_{\rm B}).
\label{tij}
\end{equation}
By taking the difference between Eq. (\ref{psieq}) and Eq. (\ref{tij}) the following (traceless) 
relation can be obtained:
\begin{equation}
\partial_{i} \partial^{j} (\phi - \psi) - \frac{1}{3} \delta_{i}^{j} \nabla^2 (\phi - \psi) = 8\pi G a^2 (\Pi_{i}^{j} + 
\tilde{\Pi}_{i}^{j}).
\label{anis0}
\end{equation}
By applying the differential operator $\partial_{j}\partial^{i}$ to both sides of Eq. (\ref{anis0}) 
we do obtain the following interesting relation:
\begin{equation}
 \nabla^4 ( \phi - \psi) = 12 \pi G a^2 [ 
(p_{\nu} + \rho_{\nu}) \nabla^2 \sigma_{\nu} +
 (p_{\gamma} + \rho_{\gamma}) \nabla^2 \sigma_{\rm B}],
 \label{anis1}
 \end{equation}
 where the  parametrization 
 \begin{equation}
 \partial_{j}\partial^{i} \Pi_{i}^{j} = ( p_{\nu} + \rho_{\nu}) \nabla^2 \sigma_{\nu}, \qquad 
 \partial_{j}\partial^{i} \tilde{\Pi}_{i}^{j} = ( p_{\gamma} + \rho_{\gamma}) \nabla^2 \sigma_{\mathrm B},
\label{parametrization}
\end{equation}
has been adopted. In Eq. (\ref{anis1}) $\sigma_{\nu}(\tau, \vec{x})$ is related with the quadrupole moment of the (perturbed) 
neutrino phase-space distribution. In Eq. (\ref{anis1})  $\sigma_{\rm B}(\tau,\vec{x})$ parametrizes 
the (normalized) magnetic anisotropic stress. 
It is relevant to remark at this point that in the MHD approximation adopted here 
the two main sources of scalar anisotropy associated with magnetic fields 
can be parametrized in terms of $\sigma_{\mathrm B}(\tau,\vec{x})$ and in terms of the 
dimensionless ratio
\begin{equation}
\Omega_{\mathrm B}(\tau,\vec{x}) = \frac{\delta\rho_{\rm B}(\tau, \vec{x})}{\rho_{\gamma}(\tau)}.
\label{OMB}
\end{equation}
Since both $\Omega_{\mathrm B}(\tau,\vec{x})$ and $\sigma_{\mathrm B}(\tau,\vec{x})$ are quadratic 
in the magnetic field intensity a non-Gaussian contribution may be expected.  $\Omega_{\rm B}(\tau,\vec{x})$ is the magnetic energy density referred to the photon energy density and it is constant to a very good approximation if magnetic flux is frozen into the plasma element. 

There is, in principle, a third contribution to the scalar problem coming from magnetic fields. 
Such a contribution arises in the evolution equation of the photon-baryon 
peculiar velocity and amounts to the divergence of the Lorentz force. While the 
mentioned equation will be derived later in this section, it is relevant to point out here 
that the MHD Lorentz force can be expressed solely in terms 
of $\sigma_{\mathrm B}(\tau,\vec{x})$ and $\Omega_{\mathrm B}(\tau,\vec{x})$.
In fact a well known vector identity stipulates that 
\begin{equation}
\partial_{i} B_{j} \partial^{j} B^{i} = \vec{\nabla} \cdot[ (\vec{\nabla} \times \vec{B})\times \vec{B} ] + \frac{1}{2} \nabla^2 B^2.
\label{A1}
\end{equation}
From the definition of $\sigma_{\rm B}$ in terms of $\tilde{\Pi}_{i}^{j}$,
i.e. Eq. (\ref{parametrization}), it is easy to show that 
\begin{equation}
\nabla^2 \sigma_{\rm B} = \frac{3}{16\pi a^4 \rho_{\gamma}}\partial_{i} B_{j} \partial^{j} B^{i} - \frac{1}{2} \nabla^2 \Omega_{\rm B}.
\label{A2}
\end{equation}
Using then Eq. (\ref{A1}) into Eq. (\ref{A2}) and recalling that 
\begin{equation}
4\pi \vec{\nabla}\cdot[ \vec{J} \times \vec{B}] = 
\vec{\nabla}\cdot[ (\vec{\nabla}\times \vec{B})\times \vec{B}],
\end{equation}
 we obtain:
 \begin{equation}
 \nabla^2 \sigma_{\rm B} = 
 \frac{3}{16\pi a^4 \rho_{\gamma}} \vec{\nabla}\cdot [
  (\vec{\nabla}\times \vec{B})
 \times \vec{B}] + 
 \frac{\nabla^2 \Omega_{\rm B}}{4}.
  \label{magndef}
 \end{equation}

\subsubsection{Curvature perturbations}
\label{ssubs423}
Two important 
 quantities must now be introduced. The first one, conventionally denoted by $\zeta$, is the 
density contrast on uniform curvature hypersurfaces \footnote{Since, as it will be discussed, $\zeta$ is
gauge-invariant, we can also interpret it as the curvature fluctuation on uniform density hypersurfaces, i.e. 
the fluctuation of the scalar curvature on the hypersurface where the total density is 
uniform.}, i.e. 
\begin{equation}
\zeta = - \psi - {\mathcal H}
\frac{(\delta \rho_{\mathrm t} + \delta \rho_{\mathrm B})}{\rho_{\mathrm t}'}.
\label{zeta1}
\end{equation}
The definition (\ref{zeta1}) is invariant under infinitesimal coordinate 
transformations. In fact, while $\delta \rho_{\mathrm B} $ is automatically 
gauge-invariant (since the magnetic field vanishes at the level 
of the background) $\psi$ and $\delta \rho_{\mathrm t}$ 
transform as \cite{f74}
\begin{eqnarray}
&& \psi \to \tilde{\psi} = \psi + {\mathcal H} \epsilon,
\nonumber\\
&& \delta \rho_{\mathrm t} \to \tilde{\delta\rho}_{\mathrm t} - \rho_{\mathrm t}' \epsilon.
\label{gaugetr}
\end{eqnarray}
for 
\begin{eqnarray}
&& \tau \to \tilde{\tau} = \tau + \epsilon^{0}
\nonumber\\\
&& x^{i} \to \tilde{x}^{i} = x^{i} + \partial^{i} \epsilon.
\label{GT}
\end{eqnarray}
Recalling Eq. (\ref{FL3}), Eq. (\ref{zeta1}) can also 
be written as 
\begin{equation}
\zeta = - \psi + \frac{\delta \rho_{\mathrm t} + \delta \rho_{\mathrm B}}{3(\rho_{\mathrm t} + p_{\mathrm t})}.
\label{zeta2}
\end{equation}

The second variable we want to introduce, conventionally denoted by ${\mathcal R}$ is the curvature perturbation on comoving orthogonal 
hypersurfaces \footnote{It is clear, from the definition (\ref{R1}) that the second term at the right
hand side is proportional, by the momentum constraint (\ref{Mom1}), to the total peculiar 
velocity of the plasma which is vanishing on comoving (orthogonal) hypersurfaces.}, i.e. 
\begin{equation}
{\mathcal R} = - \psi - \frac{{\mathcal H} ( {\mathcal H} \phi + \psi')}{{\mathcal H}^2 -{\mathcal H}'}.
\label{R1}
\end{equation}
Inserting Eq. (\ref{zeta2}) and (\ref{R1}) into Eq. (\ref{Ham1}), the Hamiltonian 
constraint takes then the form 
\begin{equation}
\zeta = {\mathcal R} + \frac{\nabla^2 \psi}{12 \pi G a^2 (p_{\mathrm t} + \rho_{\mathrm t})}.
\label{ham2}
\end{equation}
Equation (\ref{ham2}) is rather interesting in its own right and it tells that, in the long wavelength limit, 
\begin{equation}
\zeta \simeq {\mathcal R} + {\mathcal O}(k^2 \tau^2).
\label{longw}
\end{equation}
When the relevant wavelengths are larger than the Hubble radius (i.e. $k\tau \ll 1$) 
the density constrast on uniform curvature hypersurfaces and the curvature fluctuations 
on comoving orthogonal hypersurfaces coincide.
Since the ordinary Sachs-Wolfe contribution to the 
gauge-invariant temperature fluctuation is dominated by wavelengths that are larger than the 
Hubble radius after matter radiation equality (but before radiation decoupling), the calculation of $\zeta$ 
(or ${\mathcal R}$), in the long wavelength limit, will essentially give us the Sachs-Wolfe plateau. 

A  remark on the definition given in Eq. (\ref{zeta1}) is in order. The variable $\zeta$ must 
contain the {\em total} fluctuation of the energy density. This is crucial since the Hamiltonian constraint 
is sensitive to the total fluctuation of the energy density. If the magnetic energy density 
$\delta\rho_{\mathrm B}$ is correctly included in the definition of $\zeta$, then the Hamiltonian 
constraint (\ref{ham2}) maintains  its canonical form. 

Equations (\ref{ham2}) and (\ref{longw}) can be used to derive the appropriate transfer matrices, allowing, in turn, the estimate of the Sachs-Wolfe plateau. For this purpose it is important to deduce the evolution equation for 
$\zeta$. The evolution of $\zeta$ can be obtained from the evolution equation of the total density fluctuation which 
 reads, in the conformally Newtonian gauge, 
\begin{equation}
\delta\rho_{\mathrm t}' - 3 \psi' (p_{\mathrm t} + \rho_{\mathrm t}) + (p_{\mathrm t} + \rho_{\mathrm t}) \theta_{\mathrm t} + 3 {\cal H}( \delta p_{\mathrm t} + \delta\rho_{\mathrm t}) + 3 {\cal H} \delta p_{\mathrm nad} = \frac{\vec{E}\cdot \vec{J}}{a^4}. 
\label{DC}
\end{equation}
The technique is now rather simple. We can  extract $\delta\rho_{\mathrm t}$ from Eq. (\ref{zeta2}) 
\begin{equation}
\delta \rho_{\mathrm t} = 3 (\rho_{\mathrm t} + p_{\mathrm t}) ( \zeta + \psi) - \delta \rho_{\mathrm B}.
\label{deltarho}
\end{equation}
Inserting Eq. (\ref{deltarho}) into Eq. (\ref{DC}) we get to the wanted evolution equation for $\zeta$.
Before doing that it is practical to discuss the case when the relativistic fluid receives 
 contributions from different species that are simultaneously present. In the realistic case, considering that 
 the cosmological constant does not fluctuate, we will have four different species.
 
For deriving the evolution equation of $\zeta$, it is practical (and, to some extent, conventional) 
to separate the pressure fluctuation into an adiabatic 
component supplemented by a non-adiabatic contribution:
\begin{equation}
\delta p_{\mathrm t} = \biggl(\frac{\delta p_{\mathrm t}}{\delta\rho_{\mathrm t}}\biggr)_{\varsigma} \delta \rho_{\mathrm t} + \biggl(\frac{\delta p_{\mathrm t}}{\delta \varsigma}\biggr)_{\rho_{\mathrm t}} \delta \varsigma.
\label{deltap}
\end{equation}
In a relativistic description of 
gravitational fluctuations, the pressure fluctuates both because the energy density fluctuates (first term at
 the right hand side of Eq. (\ref{deltap})) of because the specific entropy of the plasma, i.e. $\varsigma$ 
 fluctuates (first term at
 the right hand side of Eq. (\ref{deltap})). The subscripts appearing in the two terms at the 
 righ-hand side of Eq. (\ref{deltap}) simply mean that the two different variations must be 
 taken, respectively, at constant $\varsigma$ (i.e. $\delta \varsigma =0$) and at constant 
 $\rho_{\mathrm t}$ (i.e. $\delta \rho_{\mathrm t} =0$). 

Here is an example of the usefulness of this decomposition.
Consider, for instance, a mixture of CDM particles and radiation. In this case the coefficient of the first term  at the right hand side of Eq. (\ref{deltap}) can be written as 
\begin{equation}
 \biggl(\frac{\delta p_{\mathrm t}}{\delta\rho_{\mathrm t}}\biggr)_{\varsigma} 
 = \frac{1}{3} \biggl(\frac{\delta\rho_{\mathrm r}}{\delta\rho_{\mathrm c} + \delta \rho_{\mathrm r}}\biggr)_{\varsigma},
\label{intermediate1} 
\end{equation}
where we simply used the fact that $\delta p_{\mathrm r} = \delta\rho_{\mathrm r}/3$ and that $\delta \rho_{\mathrm t} = \delta \rho_{\mathrm r} 
+ \delta \rho_{\mathrm c}$. Now, the quantity appearing in Eq. (\ref{intermediate1}) must be evaluated at constant $\varsigma$, i.e. for 
$\delta\varsigma =0$. The specific entropy, in the CDM radiation system, is 
given by $\varsigma = T^3/n_{\mathrm c}$ where $T$ is the temperature 
and $n_{\rm c}$ is the CDM concentration. The relative fluctuations 
of the specific entropy can then be defined and they are 
\begin{equation}
{\mathcal S} = \frac{\delta \varsigma}{\varsigma}= \frac{3}{4} \frac{\delta\rho_{\mathrm r}}{\rho_{\mathrm r}} - 
\frac{\delta \rho_{\mathrm c}}{\rho_{\mathrm c}},
\label{defentr1}
\end{equation}
where it has been used that $\rho_{\mathrm r} \simeq T^4$ and that $\rho_{\mathrm c}\simeq m \, n_{\mathrm c}$ ($m$ is here the typical mass of the CDM particle).
Requiring now that ${\mathcal S} =0$ we do get 
$\delta \rho_{\mathrm c} = (3/4) (\rho_{\mathrm c}/\rho_{\mathrm r}) 
\delta \rho_{\mathrm r}$. Thus, inserting $\delta\rho_{\mathrm c}$ into  
Eq. (\ref{intermediate1}), the following relation can be easily obtained:
\begin{equation}
  \biggl(\frac{\delta p_{\mathrm t}}{\delta\rho_{\mathrm t}}\biggr)_{\varsigma}
  = \frac{4\rho_{\mathrm r}}{3( 3 \rho_{\mathrm c} + 4 \rho_{\mathrm r})} \equiv 
  \frac{p_{\mathrm t}'}{\rho_{\mathrm t}'} = c_{\mathrm s}^2.
\label{soundspeed}
\end{equation}

The second and third equalities  in Eq. (\ref{soundspeed}) 
follow from the definition 
of the total sound speed for the CDM-radiation system.
This occurrence is general and it is not a peculiarity 
of the CDM-radiation system so that we can write, for  an 
arbitrary mixture of relativistic fluids:
\begin{equation}
  \biggl(\frac{\delta p_{\mathrm t}}{\delta\rho_{\mathrm t}}\biggr)_{\varsigma}
=  \frac{p_{\mathrm t}'}{\rho_{\mathrm t}'} = c_{\mathrm s}^2.
\label{soundspeed2}
\end{equation}
The definition of relative entropy fluctuation proposed in Eq. (\ref{defentr1})  is invariant 
under infinitesimal gauge transformations \cite{f74} and it can be generalized 
by introducing two interesting variables namely
\begin{equation}
\zeta_{\mathrm r} = - \psi - {\mathcal H}\frac{\delta \rho_{\mathrm r}}{\rho_{\rm r}'}, \qquad 
\zeta_{\mathrm c} = - \psi - {\mathcal H}\frac{\delta \rho_{\mathrm c}}{\rho_{\rm c}'}.
\label{partialzeta}
\end{equation}
Using the continuity equations for the CDM and for radiation, i.e. 
$\rho_{\mathrm r}' = - 4 {\mathcal H} \rho_{\mathrm r}$ and $\rho_{\mathrm c}' = - 3 {\mathcal H} \rho_{\mathrm c}$,
Eq. (\ref{partialzeta}) can be also written as 
\begin{equation}
\zeta_{\mathrm r} = - \psi + \frac{\delta_{\mathrm r}}{4}, \qquad 
\zeta_{\mathrm c} = - \psi + \frac{\delta_{\mathrm c}}{3},
\label{partialzeta2}
\end{equation}
where $ \delta_{\mathrm r} = \delta \rho_{\mathrm r}/\rho_{\mathrm r}$ and 
$ \delta_{\mathrm c} = \delta\rho_{\mathrm c}/\rho_{\mathrm c}$. Thus,
using Eq. (\ref{partialzeta2}), 
the relative fluctuation in the specific entropy introduced in Eq. (\ref{defentr1}) can also be written as 
\begin{equation}
{\mathcal S} = - 3 (\zeta_{\mathrm c} - \zeta_{\mathrm r}).
\label{defentr2}
\end{equation}
It is a simple exercise to verify that Eqs. (\ref{defentr1}) and (\ref{defentr2}) have indeed the same 
physical content.

Up to now the coefficient of {\em the first term} at the right-hand side of Eq. 
(\ref{deltap}) has been computed. Let us now discuss {\em the second 
term} appearing at the right hand side of Eq. (\ref{deltap}).
Conventionally, the whole second term is often denoted 
by $\delta p_{\mathrm{ nad}}$, i.e. non-adiabatic pressure 
variation.  From Eq. (\ref{defentr1})  defining the relative fluctuation in the specific entropy,
i. e. ${\mathcal S} = \delta\varsigma/\varsigma$, the following 
equation can be written: 
\begin{equation}
\delta p_{\mathrm{ nad}} = 
\biggl(\frac{\delta p_{\mathrm t}}{\delta \varsigma}\biggr)_{\rho_{\mathrm t}} \delta \varsigma \equiv 
\biggl(\frac{\delta p_{\mathrm t}}{{\mathcal S}}\biggr)_{\rho_{\mathrm t}} {\mathcal S}.
\label{deltapnad}
\end{equation}
Now,  ${\mathcal S}$ must be evaluated, inside the round bracket, 
 for $\delta\rho_{\mathrm t} =0$. The result will be  
 \begin{equation}
 \biggl(\frac{\delta p_{\mathrm t}}{{\mathcal S}}\biggr)_{\rho_{\mathrm t}} = \frac{4}{3} \frac{ \rho_{\mathrm c} 
 \,\, \rho_{\mathrm r}}{3 \rho_{\mathrm c} + 4 \rho_{\mathrm r}}.
 \label{deltapnad2}
 \end{equation}
 Recalling the definition of sound speed and using Eq. (\ref{deltapnad2}) into Eq. (\ref{deltapnad}), we do get 
 \begin{equation}
 \delta p_{\mathrm{nad}} = c_{\mathrm s}^2 \rho_{\mathrm c} {\mathcal S}.
 \label{deltapnad3}
 \end{equation}
 
If the mixture of fluids is more complicated the discussion presented so far can be 
easily generalized. If more than two fluids are present, we can still separate, formally, 
the pressure fluctuation as 
\begin{equation}
\delta p_{\mathrm t} = c_{\mathrm s}^2 \delta\rho_{\mathrm t} + \delta p_{\mathrm{nad}}.
\label{deltapnad4}
\end{equation}
However, if more than two fluids are present, the non-adiabatic pressure density 
fluctuation has a more complicated form that reduces to the one previously 
computed in the case of two fluids:
\begin{eqnarray}
&&\delta p_{\mathrm{nad}}= \frac{1}{6 {\mathcal H} \rho_{\mathrm t}'} \sum_{{\mathrm i}\,{\mathrm j}} 
\rho_{\mathrm i}'\,\rho_{\mathrm j}' (c_{\mathrm{ s\,i}}^2 -c_{\mathrm{ s\,j}}^2) {\cal S}_{\mathrm{ i\,j}},
\nonumber\\
&& {\mathcal S}_{{\mathrm i}{\mathrm j}} = - 3 (\zeta_{\mathrm i} - \zeta_{\mathrm j}),\qquad c_{\mathrm{ s\,i}}^2 = 
\frac{p_{\mathrm i}'}{\rho_{\rm i}'},
\label{deltapnad5}
\end{eqnarray}
where ${\cal S}_{\mathrm{ i\,j}}$ are the relative fluctuations in the entropy density 
that can be computed in terms 
of the density contrasts of the individual fluids. The indices ${\mathrm i}$ and ${\mathrm j}$ 
run over all the components of the plasma. Assuming a plasma formed
by photons, neutrinos, baryons and CDM particles we will have 
that various entropy fluctuations are possible. For instance 
\begin{equation}
{\mathcal S}_{\gamma{\mathrm c}} = - 3 (\zeta_{\gamma} - \zeta_{\mathrm c}),\qquad 
{\mathcal S}_{\gamma\nu}  = - 3 (\zeta_{\gamma} - \zeta_{\nu}),\qquad ....
\label{genS}
\end{equation}
where the ellipses stand for all the other possible combinations. From the definition of relative 
entropy fluctuations it appears that
 ${\mathcal S}_{\gamma\nu} = 
- {\mathcal S}_{\nu\gamma}$.  Finally, with obvious notations, while $c_{\mathrm s}^2$ denotes 
the {\em total} sound speed, $c_{{\mathrm s}\,{\mathrm i}}^2$ and $c_{{\mathrm s}\,{\mathrm i}}^2$ denote 
the sound speeds of a generic pair of fluids contributing ${\mathcal S}_{{\mathrm i}{\mathrm j}}$
to $\delta p_{\mathrm{nad}}$, i.e. 
\begin{equation}
c_{\mathrm s}^2 = \frac{p_{\mathrm t}'}{\rho_{\mathrm t}'}, \qquad
 c_{{\mathrm s}\,{\mathrm i}}^2 = \frac{p_{\mathrm i}'}{\rho_{\mathrm i}'}, \qquad c_{{\mathrm s}\,{\mathrm j}}^2 = \frac{p_{\mathrm j}'}{\rho_{\mathrm j}'}.
 \end{equation}
 In the light of Eq. (\ref{genS}), also the physical interpretation of Eq. (\ref{deltapnad4}) becomes more 
 clear. The contribution of $\delta p_{\mathrm{nad}}$ arises because of the inherent multiplicity 
 of fluid present in the plasma. 
Thanks to Eq. (\ref{deltapnad4}) using Eq. (\ref{deltarho})  in Eq. (\ref{DC}) we can obtain the evolution equation 
for $\zeta$ which becomes 
\begin{equation}
\zeta' = - \frac{{\cal H}}{p_{\rm t} + \rho_{\rm t}} \delta p_{\rm nad} + 
\frac{{\cal H}}{p_{\rm t} + \rho_{\rm t}} \biggl( c_{\rm s}^2 - \frac{1}{3}\biggr) \delta\rho_{\rm B} 
- \frac{\theta_{\rm t}}{3}.
\label{zetaevol}
\end{equation}

The evolution equation for ${\mathcal R}$ can also be directly obtained 
by taking the first time derivative of Eq. (\ref{ham2}), i.e. 
\begin{equation}
\zeta' = {\mathcal R}' + \frac{\nabla^2 \psi'}{12 \pi G a^2 ( p_{\mathrm t} + \rho_{\mathrm t})} 
+ \frac{ {\mathcal H} ( 3 c_{\mathrm s}^2 + 1) \nabla^2 \psi}{12 \pi G a^2 (p_{\mathrm t} + \rho_{\mathrm t})}.
\label{derivativezeta}
\end{equation}
By now inserting Eq. (\ref{derivativezeta}) into Eq.  (\ref{zetaevol}) and by using the momentum 
constraint of Eq. (\ref{Mom1}) to eliminate $\theta_{\mathrm t}$ we do get 
the following expression:
\begin{eqnarray}
&& {\mathcal R}' = - \frac{{\mathcal H}}{ p_{\mathrm t} + \rho_{\mathrm t}} \delta p_{\mathrm nad} +
\frac{{\mathcal H}}{p_{\mathrm t} + \rho_{\mathrm t}} \biggl( c_{\mathrm s}^2 - \frac{1}{3}\biggr) \delta\rho_{\mathrm B} 
\nonumber\\
&& - \frac{{\mathcal H} c_{\mathrm s}^2 \nabla^2\psi}{4 \pi G a^2 ( p_{\mathrm t} + 
\rho_{\mathrm t})} + \frac{{\mathcal H} \nabla^2 ( \phi - \psi)}{12 \pi G a^2 ( p_{\mathrm t} + 
\rho_{\mathrm t})}.
\label{Revol}
\end{eqnarray}
It could be finally remarked that Eq. (\ref{Revol}) can be directly derived from Eq. (\ref{tij}). For this purpose 
The definition (\ref{R1}) can be derived once with respect to $\tau$. The obtained result, once 
inserted back into Eq. (\ref{tij}) reproduces Eq. (\ref{Revol}).

\subsection{Evolution of different species}
\label{subs43}

Up to now  the global variables defining the evolution of the system have been discussed in a 
unified perspective. The evolution of the global variables 
is determined by the evolution of the density contrasts and peculiar velocities 
of the different species.  Consequently, in the following paragraphs, 
the evolution of the different species will be addressed.
\subsubsection{Photons and baryon}
\label{subs431}
The evolution equations of the lowest multipoles of the photon-baryon system 
amount, in principle, to the following two sets of equations:
\begin{eqnarray}
&&  \delta_{\mathrm b}' = 3 \psi' - \theta_{\mathrm b},
\label{deltab}\\
&& \theta_{\mathrm b}' + {\mathcal H} \theta_{\mathrm b} = - \nabla^2 \phi + 
\frac{ \vec{\nabla}\cdot[ \vec{J} \times \vec{B}]}{a^4 \rho_{\mathrm b}} + 
\frac{4}{3} \frac{\rho_{\gamma}}{\rho_{\mathrm b}} a n_{\mathrm e} x_{\mathrm e} \sigma_{\mathrm T} 
(\theta_{\gamma} - \theta_{\mathrm b}),
\label{thetab}
\end{eqnarray}
and 
\begin{eqnarray}
&& \delta_{\gamma}' = 4\psi' - \frac{4}{3} \theta_{\gamma},
\label{deltagamma}\\
&& \theta_{\gamma}' + \frac{\nabla^2 \delta_{\gamma}}{4} + \nabla^2 \phi = 
a n_{\mathrm e} x_{\mathrm e} \sigma_{\mathrm T} (\theta_{\mathrm b} - \theta_{\gamma}).
\label{thetagamma}
\end{eqnarray}
Equation (\ref{thetab}) contains, as a source term, the divergence of the Lorentz force 
that can be expressed in terms of $\sigma_{\mathrm B}(\tau, \vec{x})$ and $\Omega_{\mathrm B}(\tau,\vec{x})$, as 
already pointed out in Eqs. (\ref{magndef}).

At early times photons 
and baryons are tightly coupled by Thompson scattering, as it is clear from Eqs. (\ref{thetab}) 
and (\ref{thetagamma}) where $\sigma_{\mathrm T}$ denotes the Thompson cross section 
and $n_{\mathrm e}\,\,x_{\mathrm e}$ the concentration of ionized 
electrons.  To cast light on the physical nature of the tight coupling approximation
let us subtract Eqs. (\ref{thetagamma}) and (\ref{thetab}). The result
will be 
\begin{equation} 
(\theta_{\gamma}- \theta_{\mathrm b})' + a n_{\mathrm e} \,\, x_{\mathrm e} \biggl[ 1 + \frac{4}{3}
\frac{\rho_{\gamma}}{\rho_{\mathrm b}}\biggr] (\theta_{\gamma} - \theta_{\mathrm b}) = 
- \frac{\nabla^2 \delta_{\gamma}}{4} + {\mathcal H} \theta_{\mathrm b} - \frac{ \vec{\nabla}\cdot[\vec{J}\times 
\vec{B}]}{a^4 \rho_{\mathrm b}}.
\label{subtract}
\end{equation}
From Eq. (\ref{subtract}) it is clear that any deviation of $(\theta_{\gamma} - \theta_{\mathrm b})$ 
swiftly decays away. In fact, from Eq. (\ref{subtract}), the characteristic time for the 
synchronization of the baryon and photon velocities is of the order of 
$(x_{\mathrm e} n_{\mathrm e} \sigma_{\mathrm T})^{-1}$ which is small compared with 
the expansion time. In the limit $\sigma_{\mathrm T}\to \infty$ the tight coupling is 
exact and the photon-baryon velocity field is a unique physical entity which will be 
denoted by $\theta_{\gamma{\mathrm b}}$. From the structure of Eq. (\ref{subtract}),
the contribution of the magnetic fields in the MHD limit only enters through the Lorentz 
force while the damping term is always provided by Thompson scattering. 

To derive the evolution equations for the photon-baryon system in the tight coupling 
approximation we can add Eqs. (\ref{thetab}) and (\ref{thetagamma}) taking 
into account that $\theta_{\mathrm b} \simeq \theta_{\gamma} = \theta_{\gamma{\mathrm b}}$.
Of course, also the evolution equations of the density contrasts will depend upon 
$\theta_{\gamma{\mathrm b}}$. Consequently the full set of tightly coupled evolution equations 
for the photon-baryon fluid can be written as:
\begin{eqnarray}
&& \delta_{\gamma}' = 4\psi' - \frac{4}{3} \theta_{\gamma{\mathrm b}}
\label{pb0}\\
&& \delta_{\mathrm b}' = 3 \psi' - \theta_{\gamma{\mathrm b}},
 \label{pb1}\\
&& \theta_{\gamma{\mathrm b}}' + \frac{{\cal H} R_{\rm b}}{(1 + R_{\mathrm b})} \theta_{\gamma{\mathrm b}} + \frac{\nabla^2 \delta_{\gamma}}{4 ( 1 + R_{\rm b})} + 
\nabla^2 \phi = \frac{3}{4} \frac{\vec{\nabla}\cdot[ \vec{J} \times \vec{B}]}{a^4 
\rho_{\gamma} ( 1 + R_{\mathrm b})},
\label{pb2}
\end{eqnarray}
where 
\begin{equation}
R_{\mathrm b}(\tau) = \frac{3}{4} \frac{\rho_{\rm b}(\tau)}{\rho_{\gamma}(\tau)} = 
\biggl( \frac{698}{z + 1}\biggr) \biggl( \frac{h^2\Omega_{\rm b}}{ 0.023}\biggr).
\label{BTPR}
\end{equation}
The set of equations (\ref{pb0}), (\ref{pb1}) and (\ref{pb2}) have to be used in order 
to obtain the correct initial conditions to be imposed on the evolution 
for the integration of the brightness perturbations.

If we assume, effectively, 
that $\sigma_{\mathrm T}\to \infty$ we are working to lowest 
order in the tight coupling approximation. This means that the CMB is effectively isotropic 
in the baryon rest frame. To discuss CMB polarization in the presence of magnetic fields 
one has to go to higher order in the tight coupling expansion. However, as far as the 
problem of initial conditions is concerned, the lowest order treatment suffices, as it 
will be apparent from the subsequent discussion.

\subsubsection{Neutrinos}
\label{subs432}
After neutrino decoupling the (perturbed) neutrino phase space distribution 
evolves according to the collisionless Boltzmann equation. This 
occurrence  implies that to have a closed system of equations 
describing the initial conditions it is mandatory to {\em improve} the fluid 
description by adding to the evolution of the monopole (i.e. the neutrino density 
contrast) and of the dipole (i.e. the neutrino peculiar velocity) also the quadrupole, i.e.
the quantity denoted by $\sigma_{\nu}$ and appearing in the expression of the 
anisotropic stress of the fluid (see Eqs. (\ref{anis1}) and (\ref{parametrization})).

The derivation of the various multipoles of the perturbed neutrino phase 
space distribution is a straightforward (even if a bit lengthy) calculation 
and it has been performed, for the set of conventions employed in the present lecture, 
in Ref. \cite{f74}.  The result is, in Fourier space,
\begin{eqnarray}
&&\delta_{\nu}'= 4\psi' - \frac{4}{3} \theta_{\nu}, 
\label{deltanu}\\
&&\theta_{\nu}'  =\frac{k^2}{4} \nabla^2 \delta_{\nu} + k^2 \phi - k^2 \sigma_{\nu},
\label{thetanu}\\
&&\sigma_{\nu}' = \frac{4}{15} \theta_{\nu} - \frac{3}{10} k {\mathcal F}_{\nu\,3}.
\label{sigmanu}
\end{eqnarray}
In Eq. (\ref{sigmanu}) ${\mathcal F}_{\nu\,3}$ is the octupole of the (perturbed) neutrino
phase space distribution. 
The precise relation of the multipole moments of ${\mathcal F}_{\nu}$ with the density contrast 
and the other plasma quantities is as follows:
\begin{equation}
\delta_{\nu} = {\mathcal F}_{\nu\, 0},\qquad \theta_{\nu} = \frac{3}{4} k {\mathcal F}_{\nu\, 1},\qquad
\sigma_{\nu} = \frac{{\mathcal F}_{\nu\,2}}{2}.
\label{multipoles}
\end{equation}
For multipoles larger than the quadrupole, i.e. $\ell > 2 $ the Boltzmann hierarchy 
reads:
\begin{equation}
{\mathcal F}_{\nu\ell}' = \frac{k}{2\ell +1} [ \ell {\mathcal F}_{\nu(\ell-1)}  - (\ell+1) {\mathcal F}_{\nu (\ell+1)}].
\label{boltzmannhier}
\end{equation}
In principle, to give initial conditions we should specify, at a given time after neutrino decoupling, 
the values of {\em all} the multipoles of the neutrino phase space distribution. In practice, if the 
initial conditions are set deep in the radiation epoch, the relevant variables only extend, for 
the purpose of the initial conditions, up to the octupole. Specific examples will be given in a moment.

\subsubsection{CDM component}
\label{subs433}

The CDM component is in some sense, the easier. In the standard case the evolution equations 
do not contain neither the magnetic field contribution nor the anisotropic stress. 
The evolution of the density contrast and of the peculiar velocity are simply given, in Fourier space, by 
the following pair of equations:
\begin{eqnarray}
&& \delta_{\rm c}' = 3 \psi' - \theta_{\rm c}, 
\label{thetac}\\
&& \theta_{\rm c}' + {\cal H} \theta_{\rm c} =k^2 \phi.
\label{deltac}
\end{eqnarray}

\subsubsection{Magnetized adiabatic and non-adiabatic modes}
\label{subs434}

The evolution equations of the fluid and metric variables will 
now be solved deep in the radiation-dominated epoch 
and for wavelengths much larger than the Hubble radius, i.e. 
$|k\,\tau|\ll 1$.
In the present lecture only the magnetized adiabatic mode will 
be discussed. However, the treatment can be usefully 
extended to the other non-adiabatic modes. For this purpose 
we refer the interested reader to \cite{maxprd206} (see also 
\cite{f73}). Moreover, since this lecture has been conducted within 
the conformally Newtonian gauge, there is no reason 
to change. However, it should be noticed that fully 
gauge-invariant approaches are possible \cite{maxcqg06}.
To give the flavour of the possible simplifications 
obtainable in a gauge-invariant framework we can just 
use gauge-invariant concepts to classify more 
precisely the adiabatic and non adiabatic modes.
For this purpose, in agreement with Eq. (\ref{partialzeta}),
let us define the gauge-invariant density 
contrasts on uniform curvature hypersurfaces for the different 
species of the pre-decoupling plasma:
\begin{eqnarray}
&& \zeta_{\gamma} = - \psi + \frac{\delta_{\gamma}}{4},\qquad \zeta_{\nu} = - \psi + \frac{\delta_{\nu}}{4},
\label{zegazenu}\\
&& \zeta_{\mathrm c} = - \psi + \frac{\delta_{\mathrm c}}{3},\qquad \zeta_{\mathrm b} = - \psi + \frac{\delta_{\mathrm b}}{3}.
\label{zeczeb}
\end{eqnarray}
In terms of the variables of Eqs. (\ref{zegazenu})--(\ref{zeczeb}) the evolution equations 
for the density contrasts, i.e. Eqs. (\ref{pb0}), (\ref{deltanu}), (\ref{deltac}) and (\ref{deltac}), 
acquire a rather symmetric form:
\begin{eqnarray}
&& \zeta_{\gamma} '= - \frac{\theta_{\gamma{\rm b}}}{3},\qquad \zeta_{\nu}' = - \frac{\theta_{\nu}}{3}, 
\label{DCrad}\\
&&\zeta_{\rm c}' = - \frac{\theta_{\rm c}}{3},\qquad \zeta_{{\rm b}}' = - \frac{\theta_{\gamma{\rm b}}}{3}.
\label{DCmat}
\end{eqnarray}
From Eqs. (\ref{DCrad}) and (\ref{DCmat}) we can easily deduce a rather important 
property of fluid mixtures: in the long wavelength limit the relative fluctuations in the 
specific entropy are conserved.
Consider, for instance, the CDM-radiation mode. In this case the non vanishing entropy 
fluctuations are 
\begin{equation}
{\mathcal S}_{\gamma {\mathrm c}} = - 3 ( \zeta_{\gamma } - \zeta_{\mathrm c}), \qquad 
{\mathcal S}_{\nu{\mathrm c}} = - 3 ( \zeta_{\nu } - \zeta_{\mathrm c}).
\label{CMrad}
\end{equation}
Using Eqs. (\ref{DCrad}) and (\ref{DCmat}) the evolution equations for ${\mathcal S}_{\gamma {\mathrm c}}$
and ${\mathcal S}_{\nu {\mathrm c}}$ can be readily obtained and they are 
\begin{equation}
{\mathcal S}_{\gamma {\mathrm c}}' = -(\theta_{\gamma{\mathrm b}}- \theta_{\mathrm c}), \qquad 
{\mathcal S}_{\nu {\mathrm c}}' = - (\theta_{\nu}- \theta_{\mathrm c}).
\label{entrevol}
\end{equation}
Outside the horizon the divergence of the peculiar velocities is ${\mathcal O}(|k\tau|^2)$, so the fluctuations 
in the specific entropy are approximately constant in this limit. This conclusion implies 
that if the fluctuations in the specific entropy are zero, they will still vanish at later times. 
Such a conclusion can be evaded if the fluids of the mixture have a relevant energy-momentum 
exchange or if bulk viscous stresses are present \cite{bv1,bv2}.

A mode is therefore said to be adiabatic iff $\zeta_{\gamma} = \zeta_{\nu}= \zeta_{\mathrm c} = \zeta_{\mathrm b}$. Denoting by $\zeta_{\mathrm i}$ 
and $\zeta_{\mathrm j}$ two generic gauge-invariant density contrasts 
of the fluids of the mixture, we say that the initial conditions 
are non-adiabatic if, at least, we can find a pair of fluids for which 
$\zeta_{\mathrm i} \neq \zeta_{\mathrm j}$.

As an example, let us work out the specific form of the magnetized 
adiabatic mode. Let us consider the situation where the Universe 
is dominated by radiation after weak interactions have fallen out of thermal
equilibrium but before matter-radiation equality. 
This is the period of time where the initial conditions of CMB anisotropies 
are usually set both in the presence and in the absence of a magnetized 
contribution. Since the scale factor goes, in conformal time, as 
$a(\tau)\simeq \tau$ and ${\mathcal H} \simeq \tau^{-1}$, Eq. 
(\ref{Ham1}) can be solved for $|k\tau| \ll 1$. The density 
contrasts can then be determined, in Fourier space, to lowest order in $k\tau$ as:
\begin{eqnarray}
&&
\delta_{\gamma} = \delta_{\nu} =
 - 2\phi_{\mathrm i} - R_{\gamma} \Omega_{\rm B}, 
\nonumber\\
&& 
\delta_{\rm b} = \delta_{\rm c} = - \frac{3}{2} \phi_{\mathrm i} 
- \frac{3}{4} R_{\gamma} \Omega_{\mathrm B},
\label{DCad}
\end{eqnarray}
where the fractional contribution of photons to the radiation plasma, i.e. $R_{\gamma}$ has been introduced and it is related to $R_{\nu}$, i.e. 
the fractional contribution of massless neutrinos, as
 \begin{eqnarray}
&& R_{\gamma} = 1 - R_{\nu}, \qquad R_{\nu} = \frac{r}{1 + r},
\nonumber\\
&& r= \frac{7}{8} N_{\nu} \biggl(\frac{4}{11}\biggr)^{4/3} \equiv  0.681 \biggl(\frac{N_{\nu}}{3}\biggr).
\label{Rnu}
\end{eqnarray}
In Eq. (\ref{DCad}) $\phi_{\mathrm i}(k)$ denotes the initial value 
of the metric fluctuation in Fourier space. It is useful to remark 
that we have treated neutrinos as part of the radiation background.
If neutrinos have a mass in the meV range, they are nonrelativistic 
today, but they will be counted as radiation prior to matter-radiation 
equality. Concerning Eq. (\ref{DCad}) the last remark is that, of course, 
we just kept the lowest order in $|k\tau| <1$. It is 
possible, however, to write the solution to arbitrary 
order in $|k\tau$ as explicitly shown in Ref. \cite{f73}.
 
Let us then write Eq. (\ref{anis1}) in Fourier space and let us 
take into account that the background is dominated by radiation. The 
neutrino quadrupole is then determined to be 
\begin{eqnarray}
\sigma_{\nu} = - \frac{R_{\gamma}}{R_{\nu}} \sigma_{\mathrm B} + 
\frac{k^2 \tau^2}{6 R_{\nu}} ( \psi_{\mathrm i} - \phi_{\mathrm i}),
\label{anis2}
\end{eqnarray}
where $\psi_{\mathrm i}(k)$ is the initial (Fourier space) 
value of the metric fluctuation defined in Eq. (\ref{longfl}).

Let us then look for the evolution of the divergences of the 
peculiar velocities of the different species. Let us therefore 
write Eqs. (\ref{pb2}), (\ref{thetanu}) and (\ref{thetac}) 
in Fourier space. By direct integration the following result
can be obtained:
\begin{eqnarray}
&&\theta_{\gamma{\rm b}} = \frac{k^2 \tau}{4} [ 2 \phi_{\rm i} + R_{\nu} \Omega_{\rm B} - 4 \sigma_{\rm B} ], 
\label{VF1}\\
&& \theta_{\nu} = \frac{k^2 \tau}{2}\biggl[ \phi_{\rm i}- \frac{R_{\gamma} \Omega_{\rm B}}{2} \biggr]  
+ k^2 \tau \frac{R_{\gamma}}{R_{\nu}} \sigma_{\rm B},
\label{VF2}\\
&&\theta_{\rm c} = \frac{k^2 \tau}{2}\phi_{\rm i}.
\label{VF3}
\end{eqnarray}
As a consistency check of the solution, Eqs. (\ref{VF1}), (\ref{VF2}) 
and (\ref{VF3}) can be inserted into Eq. (\ref{Mom1}). 
Let us therefore write Eq. (\ref{Mom1}) in Fourier space 
\begin{equation}
k^2 {\mathcal H}\phi_{\mathrm i}= 4\pi G a^2 \biggl[\frac{4}{3} \rho_{\gamma}( 1  + 
\rho_{\mathrm b}) \theta_{\gamma{\mathrm b}} + \frac{4}{3} \rho_{\nu} 
\theta_{\nu} + \rho_{\mathrm c} \theta_{\mathrm c} \biggr],
\label{Mom2}
\end{equation}
where we used that $\psi_{\mathrm i}'=0$ and we also used 
the tight-coupling approximation since $\theta_{\gamma}= \theta_{\mathrm b}= 
\theta_{\gamma{\mathrm b}}$. Notice that in Eq. (\ref{Mom1}) the 
term arising from the Poynting vector has been neglected.
This approximation is rather sound within the present MHD treatment.
 In Eq. (\ref{Mom2}) 
$R_{\mathrm b} \ll 1$ (see Eq. (\ref{BTPR}) for the definition 
of $R_{\mathrm b}$) since we are well before matter-radiation equality. 
The same observation can be made for the CDM contribution which 
is negligible in comparison with the radiative contribution 
provided by photons and neutrinos.  Taking into account these 
two observations we can rewrite Eq. (\ref{Mom2}) as 
\begin{equation}
k^2 {\mathcal H} \phi_{\mathrm i} = 2 {\mathcal H}^2 (R_{\gamma} \theta_{\gamma{\mathrm b}} + R_{\nu} \theta_{\nu}),
\label{Mom3}
\end{equation}
where Eqs. (\ref{FL1}) and (\ref{FL2}) have been used.
Inserting then Eqs. (\ref{VF1}) and (\ref{VF2}) into Eq. (\ref{Mom3}) 
it can be readily obtained that the left hand side exactly equals 
the right hand side, so that the momentum constraint is enforced.

The  final equation to be solved is the one describing the evolution 
of the anisotropic stress, i.e. Eq. (\ref{sigmanu}).
Inserting Eqs. (\ref{anis2}) and (\ref{VF2}) into Eq. (\ref{sigmanu}) 
we do get an interesting constraint on the initial conditions  on the two longitudinal 
fluctuations of the geometry introduced in Eqs. (\ref{longfl}), namely:
\begin{equation}
\psi_{\rm i} = \phi_{\rm i} \biggl( 1 + \frac{2}{5} R_{\nu}\biggr) + \frac{R_{\gamma}}{5}( 4 \sigma_{\rm B} - R_{\nu} \Omega_{\rm B}).
\label{mismatch}
\end{equation}
Concerning the magnetized adiabatic mode the following 
comments are in order:
\begin{itemize}
\item{} the peculiar velocities are always suppressed, 
with respect to the other terms of the solution, by a 
factor $|k\tau|$ which is smaller than $1$ when the 
wavelength is larger than the Hubble radius;
\item{} in the limit $\sigma_{\mathrm B} \to 0$ and $\Omega_{\mathrm B}
\to 0$ the magnetized adiabatic mode presented here 
reproduces the well known standard results (see 
for instance \cite{f72});
\item{} the difference between the two longitudinal 
fluctuations of the metric is due, both to the presence 
of magnetic and fluid anisotropic stresses;
\item{} the longitudinal fluctuations 
of the geometry are both constant outside the horizon and prior to 
matter-radiation equality; this result still holds in the presence 
of a magnetized contribution as it is clearly demonstrated by 
the analytic solution presented here.
\end{itemize}

The last interesting exercise we can do with the obtained 
solution is to compute the important variables ${\mathcal R}$ 
and $\zeta$ introduced, respectively,  in Eqs. (\ref{R1}) and (\ref{zeta2}).
Since both $\psi$ and $\phi$ are constants for $|k\tau|< 1$ and 
for $\tau < \tau_{\mathrm{eq}}$, also ${\mathcal R}$ will be constant. 
In particular, by inserting Eq. (\ref{mismatch}) into Eq. (\ref{R1}), 
the following expression can be obtained:
\begin{equation}
{\mathcal R}_{\mathrm i} = - \frac{3}{2} \biggl( 1 + \frac{4}{15} R_{\nu} \biggr) \phi_{\mathrm i} - 
\frac{R_{\gamma}}{5} ( 4 \sigma_{\mathrm B} - R_{\nu} \Omega_{\mathrm B}),
\label{Ri}
\end{equation}
where ${\mathcal R}_{\mathrm i}(k)$ denotes the initial value, in Fourier space, of the curvature 
perturbations. In numerical studies it is sometimes useful to relate the initial values 
of $\phi$ and $\psi$, i.e. $\phi_{\mathrm i}$ and $\psi_{\mathrm i}$ to 
${\mathcal R}_{\mathrm i}$. This relation is expressed by the following pair of formulae 
that can be derived by inverting Eq. (\ref{Ri}) and by using 
Eq. (\ref{mismatch}):
\begin{eqnarray}
&& \phi_{\mathrm i} = - \frac{10}{15 + 4 R_{\nu}} {\mathcal R}_{\mathrm i} - \frac{2 R_{\gamma}( 4 \sigma_{\mathrm B} - R_{\gamma} \Omega_{\mathrm B})}{15 + 4 R_{\nu}},
\nonumber\\
&& \psi_{\mathrm i}  = - 
2 \frac{5 + 2 R_{\nu}}{15 + 4 R_{\nu}} {\mathcal R}_{\mathrm i} - 
\frac{2}{5} \frac{R_{\gamma} (5 + 2 R_{\nu})}{15 + 4 R_{\nu}} 
( 4 \sigma_{\mathrm B} - R_{\gamma} \Omega_{\mathrm B}).
\end{eqnarray}
From the Hamiltonian constraint written in the form (\ref{ham2}) it is easy to deduce, in the 
limit $|k\tau|\ll 1$ that $\zeta_{\mathrm i}(k) = {\mathcal R}_{\mathrm i}(k)$
The same result can be obtained through a different, but also instructive, path.
Consider the definition of $\zeta$ given either in Eq. (\ref{zeta1}) or (\ref{zeta2}).
The variable $\zeta$ can be expressed in terms of the partial density 
contrasts defined in Eqs. (\ref{zegazenu}) and (\ref{zeczeb}). More 
precisely, from the definitions of the two sets of variables it is easy to show that 
\begin{equation}
\zeta = \frac{\rho_{\nu}'  \zeta_{\nu} + \rho_{\gamma}' \zeta_{\gamma} + 
\rho_{\rm c}' \zeta_{\rm c} + \rho_{\rm b}' \zeta_{\rm b}}{\rho_{\rm t}'} + \zeta_{\rm B}
 ,\qquad \zeta_{\rm B} = \frac{\delta\rho_{\rm B}}{3( p_{\rm t} + \rho_{\rm t})}.
\label{zetapartial1}
\end{equation}
Thus, to obtain $\zeta$ it suffices to find $\zeta_{\gamma}$, $\zeta_{\nu}$, $\zeta_{\mathrm b}$ 
and $\zeta_{\mathrm c}$ evaluated at the initial time and on the adiabatic solution.
Using Eqs. (\ref{DCad}) and (\ref{mismatch}) into Eqs. (\ref{zegazenu}) and (\ref{zeczeb})
we obtain, as expected, 
\begin{equation}
\zeta_{\gamma} = \zeta_{\nu} = \zeta_{\mathrm c} = \zeta_{\mathrm b} = - \biggl( \psi_{\mathrm i} + 
\frac{\phi_{\mathrm i}}{2} \biggr) + \frac{R_{\gamma}}{4} \Omega_{\mathrm B}.
\label{zetapartial2}
\end{equation}
This result was expected, since, as previously stressed, for the adiabatic mode 
all the partial density contrasts must be equal. Inserting now Eq. (\ref{zetapartial2}) 
into Eq. (\ref{zetapartial1}) and recalling that the CDM and baryon contributions 
vanish deep in the radiation epoch, we do get
\begin{equation}
\zeta = - \biggl( \psi_{\mathrm i} + \frac{\phi_{\mathrm i}}{2} \biggr) = {\mathcal R}_{\mathrm i},
\label{zetafinal}
\end{equation}
where the last equality follows from the definition of (\ref{R1}) evaluated deep in the 
radiation epoch and for the adiabatic solution derived above.

Up to now, as explained, attention has been given to the magnetized adiabatic mode. 
There are, however, also other non adiabatic modes that can enter the game.
We will not go, in this lecture, through the derivation of the various non-adiabatic modes. 
It is however useful to give at least the result in the case of the magnetized 
CDM-radiation mode. In such a case the full solution to the same set of equations 
admitting the adiabatic solutions can be written as 
 For the case of the CDM-radiation mode the solution, in the limit 
$\tau <\tau_{1}$ and $k\tau < 1$ can be written as 
\begin{eqnarray}
&& \phi= \phi_{1} \biggl(\frac{\tau}{\tau_{1}}\biggr),\qquad  \psi= \psi_{1} \biggl(\frac{\tau}{\tau_{1}}\biggr),
\nonumber\\
&& \delta_{\gamma} = \delta_{\nu} = 4 \psi_{1} \biggl(\frac{\tau}{\tau_{1}}\biggr)- 
R_{\gamma}\Omega_{\rm B},
\nonumber\\
&& \delta_{\rm c} = - \biggl[ {\mathcal   S}_{*} + \frac{3}{4} R_{\gamma} \Omega_{\rm B}\biggr] + 3 \psi_{1} \biggl(\frac{\tau}{\tau_{1}}\biggr),
\nonumber\\
&& \delta_{\rm b} = 3 \psi_{1} \biggl(\frac{\tau}{\tau_{1}}\biggr) - \frac{3}{4} R_{\gamma} \Omega_{\rm B},
\nonumber\\
&& \theta_{\rm c} = \frac{k^2 \tau_{1}}{3} \phi_{1} \biggl(\frac{\tau}{\tau_{1}}\biggr)^2 ,
\nonumber\\
&&\theta_{\gamma{\rm b}} =\frac{k^2 \tau_1}{2} (\phi_1 + \psi_{1}) \biggl(\frac{\tau}{\tau_{1}}\biggr)^2  + \frac{k^2 \tau}{4} [ R_{\nu} \Omega_{\rm B} - 4 \sigma_{\rm B}],
\nonumber\\
&& \theta_{\nu} = \frac{k^2 \tau_1}{2} (\phi_1 + \psi_{1}) \biggl(\frac{\tau}{\tau_{1}}\biggr)^2  + \frac{k\tau}{4} \biggl(4 \frac{R_{\gamma}}{R_{\nu}} \sigma_{\rm B} - \Omega_{\rm B}\biggr),
\nonumber\\
&&  {\mathcal   F}_{\nu3} = \frac{8}{9} k\tau \biggl[ 4 \frac{R_{\gamma}}{R_{\nu}} \sigma_{\rm B} 
-\Omega_{\rm B}\biggr],
\nonumber\\
&&\sigma_{\nu} = -\frac{R_{\gamma}}{R_{\nu}} \sigma_{\rm B} + 
\frac{k^2 \tau_{1}^2}{6 R_{\nu}} ( \psi_{1} -\phi_1) \biggl(\frac{\tau}{\tau_{1}}\biggr)^3,
\label{CDMNAD1}
\end{eqnarray}
where 
\begin{eqnarray}
&&\psi_{1} = \frac{15 + 4 R_{\nu}}{8( 15 + 2 R_{\nu})}\biggl[ {\mathcal   S}_{*} + 
\frac{3}{4} R_{\gamma} \Omega_{\rm B} \biggr], 
\nonumber\\
&&\phi_{1} = \frac{15 - 4 R_{\nu}}{8( 15 + 2 R_{\nu})}\biggl[ {\mathcal   S}_{*} + 
\frac{3}{4} R_{\gamma} \Omega_{\rm B} \biggr].
\label{CDMNAD2}
\end{eqnarray}
In Eq. (\ref{CDMNAD1}) the following notation for the non-vanishing 
entropy fluctuations has been employed:
\begin{equation}
{\mathcal S}_{{\mathrm c}\gamma}= {\mathcal S}_{{\mathrm c}\nu} = {\mathcal S}_{*}.
\label{defS}
\end{equation}

In deriving Eq. (\ref{CDMNAD1}) it is practical to use a form of the scale factor (obtained by solving Eqs. 
(\ref{FL1}), (\ref{FL2}) and (\ref{FL3}) for a mixture of matter and radiation)
which explicitly interpolates between a radiation-dominated regime and a matter-dominated regime:
\begin{equation}
a(\tau) = a_{\rm eq} \biggl[ \biggl(\frac{\tau}{\tau_1}\biggr)^2 + 2 \biggl(\frac{\tau}{\tau_{1}}\biggr) \biggr], \qquad 1 + z_{\rm eq} = \frac{1}{a_{\rm eq}} = \frac{h^2 \Omega_{{\rm m}0}}{h^2 \Omega_{{\rm r}0}},
\label{SF}
\end{equation}
 where $\Omega_{{\rm m}0}$ and $\Omega_{{\rm r}0}$ are evaluated at the present time and the 
 scale factor is normalized in such a way that $a_0= 1$. In Eq. (\ref{SF}) $\tau_{1} = (2/H_{0}) 
 \sqrt{a_{\rm eq}/\Omega_{{\rm m}0}}$. In terms of $\tau_{1}$ the equality time is 
 \begin{equation}
 \tau_{\rm eq} = (\sqrt{2} -1) \tau_{1} = 119.07 \,\, \biggl( \frac{h^2 \Omega_{{\rm m}0}}{0.134}\biggr)^{-1}\,\, {\rm Mpc},
 \label{SF1}
 \end{equation}
 i.e.  $2 \tau_{\rm eq} \simeq \tau_{1}$.
 In this framework the total optical depth from the present to the critical recombination epoch, i.e. $ 800< z < 1200$
 can be approximated analytically,  as discussed in \cite{wh}.  By defining the redshift of decoupling as the one where 
 the total optical depth is of order 1, i.e. $\kappa(z_{\rm dec},0) \simeq 1$, we will have, approximately 
 \begin{equation}
 z_{\rm dec}  \simeq 1139 \biggl(\frac{ \Omega_{\rm b}}{0.0431}\biggr)^{- \alpha_{1}}, \,\,\,\,\,\,\,\,\alpha_{1} = \frac{0.0268}{0.6462 + 0.1125\ln{(\Omega_{\rm b}/0.0431)}},
 \label{zdec}
\end{equation}
 where $h = 0.73$.  From Eqs. (\ref{zdec}) and (\ref{SF}) it follows 
 that for $ 1100 \leq z_{\rm dec} \leq 1139$, $ 275\,\, \mathrm{Mpc} \leq \tau_{\rm dec} 
 \leq 285\,\, \mathrm{Mpc}$. 
 
Equations (\ref{SF}) and (\ref{SF1}) will turn out to be relevant for the effective numerical integration 
of the brightness perturbations which will be discussed later on.
For numerical purposes  the late-time cosmological parameters will be fixed, for a spatially flat Universe, as
\footnote{The values of the cosmological parameters introduced in Eq. (\ref{par})
are compatible with the ones estimated from WMAP-3 \cite{wmap3,wmap1,wmap2} 
in combination with the ``Gold" sample of SNIa \cite{riess} consisting of $157$ 
supernovae (the furthest being at redshift z = 1.75). We are aware of the fact 
that WMAP-3 data alone seem to favour a slightly smaller value of $\omega_{\rm m}$ 
(i.e. $0.126$). Moreover, WMAP-3 data may also have slightly 
different implications if combined with supernovae of the SNLS project \cite{astier}.
The values given in Eq. (\ref{par}) will just be used for a realistic numerical illustration
of the methods developed in the present investigation.}
\begin{equation}
\omega_{\gamma} = 2.47\times 10^{-5},\qquad \omega_{\rm b} = 0.023,\qquad 
\omega_{\rm c} = 0.111, \qquad \omega_{\rm m} = \omega_{\rm b} + \omega_{\rm c},
\label{par}
\end{equation} 
where $\omega_{X} = h^2 \Omega_{X}$ and 
$\Omega_{\Lambda}= 1- \Omega_{\rm m}$; the present 
value of the Hubble parameter $H_{0}$ will be fixed, for numerical 
estimates,  to $73$ in units 
of ${\rm km}/({\rm sec}\, {\rm Mpc})$.

\subsubsection{Transfer matrix and Sachs-Wolfe plateau}
\label{subs424}

Before presenting some numerical approaches suitable for the analysis of magnetized CMB anisotropies 
it is useful to discuss a class of analytical estimates that allow the calculation of the so-called 
Sachs-Wolfe plateau. 
The idea, in short, is very simple. We have the evolution equation for $\zeta$ given in Eq. 
(\ref{zetaevol}). This evolution 
equation can be integrated across the matter-radiation transition using the interpolating 
form of the scale factor proposed in Eq. (\ref{SF}).

Consider, first, the case of the magnetized adiabatic mode where $\delta p_{\mathrm{nad}} =0$.
Deep in the radiation-dominated epoch, for $\tau \ll \tau_{\rm eq}$, 
$c_{\rm s}^2 \to 1/3$ and, from Eq. (\ref{zetaevol}), $\zeta'=0$, so that 
\begin{equation}
\zeta = \zeta_{\rm i} \simeq {\cal R}_{\rm i},\qquad \zeta_{\rm i} = - \frac{3}{2}\phi_{\rm i}\biggl( 1 + \frac{4}{15} R_{\nu}\biggr) - \frac{R_{\gamma}}{5}( 4 \sigma_{\rm B} - R_{\nu} \Omega_{\rm B}).
\end{equation}
When the Universe becomes matter-dominated, after $\tau_{\rm eq}$,
$c_{\rm s}^2 \to 0$ and the second term at the right hand side of Eq. 
(\ref{zetaevol}) does contribute significantly at decoupling (recall that 
for $h^2 \Omega_{\rm matter} = 0.134$, $\tau_{\rm dec} = 2.36 \, \tau_{\rm eq}$).
Consequently, from Eq. (\ref{zetaevol}), recalling that $c_{\rm s}^2 = 
4 a_{\rm eq}/[ 3 ( 3 a + 4 a_{\rm eq})]$, we obtain 
\begin{equation}
 \zeta_{\rm f} = \zeta_{\rm i} - \frac{3 \,a\,R_{\gamma}\, \Omega_{\rm B}}{4 ( 3 a + 4 a_{\rm eq})},\qquad \Omega_{{\rm B}\,{\rm f}} = \Omega_{{\rm B}\,{\rm i}}. 
 \label{ZF}
\end{equation}
The inclusion of one (or more) non-adiabatic modes changes 
the form of Eq. (\ref{zetaevol}) and, consequently, the related solution 
(\ref{ZF}). For instance, in the case of the CDM-radiation non-adiabatic mode 
the relevant terms arising in the sum (\ref{deltapnad5}) are 
${\cal S}_{{\rm c}\gamma} = 
{\cal S}_{{\rm c}\nu} = {\cal S}_{\rm i}$ where ${\cal S}_{i}$ is the (constant) 
fluctuation in the relative entropy density initially present 
(i.e. for $\tau \ll \tau_{\rm eq}$). If  this is the case
 $\delta p_{\rm nad} =  c_{\rm s}^2 \rho_{\rm c} {\cal S}_{i}$ and Eq.
(\ref{zetaevol}) can be easily solved. The transfer matrix for magnetized CMB anisotropies can then be written as 
\begin{equation}
\pmatrix{
 \zeta_{{\rm f}} \cr
{\cal S}_{{\rm f}}\cr
\Omega_{{\rm B}\,{\rm f}}} = 
\pmatrix{{\cal M}_{\zeta \zeta} & {\cal M}_{\zeta{\cal S}} & 
{\cal M}_{\zeta {\rm B}}\cr
0 & {\cal M}_{{\cal S}{\cal S}}& {\cal M}_{{\cal S} {\rm B}}\cr
0 & 0& {\cal M}_{{\rm B} {\rm B}} }
\pmatrix{
 \zeta_{{\rm i}}\cr
{\cal S}_{{\rm i}}\cr
\Omega_{{\rm B}\,{\rm i}}}.
 \label{MAT1}
\end{equation}
In the case of a mixture of (magnetized) adiabatic and CDM-radiation
modes, we find, for $a > a_{\rm eq}$ 
\begin{eqnarray}
&&{\cal M}_{\zeta \zeta} \to 1, \qquad {\cal M}_{\zeta {\cal S}} \to - \frac{1}{3},\qquad {\cal M}_{\zeta{\rm B}}
\to - \frac{R_{\gamma}}{4},
\nonumber\\
&&{\cal M}_{{\cal S}{\cal S}} \to 1,\qquad {\cal M}_{{\cal S}{\rm B}}\to 0,
\label{MAT2}
\end{eqnarray}
and  ${\cal M}_{{\rm B}{\rm B}} \to 1$.
Equations (\ref{MAT1}) and (\ref{MAT2}) may be used, for instance, 
to obtain the magnetized curvature and entropy fluctuations 
at photon decoupling in terms of the same quantities evaluated 
for $\tau \ll \tau_{\rm eq}$. A full numerical analysis of the problem 
confirms the analytical results summarized by Eqs. (\ref{MAT1}) and 
(\ref{MAT2}). The most general initial condition for CMB anisotropies will then be a combination of (correlated) fluctuations receiving contribution 
from $\delta p_{\rm nad}$ and from the fully inhomogeneous 
magnetic field. To illustrate this point, the form of the Sachs-Wolfe (SW) plateau in the sudden decoupling limit will now be discussed. 

To compute the SW contribution we need 
to solve the evolution equation of the monopole of the temperature 
fluctuations in the tight coupling limit, i.e. from Eqs. (\ref{pb1}) and 
(\ref{pb2}),  
\begin{equation}
\delta_{\gamma}''+ \frac{{\cal H} R_{\rm b}}{ 1 + R_{\rm b}} \delta_{\gamma}' + 
\frac{k^2}{3} \frac{\delta_{\gamma}}{1 + R_{\rm b}} = 4 \psi'' + 
\frac{4 {\cal H} R_{\rm b}}{1 + R_{\rm b}} \psi' - \frac{4}{3} k^2 \phi - 
\frac{k^2}{3 ( 1 + R_{\rm b})} ( \Omega_{\rm B} - 4 \sigma_{\rm B}).
\label{monopole}
\end{equation}
In the sudden decoupling approximation the visibility function, i.e. ${\cal K}(\tau) = \kappa'(\tau) e^{-\kappa(\tau)}$  and the optical depth, i.e. $\epsilon^{-\kappa(\tau)}$ are approximated, respectively, by 
$\delta(\tau - \tau_{\rm dec})$ and by $\theta(\tau - \tau_{\rm dec})$ (see \cite{old2,old3} for an estimate of the 
width of the last scattering surface). 
The power spectra of $\zeta$, ${\cal S}$ and $\Omega_{\rm B}$ are given, respectively, by:
\begin{eqnarray}
&&{\cal P}_{\zeta}(k) = {\cal A}_{\zeta} \biggl(\frac{k}{k_{\rm p}}\biggr)^{n_{r}-1},\qquad
{\cal P}_{{\cal S}}(k) 
 = {\cal A}_{\cal S} \biggl(\frac{k}{k_{\rm p}}\biggr)^{n_{s}-1},
\label{RPS}\\
&&{\cal P}_{\Omega}(k) = {\cal F}(\varepsilon) \overline{\Omega}_{{\rm B}\, L}^2 \biggl(\frac{k}{k_{L}}\biggr)^{2 \varepsilon},
\label{OMPS}
\end{eqnarray}
where ${\cal A}_{\zeta}$, ${\cal A}_{{\cal S}}$ and $\overline{\Omega}_{{\rm B}\,L}$ 
are  constants and 
\begin{eqnarray}
&&{\cal F}(\varepsilon) = \frac{4(6 - \varepsilon) ( 2 \pi)^{ 2 \varepsilon}}{\varepsilon ( 3 - 2 \varepsilon)
 \Gamma^2(\varepsilon/2)},
\nonumber\\ 
&&\overline{\Omega}_{{\rm B}\,\, L} = \frac{\rho_{{\rm B}\,L}}{\overline{\rho}_{\gamma}}, 
\qquad \rho_{{\rm B}\,\, L}=\frac{ B_{L}^2}{8\pi},\qquad \overline{\rho}_{\gamma}= a^{4}(\tau) \rho_{\gamma}(\tau).
\label{DEFB}
\end{eqnarray}
To deduce Eqs. (\ref{RPS}), (\ref{OMPS}) and (\ref{DEFB}) the magnetic field has been regularized, according to a common practice \cite{f3,f4,f4a},  
over a typical comoving scale $L = 2\pi/k_{L}$  with a Gaussian window function and it has been assumed that the magnetic field intensity is stochastically distributed as 
\begin{equation}
\langle B_{i}(\vec{k},\tau) B^{j}(\vec{p},\tau) \rangle =  \frac{2\pi^2}{k^3} \,P_{i}^{j}(k)\, P_{\rm B}(k,\tau)\, \delta^{(3)}(\vec{k} + \vec{p}),
\label{Bcorr}
\end{equation}
where
\begin{equation}
P_{i}^{j}(k) = \biggl(\delta_{i}^{j} - \frac{k_{i}k^{j}}{k^2} \biggr),\qquad P_{\rm B}(k,\tau) = A_{\rm B} 
\biggl(\frac{k}{k_{\rm p}}\biggr)^{\varepsilon}.
\label{MPS}
\end{equation}
As a consequence of Eq. (\ref{Bcorr}) the magnetic field does not break
the spatial isotropy of the background geometry.
The quantity $k_{\rm p}$ appearing in Eqs. (\ref{RPS}) and (\ref{MPS}) is 
conventional pivot scale that is $0.05\, {\rm Mpc}$(see \cite{f60,f61,f62} for a discussion of other possible choices). Equations (\ref{OMPS}) and (\ref{DEFB}) hold 
for $0<\varepsilon < 1$. In this limit the ${\cal P}_{\Omega}(k)$ (see Eq. (\ref{OMPS})) is nearly scale-invariant 
(but slightly blue). This means that the effect of the magnetic and thermal 
diffusivity scales (related, respectively, to the finite value of the 
conductivity and of the thermal diffusivity coefficient) do not affect 
the spectrum \cite{f4a}. In the opposite limit, i.e. $\varepsilon \gg 1$ the value of the mode-coupling integral appearing in the two-point function of the magnetic energy density (and of the magnetic anisotropic stress) is dominated by ultra-violet effects related to the mentioned diffusivity scales \cite{f4a}. Using then Eqs. (\ref{RPS}),(\ref{OMPS}) and (\ref{DEFB}) 
the $C_{\ell}$ can be computed for the region of the SW plateau (i.e.  for multipoles $\ell < 30$):
\begin{eqnarray}
&&C_{\ell} = \biggl[ \frac{{\cal A}_{\zeta}}{25} \,{\cal Z}_{1}(n_{r},\ell)  +
\frac{9}{100} \, R_{\gamma}^2  \overline{\Omega}^{2}_{{\rm B}\,L} {\cal Z}_{2}(\epsilon,\ell) - 
\frac{4}{25} \sqrt{{\cal A}_{\zeta} {\cal A}_{{\cal S}}} {\cal Z}_{1}(n_{r s},\ell) \cos{\gamma_{r s}}
\nonumber\\
&& +\frac{4}{25} \,{\cal A}_{{\cal S}} \,
{\cal Z}_{1}(n_{s},\ell)- 
\frac{3}{25} \sqrt{ {\cal A}_{\zeta}} \, R_{\gamma} \,\overline{\Omega}_{{\rm B}\, L}\,{\cal Z}_{3} (n_{r},\varepsilon, \ell) \cos{\gamma_{br}} 
\nonumber\\
&& + \frac{6}{25} \sqrt{ {\cal A}_{{\cal S}}}\,R_{\gamma} \overline{\Omega}_{{\rm B}\,L}\, {\cal Z}_{3}(n_{s},\varepsilon, \ell)\cos{\gamma_{b s}} \biggr],
\label{SWP}
\end{eqnarray}
where the functions ${\cal Z}_{1}$, ${\cal Z}_{2}$ and ${\cal Z}_{3}$ 
\begin{eqnarray}
&&{\cal Z}_{1}(n,\ell) = \frac{\pi^2}{4} \biggl(\frac{k_0}{k_{\rm p}}\biggl)^{n-1} 2^{n} \frac{\Gamma( 3 - n) \Gamma\biggl(\ell + 
\frac{ n -1}{2}\biggr)}{\Gamma^2\biggl( 2 - \frac{n}{2}\biggr) \Gamma\biggl( \ell + \frac{5}{2} - \frac{n}{2}\biggr)},
\label{Z1}\\
&&{\cal Z}_{2}(\varepsilon,\ell) = \frac{\pi^2}{2} 2^{2\varepsilon} {\cal F}(\varepsilon) \biggl( \frac{k_{0}}{k_{L}}\biggr)^{ 2 \varepsilon} \frac{ \Gamma( 2 - 2 \varepsilon) 
\Gamma(\ell + \varepsilon)}{\Gamma^2\biggl(\frac{3}{2} - \varepsilon\biggr) \Gamma(\ell + 2 -\varepsilon)},
\label{Z2}\\
&&{\cal Z}_{3}(n,\varepsilon,\ell) =\frac{\pi^2}{4} 2^{\varepsilon} 2^{\frac{n +1}{2}} \,\sqrt{{\cal F}(\varepsilon)}\, \biggl(\frac{k_{0}}{k_{L}}\biggr)^{\varepsilon} \biggl(\frac{k_{0}}{k_{\rm p}}\biggr)^{\frac{n + 1}{2}} \times
\nonumber\\
&& \times \frac{ \Gamma\biggl(\frac{5}{2} - \varepsilon - \frac{n}{2}\biggr) \Gamma\biggl( \ell + 
\frac{\varepsilon}{2} + \frac{n}{4} - \frac{1}{4}\biggr)}{\Gamma^2\biggl(\frac{7}{4} - \frac{{\varepsilon}}{2} - \frac{n}{4}\biggr)
\Gamma\biggl( \frac{9}{4} + \ell - \frac{\varepsilon}{2} - \frac{n}{4} \biggr)},
\label{Z3}
\end{eqnarray}
are defined in terms of the magnetic tilt $\varepsilon$ and of a generic spectral index $n$ which may correspond, depending on the specific contribution, either to $n_{r}$ (adiabatic spectral index), or to $n_{s}$(non-adiabatic spectral index)  or even to $n_{rs} = (n_{r} + n_{s})/2$ (spectral index of the cross-correlation).  In Eq. (\ref{SWP}) $\gamma_{rs}$, $\gamma_{br}$ and $\gamma_{s b}$ are the correlation angles. In the absence of magnetic and non-adiabatic contributions and for Eqs. (\ref{SWP}) and Eq. (\ref{Z1}) imply that for $n_{r}=1$ (Harrison-Zeldovich spectrum) $ \ell (\ell +1) C_{\ell}/2\pi = 
{\cal A}_{\zeta}/25$ and WMAP data \cite{f5} would imply that 
${\cal A}_{\zeta}= 2.65 \times 10^{-9}$. Consider then the 
physical situation where on top of the adiabatic mode there is a magnetic contribution. If there is no correlation 
between the magnetized contribution and the adiabatic contribution, i.e. 
$\gamma_{b r} =\pi/2$, the SW plateau will be enhanced in comparison 
with the case when magnetic fields are absent.  The same situation 
arises when the two components are anti-correlated (i.e. $\cos{\gamma_{br}}<0$).
However, if the fluctuations are positively correlated (i.e.  
$\cos{\gamma_{b r}}>0$) the cross-correlation adds negatively to the 
sum of the two autocorrelations of $\zeta$ and $\Omega_{\rm B}$ so that 
the total result may be an  overall reduction of the power  with respect
to the case  $\gamma_{br} =\pi/2$. In Eq. (\ref{Z1}),(\ref{Z2}) and (\ref{Z3})
$k_{0} = \tau_{0}^{-1}$ where $\tau_{0}$ is the present observation time.

\subsection{Numerical analysis}
\label{subs44}
The main idea of the numerical analysis is rather simple. 
Its implementation, however, may be rather complicated. In order to capture the simplicity out of the possible 
complications we will proceed as follows. We will first discuss a rather naive 
approach to the integration of CMB anisotropies. Then, building up on this example,
 the results obtainable in the case of magnetized scalar modes will be illustrated.

\subsubsection{Simplest toy model}
\label{subs451}
Let us therefore apply the Occam razor and let us consider the simplest 
situation we can imagine, that is to say the case where 
\begin{itemize}
\item{} magnetic fields are absent;
\item{} neutrinos are absent;
\item{} photons and baryons are described within the tight-coupling approximation to lowest order (i.e. $\sigma_{\mathrm T} \to \infty$);
\item{} initial conditions are set either from the adiabatic mode or from the CDM-radiation mode.
\end{itemize}
This is clearly the simplest situation we can envisage. Since neutrinos are absent there 
is no source of anisotropic stress and the two longitudinal fluctuations of the metric are 
equal, i.e. $\phi = \psi$. Consequently, the system of equations to be solved becomes 
\begin{eqnarray}
&& {\cal R}' = \frac{k^2 c_{s}^2 {\cal H}}{{\cal H}^2 - {\cal H}'} \psi - \frac{{\cal H}}{p_{\rm t} + \rho_{\rm t}} \delta p_{\rm nad},
\label{simple1}\\
&& \psi' = - \biggl( 2 {\cal H} -\frac{ {\cal H}'}{{\cal H}} \biggr)\psi - \biggl({\cal H} - \frac{{\cal H}'}{{\cal H}}\biggl) {\cal R},
\label{simple2}\\
&& \delta_{\gamma}' = 4 \psi' - \frac{4}{3} \theta_{\gamma{\rm b}},
\label{simple3}\\
&& \theta_{\gamma{\rm b}}' = - \frac{{\cal H} R_{\rm b}}{R_{\rm b} + 1} \theta_{\gamma{\rm b} }+ 
\frac{k^2 }{4 ( 1 + R_{\rm b})} \delta_{\gamma} + k^2 \psi,
\label{simple4}\\
&& \delta_{\mathrm  c}' = 3 \psi' - \theta_{\mathrm c},
\label{simple5}\\
&& \theta_{\mathrm c}'  = - {\mathcal H} \theta_{\mathrm c} + k^2 \psi.
\label{simple6}
\end{eqnarray}
We can now use the explicit form of the scale factor discussed in Eq. (\ref{SF}) which implies:
\begin{eqnarray}
&&{\cal H} = \frac{1}{\tau_1} \frac{2 ( x +1)}{x ( x + 2)}, 
\nonumber\\
&& {\cal H}'  = - \frac{2}{\tau_1^2} \frac{ x^2 + 2 x + 4}{x^2 ( x + 2)^2},
\nonumber\\
&& {\cal H}^2 - {\cal H}' = \frac{1}{\tau_{1}^2} \frac{ 2 ( 3 x^2 + 6 x + 4)}{x^2 ( x + 2)^2},
\label{hubbles}
\end{eqnarray}
where $x = \tau/\tau_{1}$.
With these specifications the evolution equations given in (\ref{simple1})--(\ref{simple6}) 
become
\begin{eqnarray}
&&\frac{d {\mathcal R}}{d x} = \frac{4}{3} \frac{ x ( x+ 1) ( x + 2)}{( 3 x^2 + 6 x + 4)^2} \kappa^2 \psi ,
\label{simpleex1}\\
&& \frac{d \psi }{d x} = - \frac{3x^2 + 6 x + 4}{x ( x + 1 )( x + 2)} {\mathcal R} - 
\frac{ 5 x^2 + 10 x + 6}{x (x + 1) ( x + 2)} \psi,
\label{simpleex2}\\
&& \frac{ d \delta_{\gamma}}{d x} = - \frac{ 4 ( 3 x^2 + 6 x + 4)}{x ( x + 1) ( x +2)} {\mathcal R} - \frac{4( 5 x^2 + 10 x + 6)}{x (x + 1) ( x + 2)} \psi - \frac{4}{3} \tilde{\theta}_{\gamma{\rm b}},
\label{simpleex3}\\
&& \frac{ d \tilde{\theta}_{\gamma{\rm b}}}{d x} = - \frac{ 2 R_{\rm b}}{R_{\rm b} + 1} \frac{ ( x+ 1)}{x ( x + 2)} + \frac{\kappa^2 }{4 ( 1 + R_{\rm b})} \delta_{\gamma} + \kappa^2 \psi,
\label{simpleex4}\\
&& \frac{d\delta_{\mathrm c}}{dx} =  - \frac{3(3x^2 + 6 x + 4)}{x ( x + 1 )( x + 2)} {\mathcal R} - 
\frac{3( 5 x^2 + 10 x + 6)}{x (x + 1) ( x + 2)} \psi - \tilde{\theta}_{\mathrm c},
\label{simpleex5}\\
&& \frac{d \tilde{\theta}_{\mathrm c}}{d x} =  -\frac{2 ( x +1)}{x ( x + 2)}\tilde{\theta}_{\mathrm c} + \kappa^2 \psi.
\label{simpleex6}
\end{eqnarray}
In Eqs. (\ref{simpleex1})--(\ref{simpleex6}) the following rescalings have been used:
\begin{equation}
\kappa = k \tau_{1}, \qquad \tilde{\theta}_{\gamma{\mathrm b}} = \tau_{1} \theta_{\gamma{\mathrm b}},\qquad \tilde{\theta}_{{\mathrm c}} = \tau_{1} \theta_{{\mathrm c}}.
\end{equation}
The system of equations (\ref{simpleex1})--(\ref{simpleex6}) 
can be readily integrated by giving initial conditions for at $x_{\mathrm i} \ll 1$. In 
the case of the adiabatic mode (which is the one contemplated by Eqs. (\ref{simpleex1})--(\ref{simpleex6}) since we set $\delta p_{\mathrm nad}=0$) the initial conditions are as 
follows
\begin{eqnarray}
&&{\mathcal R}(x_{\mathrm i}) = {\mathcal R}_{*},\qquad \psi(x_{\mathrm i}) = 
- \frac{2}{3} {\mathcal R}_{*} ,
\nonumber\\
&&\delta_{\gamma}(x_{\mathrm i}) = -2 \psi_{*}, \qquad \tilde{\theta}_{\gamma{\rm b}}(x_{\mathrm i}) =0,
\nonumber\\
&& \delta_{\mathrm c}(x_{\mathrm i}) = - \frac{3}{2}\psi_{*},\qquad \tilde{\theta}_{{\rm c}}(x_{\mathrm i})=0.
\label{adinsimpl}
\end{eqnarray}
It can be shown by direct numerical integration that the system (\ref{simpleex1})--(\ref{simpleex6}) 
gives a reasonable semi-quantitative description of the acoustic oscillations. To simplify initial 
conditions even further we can indeed assume a flat Harrison-Zeldovich spectrum and set ${\mathcal R}_{*} = 1$.

The same philosophy used to get to this simplified form can be used to integrate the full system.
In this case, however, we would miss the important contribution of polarization 
since, to zeroth order in the tight-coupling expansion, the CMB is not polarized. 

\subsubsection{Integration of brightness perturbations}
\label{subss452}
To discuss the polarization, we have to go (at least) to first-order in the tight 
coupling expansion \cite{TC1,TC2,TC3}. For this purpose, it is appropriate to introduce 
the evolution equations of the brightness perturbations of the $I$, $Q$ and $U$ Stokes parameters characterizing the radiation field. Since the Stokes parameters $Q$ and $U$ are not invariant under rotations about the axis 
of propagation the degree of polarization $P = (Q^2 + U^2)^{1/2}$ is customarily introduced \cite{TC3,CH}. The relevant  brightness 
perturbations will then be denoted as $\Delta_{\rm I}$, $\Delta_{\rm P}$. This description, reproduces, to zeroth order 
in the tight coupling expansion, the fluid equations that have been presented 
before to set initial conditions prior to equality. For 
instance, the photon density contrast and the divergence of the photon 
peculiar velocity are related, respectively, to the 
monopole and to the dipole of the brightness perturbation of the intensity field, 
i.e. $\delta_{\gamma} = 4 \Delta_{{\rm I}0}$ 
and  $\theta_{\gamma} = 3 k \Delta_{{\rm I}1}$. 
The evolution equations of the brightness perturbations can then 
be written, within the conventions set by Eq. (\ref{longfl}) 
\begin{eqnarray}
&& \Delta_{\rm I}' + (i k \mu + \kappa') \Delta_{\rm I} + i k \mu \phi= 
\psi' + \kappa' \biggl[ \Delta_{{\rm I}0} + \mu v_{\rm b} - 
\frac{1}{2} P_{2}(\mu) S_{\rm P}\biggr],
\label{DI}\\
&& \Delta_{\rm P}' + ( i k \mu + \kappa') \Delta_{\rm P} = \frac{\kappa'}{2} 
[ 1 - P_{2}(\mu)] S_{\rm P},
\label{DQ}\\
&& v_{\rm b}' + {\mathcal   H} v_{\rm b} + i k \phi + \frac{i k}{4 R_{\rm b}} [\Omega_{\rm B} - 4 \sigma_{\rm B}] + \frac{\kappa'}{R_{\rm b}} ( v_{\rm b} + 3 i \Delta_{{\rm I}1})=0.
\label{vb}
\end{eqnarray}
Equation (\ref{vb}) is nothing but the second relation obtained in  Eq. (\ref{thetab})
having introduced the quantity $i k v_{\rm b} = \theta_{\rm b}$. The source 
terms appearing in Eqs. (\ref{DI}) and (\ref{DQ}) include a dependence 
on $P_{2}(\mu)= (3 \mu^2 -1)/2$ ( $P_{\ell}(\mu)$ denotes, in this 
framework, the $\ell$-th Legendre polynomial);
 ; $\mu = \hat{k} \cdot\hat{n}$ is 
simply
the projection of the Fourier wave-number on the direction of the photon momentum. In Eqs. (\ref{DI}) and (\ref{DQ}) the source 
term $S_{\rm P}$ is defined as
\begin{equation}
S_{\rm P}(k,\tau) = \Delta_{{\rm I}2}(k,\tau) + \Delta_{{\rm P}0}(k,\tau) + \Delta_{{\rm P}2}(k,\tau).
\end{equation}

The evolution equations in the tight coupling approximation will now be integrated 
numerically. More details on the tight coupling expansion in the presence 
of a magnetized contribution can be found in \cite{maxprd206}.

The normalization of the numerical calculation is enforced by evaluating, 
analytically, the Sachs-Wolfe plateau and by deducing, for a given set 
of spectral indices of curvature and entropy perturbations, the 
amplitude of the power spectra at the pivot scale.  Here is an 
example of this strategy.
The Sachs-Wolfe (SW) plateau can  be estimated analytically 
from the evolution equation of ${\mathcal   R}$ (or $\zeta$) by using the technique of the transfer 
matrix appropriately generalized to the case where, on top of the adiabatic and non-adiabatic contributions the magnetic fields are consistently taken into account.  The main result 
is expressed by Eq. (\ref{SWP}).

\begin{figure}
\centering
\includegraphics[height=5cm]{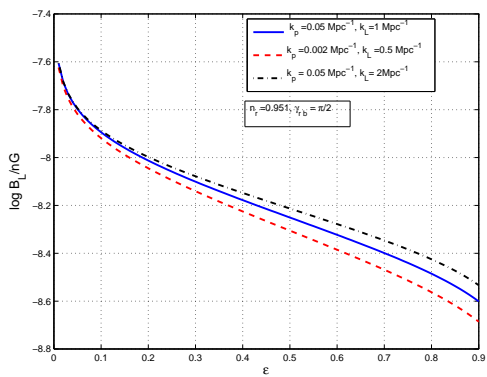}
\caption{Bounds on the protogalactic field intensity as a function of the magnetic spectral index $\varepsilon$ for 
different values of the parameters defining the adiabatic contribution to the SW plateau.}
\label{F3}      
\end{figure}

\begin{figure}
\centering
\includegraphics[height=5cm]{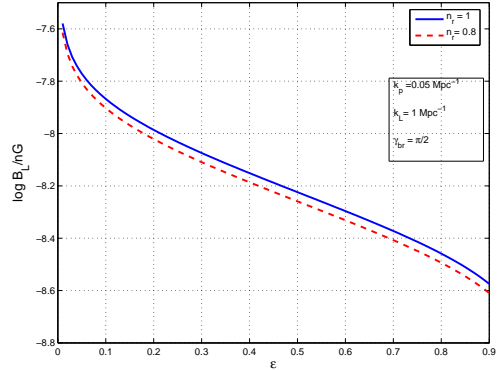}
\caption{Same plot as in Fig. \ref{F3} but with emphasis on the variation 
of $n_{\rm r}$.}
\label{F4}      
\end{figure}

If the SW plateau is determined by an adiabatic component supplemented 
by a (subleading) non-adiabatic contribution both correlated with the magnetic field intensity the 
obtainable bound may not be so constraining 
(even well above the nG range) due to the proliferation of parameters. 
A possible strategy is therefore to fix the parameters of the adiabatic mode 
to the values determined by WMAP-3 and then explore the effect of a magnetized contribution 
which is not correlated with the adiabatic mode. This implies, in Eq. (\ref{SWP}) that ${\mathcal A}_{{\mathcal S}}=0$
and $\gamma_{\rm br} = \pi/2$. 
Under this assumption, in  Figs. \ref{F3} and \ref{F4}  the bounds on $B_{L}$ are illustrated. The nature of the 
constraint depends, in this case, both on the amplitude of the protogalactic
field  (at the present epoch and smoothed over a typical comoving scale 
$L = 2\pi/k_{L}$) and upon its spectral slope, i.e. $\varepsilon$. In the case $\varepsilon < 0.5$ the magnetic energy 
spectrum is nearly scale-invariant. In this case, diffusivity effects are negligible (see, for instance, \cite{f1,f4}). 
As already discussed, if $\varepsilon \gg 1$ the diffusivity effects (both thermal and magnetic)  dominate the mode-coupling integral that lead to the magnetic energy 
spectrum \cite{f1,f4}.

In Fig. \ref{F3} the magnetic field intensity should be below the different curves if the 
adiabatic contribution dominates the SW plateau. Different choices of the pivot scale $k_{\rm p}$ and of the smoothing scale $k_{L}$, are also illustrated. 
In Fig. \ref{F3}  the scalar spectral index is fixed to $n_{r} = 0.951$ \cite{wmap3}.
In Fig. \ref{F4} the two curves corresponding, respectively, to $n_{r}=0.8$ and $n_{r} =1$ are reported.

If $\varepsilon <0.2$ the bounds are comparatively less restrictive than in the case $\varepsilon \simeq 0.9$. 
The cause of this occurrence is that we are here just looking at the largest wavelengths of the problem. As it will become 
clear in a moment, intermediate scales will be  more sensitive to the presence of fully inhomogeneous magnetic fields. 

According to Figs. \ref{F3} and \ref{F4} for a given value of the magnetic spectral index and of the scalar spectral index 
the amplitude of the magnetic field has to be sufficiently small not to affect the dominant adiabatic nature of the SW plateau. 
Therefore Figs. \ref{F3} and \ref{F4} (as well as other similar plots) can be  used 
to normalize the numerical calculations for the power spectra of the brightness perturbations, i.e. 
\begin{equation}
\frac{k^{3}}{2\pi^2} |\Delta_{\rm I}(k,\tau)|^2,\qquad \frac{k^{3}}{2\pi^2} |\Delta_{\rm P}(k,\tau)|^2,\qquad
\frac{k^{3}}{2\pi^2} |\Delta_{\rm I}(k,\tau) \Delta_{\rm P}(k,\tau)|.
\label{defPS}
\end{equation}
Let us then assume, for consistency with the cases reported in Figs. \ref{F3} and \ref{F4}, that we are dealing with the situation 
where the magnetic field is not correlated with the adiabatic mode.
 It is then possible to choose 
a definite value of the magnetic spectral index (for instance $\epsilon = 0.1$) and a definite value 
of the adiabatic spectral index, i.e. $n_{r}$ (for instance $n_{r} =0.951$, in agreement 
with \cite{wmap3}).  By using the SW plateau the normalization can be chosen in such a way the the adiabatic mode dominates 
over the magnetic contribution. In the mentioned case, Fig. \ref{F3} implies $B_{L} < 1.14 \times 10^{-8}\,\,{\rm G}$ for a pivot 
scale $k_{p} = 0.002\, {\mathrm{Mpc}}^{-1}$. Since the relative weight of the power spectra given in Eqs. (\ref{RPS}) 
and (\ref{OMPS}) is fixed, it is now possible to set initial conditions for the adiabatic mode according to Eqs. 
(\ref{DCad})--(\ref{anis2}), (\ref{VF1})--(\ref{VF3}) and (\ref{mismatch})  deep in the radiation-dominated phase.  The initial time 
of integration will be chosen as $\tau_{\rm i} = 10^{-6} \tau_{1}$ in the notations discussed in Eq. (\ref{SF}). According to Eq. (\ref{SF1}), this choice implies that $\tau_{\rm i} \ll \tau_{\rm eq}$.

The power spectra of the brightness perturbations, i.e. 
Eq. (\ref{defPS}), can be then computed by numerical integration. Clearly the calculation will depend 
upon the values of $\omega_{\rm m}$, $\omega_{\rm b}$, $\omega_{\rm c}$ and $R_{\nu}$. We will simply 
fix these parameters to their fiducial values reported in Eqs. (\ref{par}) (see also (\ref{BTPR})) and we will take $N_{\nu} = 3$ 
in Eq. (\ref{Rnu}) determining, in this way the fractional contribution of the neutrinos to the radiation plasma.

\begin{figure}
\centering
\includegraphics[height=5cm]{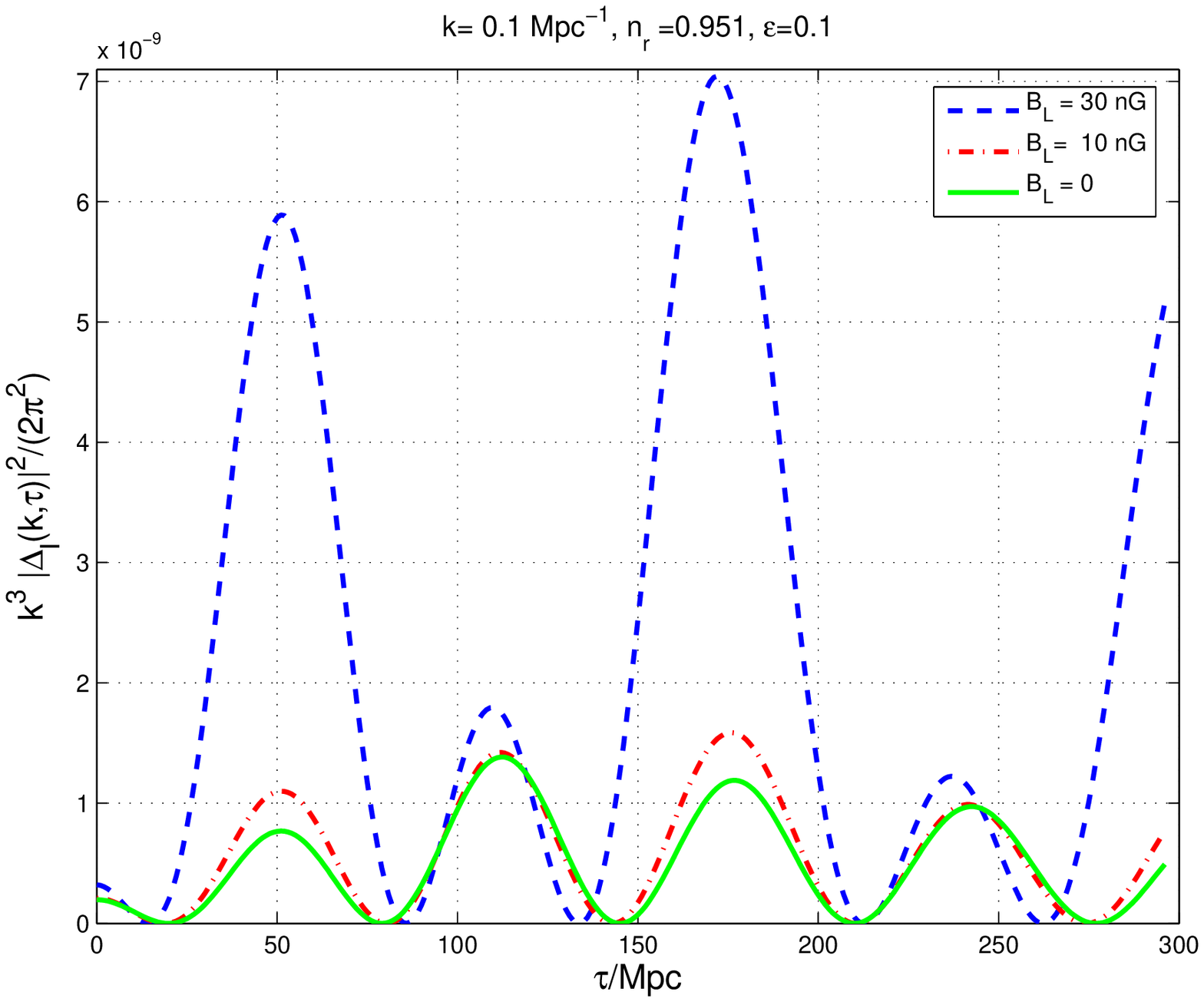}
\includegraphics[height=5cm]{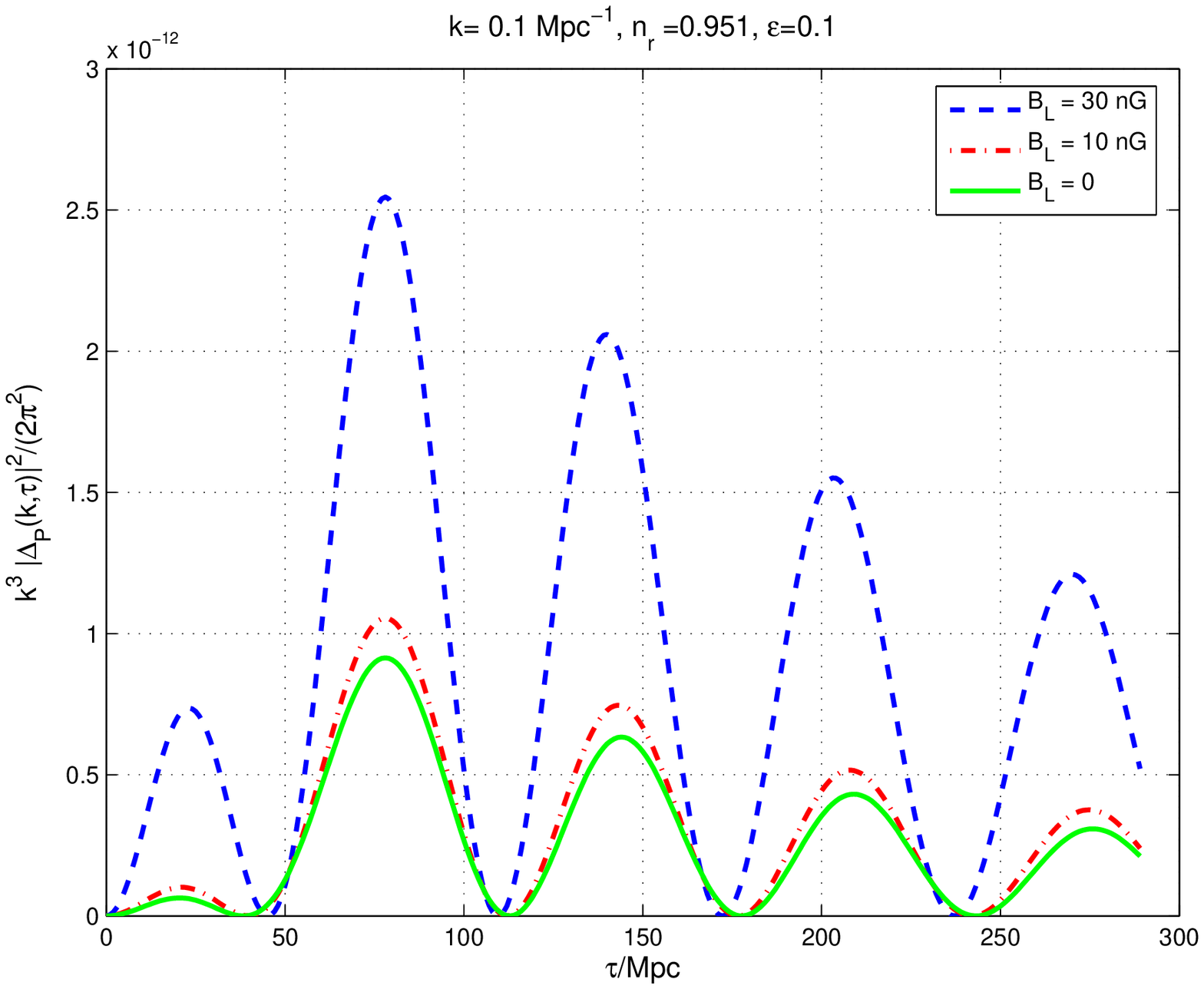}
\includegraphics[height=5cm]{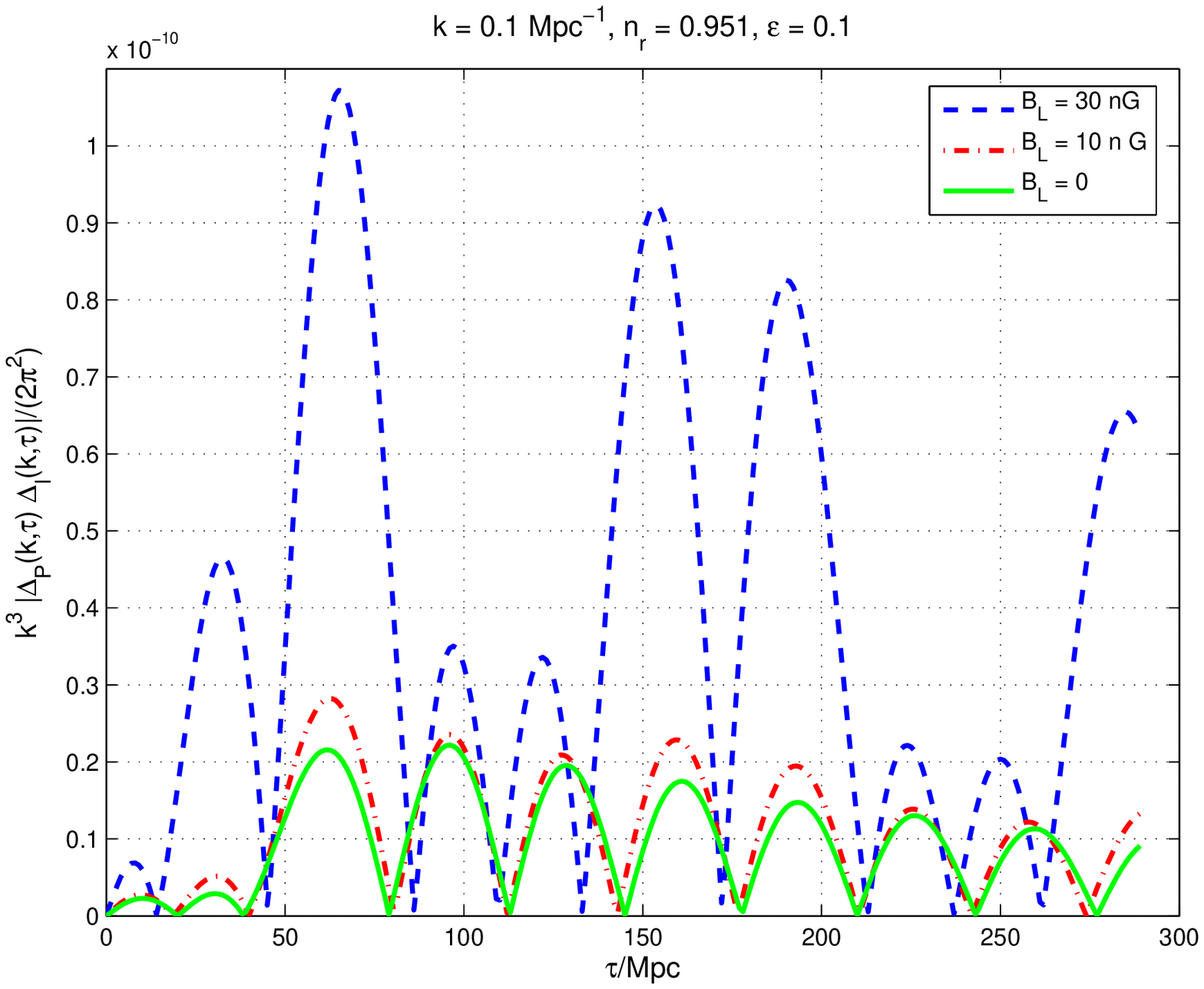}
\caption{
The power spectra of the brightness perturbations for a typical wave-number $k = 0.1 {\mathrm Mpc}^{-1}$.
 The values 
of the parameters are specified in the legends. 
The pivot scale is $k_{\rm p} = 0.002\,\, {\mathrm Mpc}^{-1}$ and the 
smoothing scale is $k_{L} = {\mathrm{ Mpc}}^{-1}$ (see Figs. \ref{F3} and \ref{F4}).}
\label{F6}      
\end{figure}

The first interesting exercise, for the present purposes, is reported in Fig. \ref{F6} where the 
power spectra of the brightness perturbations are illustrated for a wave-number $ k = 0.1\,\,{\mathrm Mpc}^{-1}$.
Concerning the results reported in Fig. \ref{F6} different comments are in order:
\begin{itemize}
\item{} for $\varepsilon = 0.1$ and $n_{r} =0.951$, the SW plateau imposes $B_{L} < 1.14 \times 10^{-8} \,\,{\mathrm G}$; from Fig. \ref{F6} it follows that a magnetic field of only $30\,\,{\rm nG}$ (i.e. marginally incompatible with the 
SW bound) has a large effect on the brightness perturbations as it can be argued by comparing, in Fig. \ref{F6}, the 
dashed curves (corresponding to $30\,\,{\rm nG}$ ) to the full curves which illustrate the case of vanishing 
magnetic fiels;
\item{} the situation where $B_{L} > {\mathrm nG}$ cannot be simply summarized by saying that the amplitudes of the power spectra 
get larger since there is a combined effect which both increases the amplitudes and shifts slightly the phases of the oscillations;
\item{} from the qualitative point of view, it is still true that the intensity oscillates as a cosine, the polarization as a sine;
\item{} the phases of the cross-correlations are, comparatively, the most affected by the presence of the magnetic field.
\end{itemize}
\begin{figure}
\centering
\includegraphics[height=5cm]{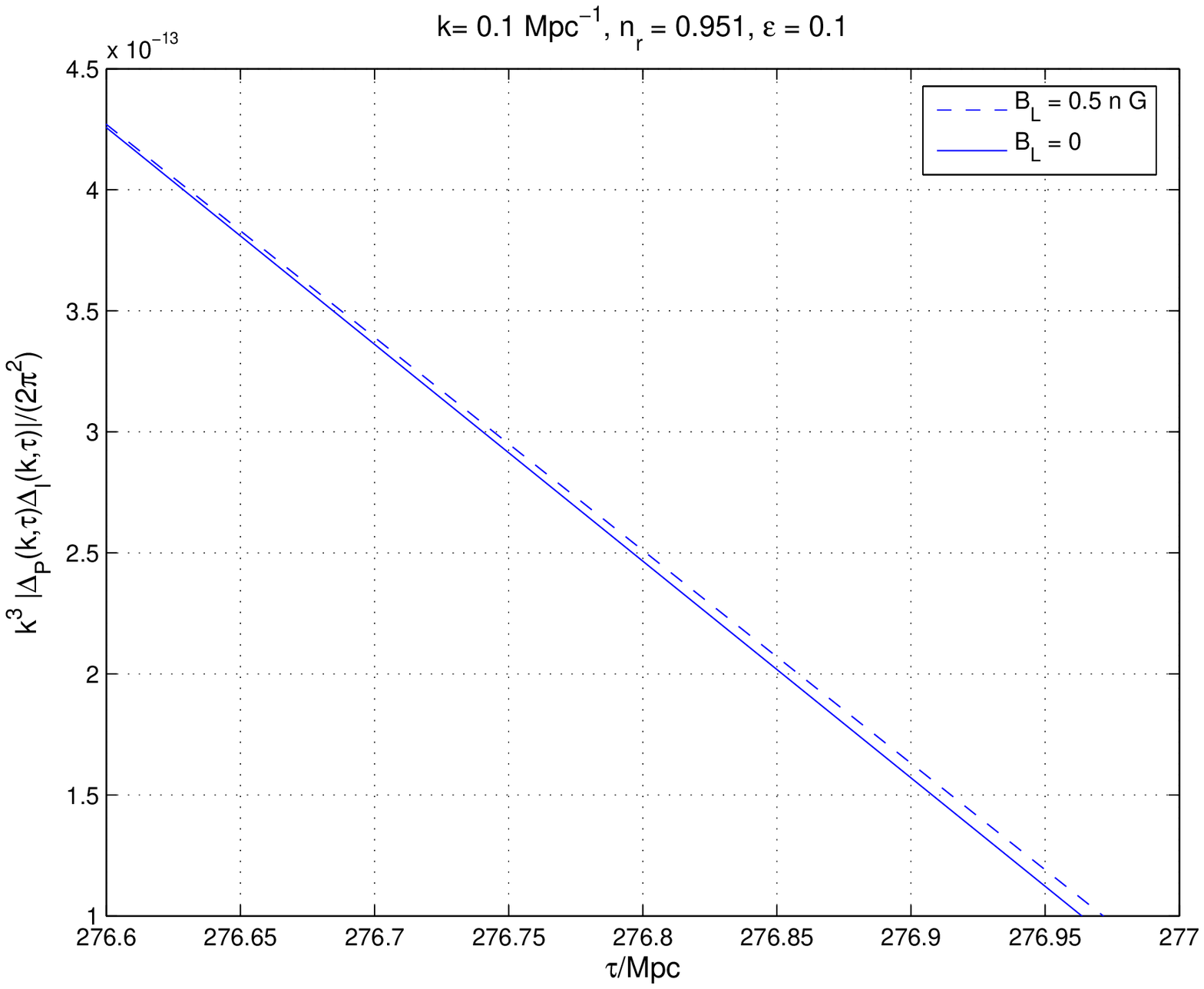}
\includegraphics[height=5cm]{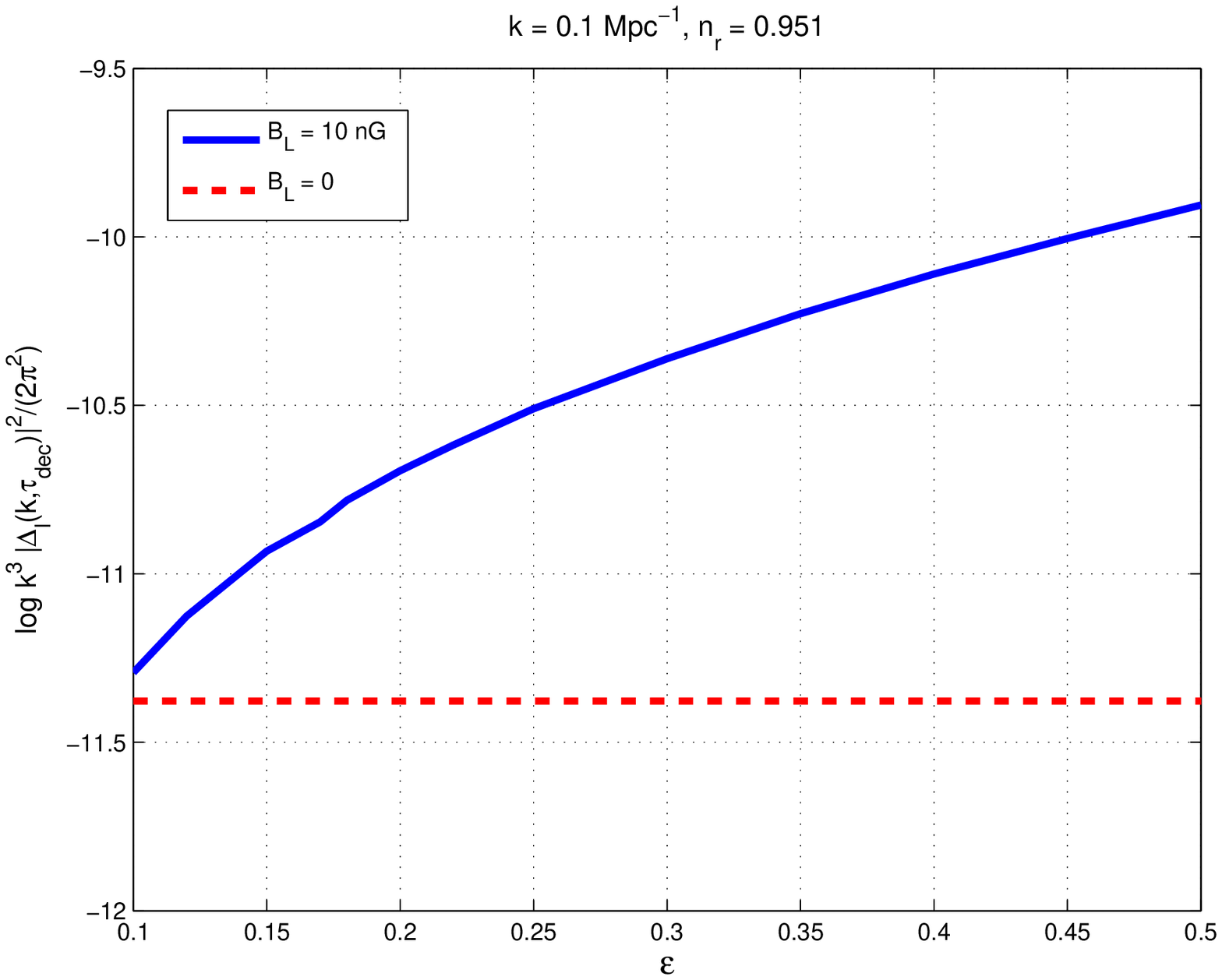}
\caption{A detail of the cross-correlation (top). The autocorrelation of the intensity at $\tau_{\rm dec}$ as a function 
of $\varepsilon$, i.e. the magnetic spectral index  (bottom).}
\label{F7}      
\end{figure}
The features arising in Fig. \ref{F6} can be easily illustrated 
for other values of $\epsilon$ and for different choices of the pivot or smoothing 
scales. The general lesson that can be drawn is that 
the constraint derived only by looking at the SW plateau are only a necessary condition on the strength of the magnetic 
field.  They are, however, not sufficient to exclude observable effects at smaller scales. This aspect is illustrated in the plot at the left in Fig. \ref{F7} which captures 
a detail of the cross-correlation. The case when $B_{L} =0$ can be still distinguished from the case $B_{L} = 0.5\,\,{\mathrm nG}$. 
Therefore, recalling that for the same choice of parameters the SW plateau implied that $B_{L} < 11.4\,\,{\mathrm nG}$, 
it is apparent that the intermediate scales lead to more stringent conditions even for nearly scale-invariant spectra 
of magnetic energy density. For the range of parameters of Fig. \ref{F7} we will have 
that $B_{L} < 0.5\,\,{\mathrm nG}$ which is more stringent than the condition 
deduced from the SW plateau by, roughly, one order of magnitude.

If $\varepsilon$ increases to higher values (but always with $\varepsilon <0.5$)
by keeping fixed $B_{L}$ (i.e. the strength of the magnetic field smoothed 
over a typical length scale $L = 2\pi/k_{L}$) the amplitude of the brightness perturbations gets larger in comparison 
with the case when the magnetic field is absent. This aspect is illustrated in the bottom plot of Fig. \ref{F7} where the logarithm (to base $10$) of the intensity autocorrelation is evaluated at a fixed wave-number (and at $\tau_{\rm dec}$) as a function of 
$\varepsilon$. The full line (corresponding to a $B_{L} = 10\,\,{\mathrm nG}$) is progressively divergent 
from the dashed line (corresponding to $B_{L} =0$) as $\varepsilon$ increases.  

\begin{figure}
\centering
\includegraphics[height=4cm]{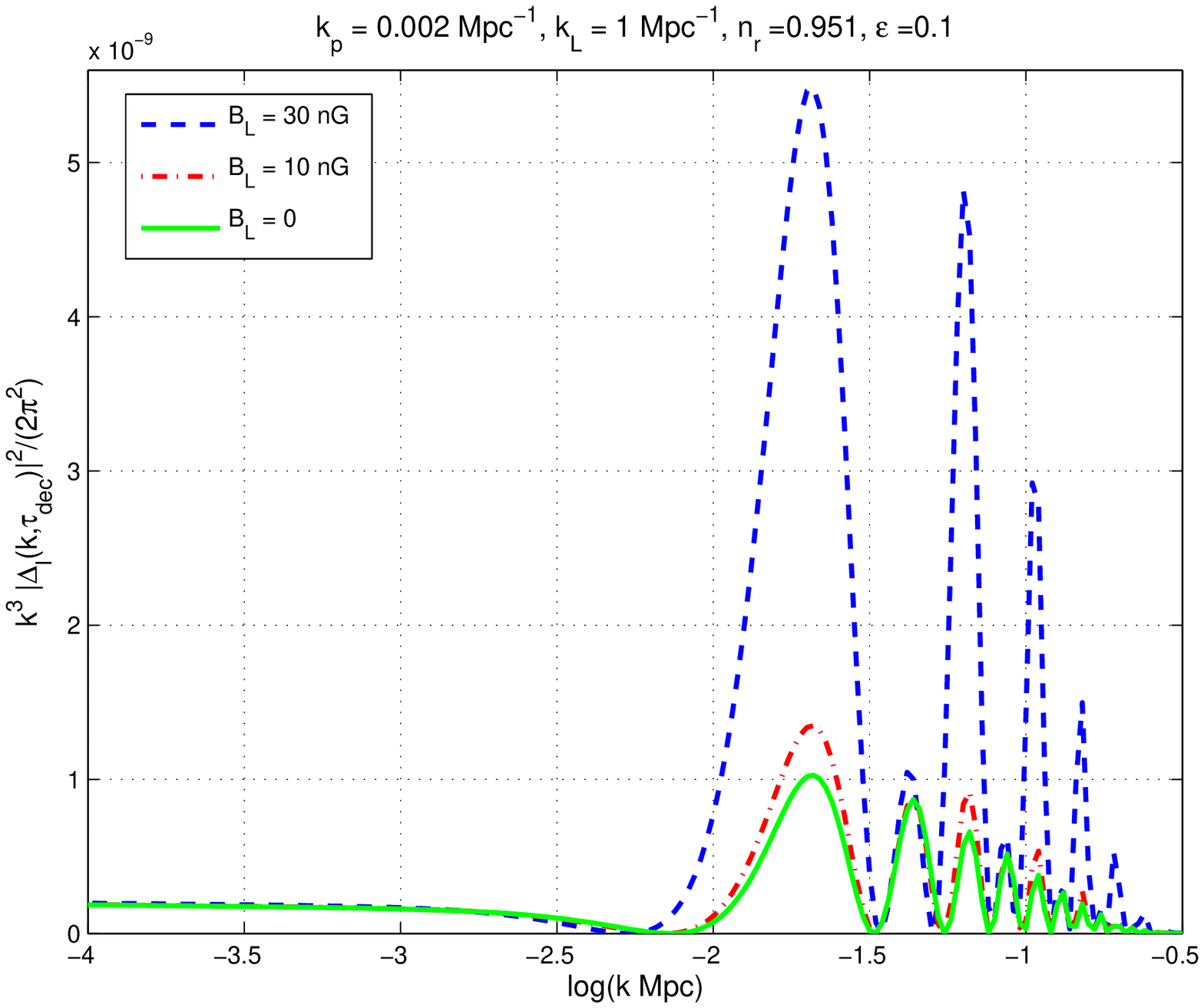}
\includegraphics[height=4cm]{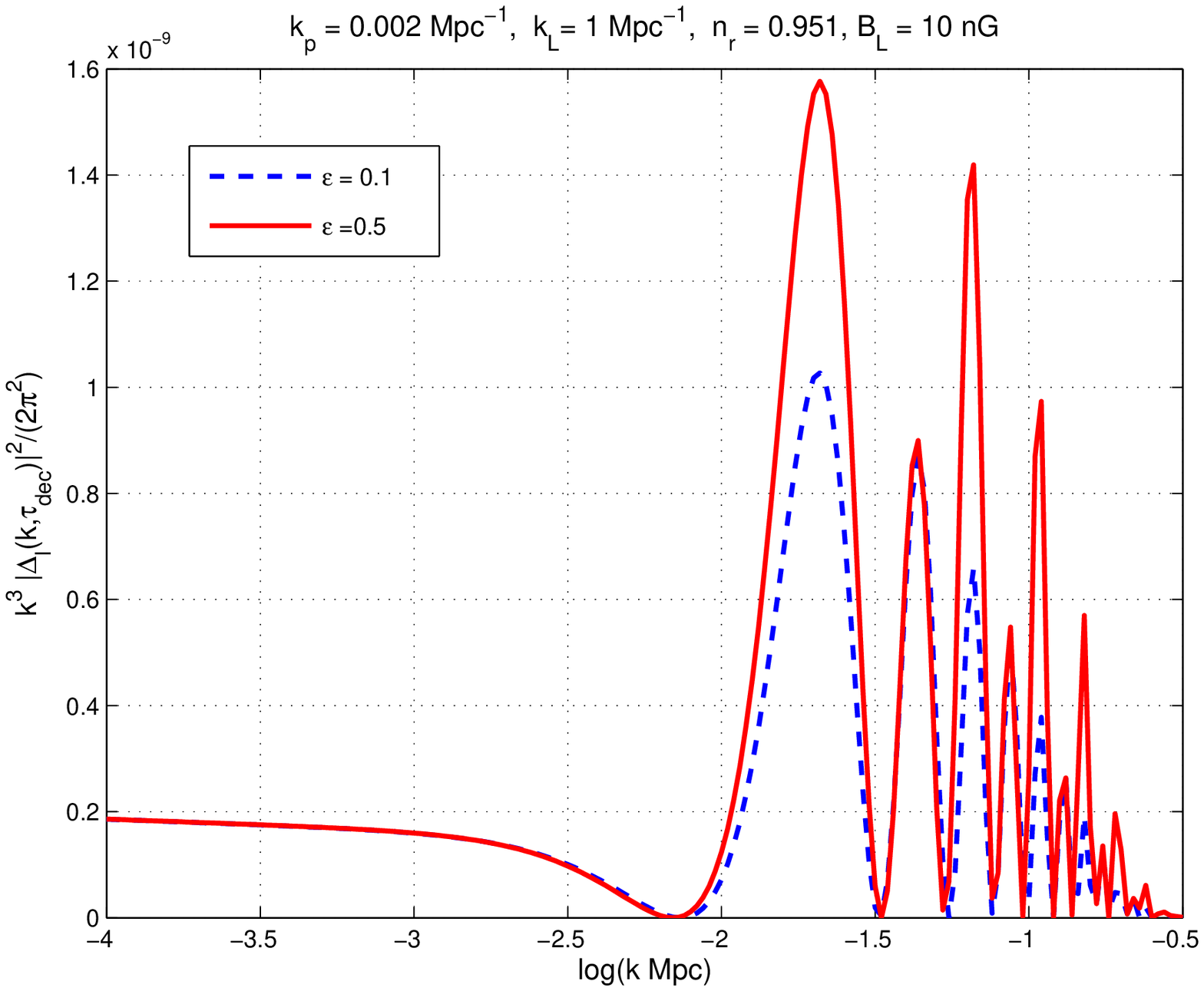}
\includegraphics[height=4cm]{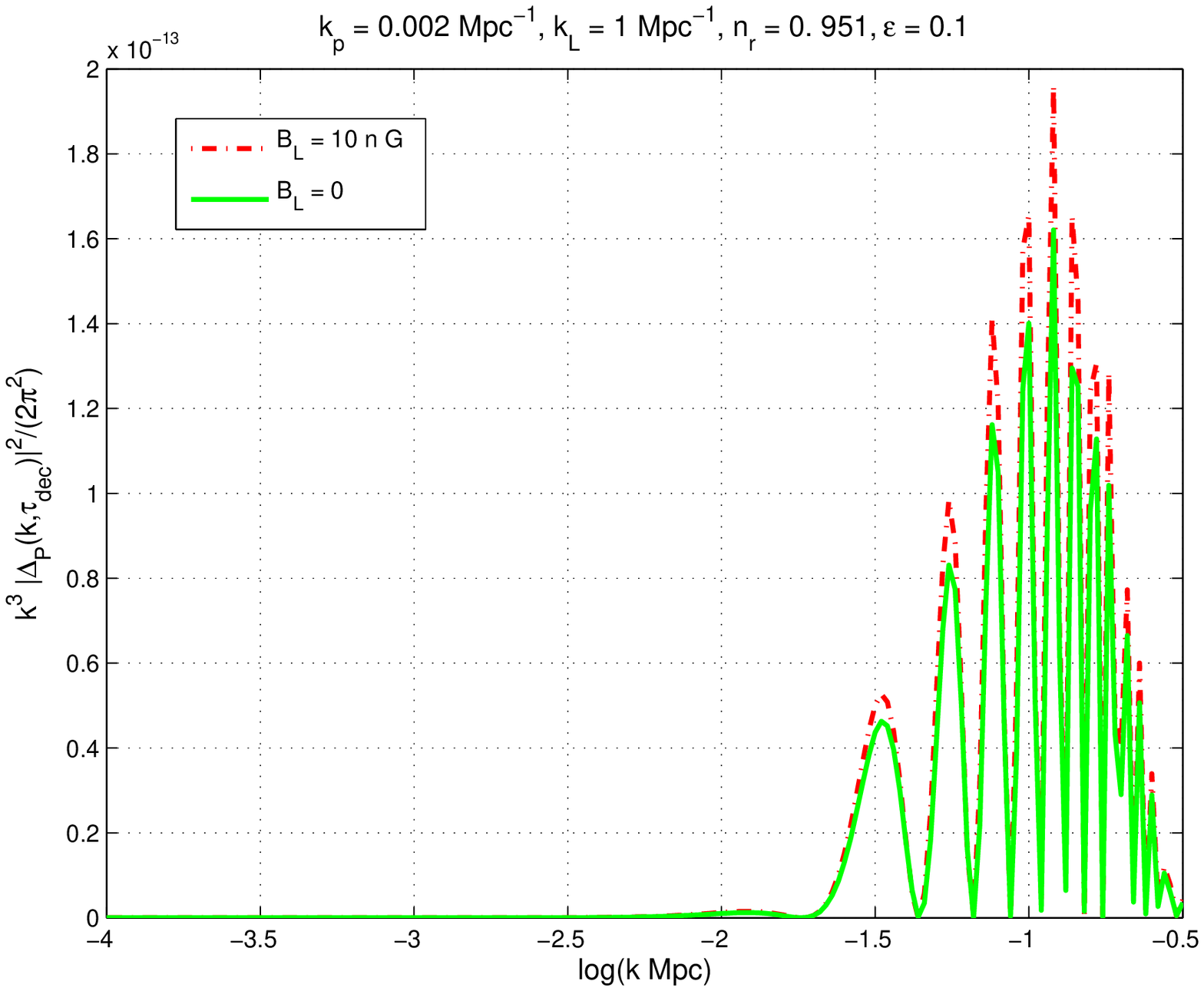}
\includegraphics[height=4cm]{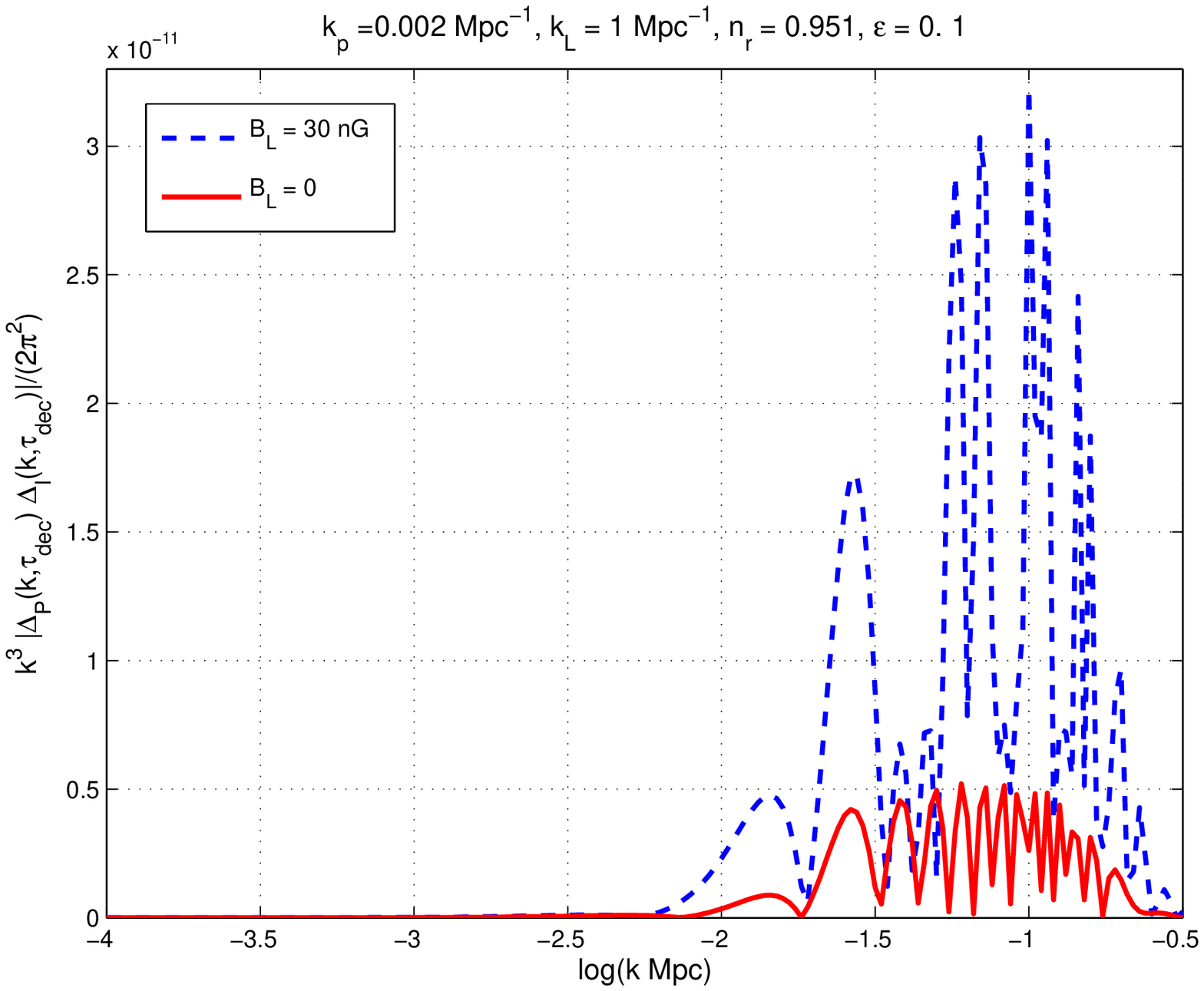}
\caption{The power spectra of the brightness perturbations at $\tau_{\rm dec}$ for the parameters reported in the legends.}
\label{F8}      
\end{figure}

In Fig. \ref{F8} the power spectra of the brightness perturbations are reported at $\tau_{\rm dec}$ and as a function of $k$.
In the two plots at the top the autocorrelation of the intensity is reported for different values of $B_{L}$ (left plot) and 
for different values of $\varepsilon$ at fixed $B_{L}$ (right plot). In the two plots at the bottom the polarization 
power spectra are reported always at $\tau_{\rm dec}$ and for different values of $B_{L}$ at fixed $\varepsilon$.
The position of the first peak of the autocorrelation of the intensity is, approximately,  $k_{\rm d}
 \simeq 0.017 \,\,{\mathrm Mpc}^{-1}$. 
 The position of the first peak of the cross-correlation is, approximately, 
 $3/4$ of  $k_{\rm d}$. From this consideration, again, we can obtain that $B_{L} < 0.3 \,\, {\mathrm nG}$ which is more constraining 
 than the SW condition.
 
Up to now the adiabatic mode has been considered in detail. We could easily add, however, non-adiabatic modes that 
are be partially correlated with the adiabatic mode. It is rather plausible, in this situation, that by 
adding new parameters, also the allowed value of the magnetic field may increase.
Similar results can be achieved by deviating from the assumption that the magnetic field and the 
curvature perturbations are uncorrelated. This aspect can be understood already 
from the analytical form of the SW plateau (\ref{SWP}). 
If there is no correlation 
between the magnetized contribution and the adiabatic contribution, i.e. 
$\gamma_{b r} =\pi/2$, the SW plateau will be enhanced in comparison 
with the case when magnetic fields are absent.  The same situation 
arises when the two components are anti-correlated (i.e. $\cos{\gamma_{br}}<0$).
However, if the fluctuations are positively correlated (i.e.  
$\cos{\gamma_{b r}}>0$) the cross-correlation adds negatively to the 
sum of the two autocorrelations of ${\mathcal R}$ and $\Omega_{\rm B}$ so that 
the total result may be an  overall reduction of the power  with respect
to the case  $\gamma_{br} =\pi/2$.

From Fig. \ref{F8} various features can be appreciated.
The presence of magnetic fields, as already pointed out, does not 
affect only the amplitude but also the phases of oscillations 
of the various brightness perturbations. Moreover, an increase 
in the spectral index $\varepsilon$ also implies 
a quantitative difference in the intensity autocorrelation.

\section{Concluding remarks}
\label{sec5}
There is little doubts that large-scale magnetic  exist in nature.
These fields have been observed in a number of different astrophysical systems.
The main question concerns therefore their origin. 
String cosmological models of pre-big bang type still represent a viable 
and well motivated theoretical option. 

Simple logic dictates that  if the origin of the large-scale magnetic fields is 
 primordial (as opposed to astrophysical) it is plausible to 
  expect the presence of magnetic fields in the primeval plasma also {\em before }
 the decoupling of radiation from matter.  CMB anisotropies
 are germane to several aspect of large-scale magnetization. 
 CMB physics may be the tool that will finally enable us 
either to confirm or to rule out the primordial nature  of galactic and clusters 
 magnetic fields seeds. 
 In the next five to ten years the forthcoming  CMB precision polarization experiments 
 will be sensitive in, various frequency channels between 30 GHz and, roughly 900 GHz. The observations will be conducted both via satellites (like the  Planck satellite) 
 and via ground based detectors (like in the case of the QUIET arrays).
 In a complementary view, the SKA telescope will provide full-sky surveys
 of Faraday rotation that may even get close to 20 GHz.
 
 In an optimistic perspective the forthcoming experimental data together 
 with the steady progress in the understanding of the dynamo theory will  
 hopefully explain the rationale for the ubiquitous nature of large-scale 
 magnetization.  In a pessimistic perspective, the primordial nature 
 of magnetic seeds will neither be confirmed nor ruled out. 
It is wise to adopt a model-independent approach by 
 sharpening those theoretical tools that 
 may allow, in the near future, a direct observational test of the effects 
 of large-scale magnetic fields on CMB anisotropies. 
 Some efforts along this perspective have been 
 reported in the present lecture. In particular, the following 
 results have been achieved:
 \begin{itemize}
 \item{} scalar CMB anisotropies have been described in the presence of a fully 
 inhomogeneous magnetic field;
 \item{} the employed formalism allows the extension of the usual CMB 
 initial conditions to the case when large-scale magnetic fields are present 
 in the game;
 \item{} by going to higher order in the tight coupling expansion 
 the evolution of the brightness perturbations has been computed 
 numerically;
 \item{} it has been shown that the magnetic fields may affect not only the 
 amplitude but also the relative phases of the Doppler oscillations;
 \item{} from the analysis of the cross-correlation power spectra 
 it is possible to distinguish, numerically, the effects of a magnetic field 
 as small as $0.5\,\,\mathrm{n\,G}$.
 \end{itemize}
 It is interesting to notice that a magnetic field in the range $10^{-10}$--$10^{-11}$ G 
 is still viable according to the present considerations. It is, therefore, not excluded
 that large-scale magnetic fields may come from a primordial field of the order 
 of $0.1$-$0.01$ nG present prior to gravitational collapse 
 of the protogalaxy. Such a field, depending upon the details of the gravitational 
 collapse may be amplified to the observable level by compressional amplification.
 The present problems in achieving a large dynamo amplification may therefore
 be less relevant than for the case when the seed field is in the range $10^{-9} \mathrm{nG} =
 10^{-18}\mathrm{nG}$. To confirm this type of scenario it will be absolutely 
 essential to introduce the magnetic field background into the current strategies 
 of parameter extraction.
 
 The considerations reported in the present lecture provide already the framework 
 for such an introduction. In particular, along a minimalist perspective, the inclusion 
 of the magnetic field background boils down to add two new extra-parameters: the spectral 
 slope and amplitude of the magnetic field (conventionally smoothed over a typical 
 comoving scale of Mpc size). The magnetic field contribution will then slightly 
 modify the adiabatic paradigm by introducing, already at the level 
 of initial conditions, a subleading non-Gaussian (and quasi-adiabatic) correction.

\printindex

\begin{thebibliography}{99.}

\bibitem{alv1} H. Alfv\'en : Arkiv. Mat. F. Astr., o. Fys. \textbf{29 B}, 2 (1943).

\bibitem{fermi} E. Fermi:  Phys. Rev. \textbf{75}, 1169 (1949).

\bibitem{alv2}  H. Alfv\'en: Phys. Rev. \textbf{75}, 1732 (1949).

\bibitem{alv3}  R. D. Richtmyer, E. Teller:  Phys. Rev. \textbf{ 75}, 1729 (1949).

\bibitem{hiltner} W. A. Hiltner:  Science \textbf{109}, 165 (1949).

\bibitem{hall} J. S. Hall:  Science \textbf{109}, 166 (1949).

\bibitem{davis} L. J. Davis  J. L. Greenstein: Astrophys. J. \textbf{114}, 206 (1951).

\bibitem{fermi2} E. Fermi, S. Chandrasekar:  Astrophys. J. \textbf{118}, 113 (1953).

\bibitem{fermi3} E. Fermi, S. Chandrsekar: Astrophys. J. \textbf{118}, 116 (1953).

\bibitem{wiel} R. Wielebinski, J. Shakeshaft:  Nature\textbf{195}, 982 (1962).

\bibitem{lyne} A. G. Lyne, F. G. Smith: Nature \textbf{ 218}, 124 (1968).

\bibitem{lynebook}  A. G. Lyne, F. G. Smith: \textit{Pulsar Astronomy}, 
(Cambridge University Press, Cambridge, UK, 1998).

\bibitem{heiles} C. Heiles:  Annu. Rev. Astron. Astrophys. \textbf{ 14}, 1 (1976).

\bibitem{kro} P. ~P. Kronberg:  Rep. Prog. Phys. \textbf{ 57}, 325 (1994).

\bibitem{gov}  F.~Govoni, L.~Feretti :  Int.\ J.\ Mod.\ Phys.\ D \textbf{13}, 1549 (2004).

\bibitem{fer} B.M. Gaensler, R. Beck, L. Feretti: New Astron.\ Rev.\  \textbf{ 48}, 1003 (2004).

\bibitem{kro2} Y.~Xu, P.~P.~Kronberg, S.~Habib, Q.~W.~Dufton:  Astrophys.\ J.\  \textbf{637}, 19 (2006).

\bibitem{kro3} P.~ P.~Kronberg :  Astron.\ Nachr. \textbf{327}, 517 (2006).

\bibitem{f1}   M.~Giovannini: Int.\ J.\ Mod.\ Phys.\ D \textbf{13}, 391 (2004).

\bibitem{SKA} See http://www.skatelescope.org for more informations.

\bibitem{planck} See http://www.rssd.esa.int for more informations.

\bibitem{maxcqg} M.~Giovannini: Class.\ Quant.\ Grav.\  \textbf{23}, R1 (2006).

\bibitem{bernstein} J.~Bernstein, L.~S.~Brown and G.~Feinberg:
  Rev.\ Mod.\ Phys.\  \textbf{ 61}, 25 (1989).

\bibitem{boyd} T.~J.~M Boyd, J.~J.~Serson: \textit{ The physics 
of plasmas}, (Cambridge University Press, Cambridge, UK, 2003).

\bibitem{krall} N. A. Krall, A. W. Trivelpiece: \textit{ Principles of
Plasma Physics}, (San Francisco Press, San Francisco 1986).

\bibitem{chen} F. Chen: \textit{ Introduction to Plasma Physics}, (Plenum 
Press, New York 1974).

\bibitem{biskamp}  D. Biskamp: \textit{ Non-linear Magnetohydrodynamics}
(Cambridge University Press, Cambridge, 1994).

\bibitem{vlasov} A. Vlasov: Zh. \'Eksp. Teor. Fiz. \textbf{ 8}, 291 (1938);
J. Phys. \textbf{ 9}, 25 (1945).

\bibitem{landau} L. D. Landau: J. Phys. U.S.S.R. \textbf{10}, 25 (1945).

\bibitem{maxbir} M.~Giovannini: 
  Phys.\ Rev.\ D \textbf{71}, 021301 (2005).
  
\bibitem{maxknot} M.~Giovannini:
  Phys.\ Rev.\ D \textbf{ 58}, 124027 (1998).

\bibitem{parker} E. N. Parker: \textit{ Cosmical Magnetic Fields} (Clarendon Press, 
Oxford, 1979).

\bibitem{zeldovich} Ya. B. Zeldovich, A. A. Ruzmaikin, D.D. Sokoloff:
\textit{Magnetic Fields in Astrophysics} (Gordon  Breach Science, New York, 
1983).

\bibitem{ruzmaikin} A. A. Ruzmaikin, A. M. Shukurov, D. D. Sokoloff : \textit{Magnetic 
Fields of Galaxies}, (Kluwer Academic Publisher, Dordrecht, 1988).

\bibitem{kulsrud1} R. Kulsrud: Annu. Rev. Astron. Astrophys. \textbf{ 37}, 37 (1999).

\bibitem{lazarian} A. Lazarian, E. Vishniac, J. Cho:  Astrophys. J. \textbf{603}, 180 
(2004); Lect. Notes Phys. :  \textbf{614}, 376 (2003).

\bibitem{brandenburg} A.~Brandenburg, K.~Subramanian: Phys.\ Rept.\  \textbf{417}, 1 (2005).

\bibitem{bs1} A.~Brandenburg, A.~Bigazzi, K.~Subramanian:
  Mon.\ Not.\ Roy.\ Astron.\ Soc.\  \textbf{325}, 685 (2001).

\bibitem{bs2}  K.~Subramanian, A.~Brandenburg:  Phys.\ Rev.\ Lett.\  \textbf{93}, 205001 (2004).

\bibitem{bs3} A.~Brandenburg, K.~Subramanian: Astron.\ Astrophys.\  \textbf{439}, 835 (2005).

\bibitem{vains} S. I. Vainshtein, Ya. B. Zeldovich:  Usp. Fiz. Nauk. \textbf{ 106}, 431 (1972).

\bibitem{matt} W. H. Matthaeus, M. L. Goldstein, S. R. Lantz:  Phys. Fluids \textbf{29}, 1504 (1986).

\bibitem{rees1} M. J. Rees:  Lect. Notes Phys. \textbf{664}, 1 (2005).

\bibitem{subr1} K. Subramanian, D. Narashima, S. Chitre:  Mon.\ Not.\ Roy.\ Astron.\ Soc.\  \textbf{271}, 
L15 (1994).

\bibitem{zweibel} N.~Y.~Gnedin, A.~Ferrara, E.~G.~Zweibel:
  Astrophys.\ J.\  \textbf{539}, 505 (2000).

\bibitem{kulsrud2} R. Kulsrud, S. erson:  Astrophys. J. \textbf{396}, 606 (1992).

\bibitem{zel1} Ya. Zeldovich, I. Novikov: \textit{ The structure  evolution of the Universe} (Chicago
University Press, Chicago, 1971), Vol. 2.

\bibitem{zel2} Ya. Zeldovich:  Sov. Phys. JETP \textbf{ 21}, 656 (1965).

\bibitem{harrison1} E. Harrison: Phys. Rev. Lett. \textbf{18}, 1011 (1967).

\bibitem{harrison2}  E. Harrison: Phys. Rev. \textbf{167}, 1170 (1968).

\bibitem{harrison3} E. Harrison: Mon. Not. R. Astr. Soc. \textbf{147}, 279 (1970).

\bibitem{biermann} L. Biermann:  Z. Naturf. \textbf{5A}, 65 (1950).

\bibitem{mishustin} I. Mishustin, A. Ruzmaikin:  Sov. Phys. JETP \textbf{34}, 223 (1972).

\bibitem{hmk1} M.~Giovannini: Phys.\ Rev.\ D \textbf{ 61}, 063004 (2000).

\bibitem{hmk2}   M.~Giovannini: Phys.\ Rev.\ D \textbf{61}, 063502 (2000).

\bibitem{hmk3} G.~Piccinelli, A.~Ayala: Lect.\ Notes Phys.\  \textbf{ 646}, 293 (2004).

\bibitem{bdv1} D.~Boyanovsky, H.~J.~de Vega, M.~Simionato:
Phys.\ Rev.\ D \textbf{67}, 123505 (2003).

\bibitem{bdv2} D.~Boyanovsky, M.~Simionato, H.~J.~de Vega:
  Phys.\ Rev.\ D \textbf{67}, 023502 (2003).

\bibitem{anomaly}  M.~Giovannini, M.~E.~Shaposhnikov,
  Phys.\ Rev.\ D \textbf{ 57}, 2186 (1998).

\bibitem{bamba0}  K.~Bamba: arXiv:hep-ph/0611152.

\bibitem{angel} A.~Sanchez, A.~Ayala, G.~Piccinelli:
  arXiv:hep-th/0611337.

\bibitem{gioshap} M.~Giovannini, M.~E.~Shaposhnikov:
  Phys.\ Rev.\ D \textbf{62}, 103512 (2000).

\bibitem{mgs2} M. Giovannini,
M. Shaposhnikov: {\it Proc. of CAPP2000} (July 2000, Verbier Switzerland)
  eprint Archive [hep-ph/0011105]. 

\bibitem{cal} E. Calzetta, A. Kus, F. Mazzitelli: Phys. Rev. D, \textbf{ 57}, 7139 (1998).

\bibitem{cal1} A. Kus,  E. Calzetta, F. Mazzitelli, C. Wagner:  Phys.Lett. B \textbf{ 472},  
287 (2000).

\bibitem{turnerwidrow} M. S. Turner, L. M. Widrow:  Phys. Rev. D \textbf{ 37}, 2734 (1988).

\bibitem{ratra} B. Ratra:  Astrophys. J. Lett. \textbf{391}, L1 (1992).

\bibitem{dolgov} A. Dolgov:  Phys. Rev. D \textbf{ 48}, 2499 (1993).

\bibitem{drummond} I. Drummond, S. Hathrell: Phys.Rev. D \textbf{ 22}, 343 (1980).

\bibitem{carroll1}  S. Carroll, G. Field, R. Jackiw: Phys. Rev. D \textbf{ 41},  1231 (1990).

\bibitem{carroll2}  W. D. Garretson, G. Field, S. Carroll:  Phys. Rev. D \textbf{46}, 5346 (1992).

\bibitem{carroll3} G. Field, S. Carroll  Phys. Rev. D: \textbf{ 62}, 103008 (2000).

\bibitem{variation} M. Giovannini:  Phys. Rev. D  \textbf{ 64}, 061301 (2001).

\bibitem{bamba} K.~Bamba,  J.~Yokoyama, e-print Archive [astro-ph/0310824].

\bibitem{gravphot} M. Gasperini, {\it Phys. Rev. D} {\bf 63}, 047301 (2001)

\bibitem{okun} L. Okun, {\it Sov. Phys. JETP} {\bf 56}, 502 (1982).

\bibitem{bertolami} O. Bertolami  D. Mota, {\it Phys. Lett. B} {\bf 455}, 96 (1999).

\bibitem{mgint} M. Giovannini, {\it Phys. Rev. D} {\bf 62}, 123505 (2000).

\bibitem{ford} L. H. Ford, Phys.Rev. D \textbf{ 31}, 704 (1985).

\bibitem{GAB1} M.~Gasperini, M.~Giovannini, G.~Veneziano:
  Phys.\ Rev.\ Lett.\  \textbf{ 75}, 3796 (1995).

\bibitem{GAB2}  M.~Gasperini, M.~Giovannini, G.~Veneziano:
  Phys.\ Rev.\ D \textbf{ 52}, 6651 (1995).
  
\bibitem{gg}  M.~Gasperini,  M.~Giovannini: Phys.\ Rev.\ D \textbf{47}, 1519 (1993).

\bibitem{yuen} H. Yuen: Phys. Rev. A \textbf{13}, 2226 (1976).

\bibitem{luciano} A.~O.~Barut, L.~Girardello: Commun.\ Math.\ Phys.\  \textbf{21}, 41 (1971).
 
 \bibitem{stoler} D. Stoler: Phys. Rev. D \textbf{1}, 3217 (1970);
 D. Stoler: Phys. Rev. D \textbf{4}, 2309 (1971).

 \bibitem{mol} S.~Fubini, A.~Molinari:  Nucl.\ Phys.\ Proc.\ Suppl.\  \textbf{33C}, 60 (1993).

\bibitem{loudon}  R. Loudon: J. Mod. Opt. \textbf{34}, 709 (1987).

\bibitem{loudon2} R. Loudon: \textit{The quantum theory of light}
(Clarendon Press, Oxford, 1983).

\bibitem{schumi} B. L. Schumaker: Phys. Rept. \textbf{135}, 318 (1986).

\bibitem{mandel} L. Mandel, E. Wolf: \textit{Optical coherence and quantum optics}, (Cambridge University Press, Cambridge 
UK, 1995).

\bibitem{PBB1} G.~Veneziano:  Phys.\ Lett.\ B \textbf{265}, 287 (1991).

\bibitem{PBB2} M.~Gasperini, G.~Veneziano: Astropart.\ Phys.\  \textbf{ 1}, 317 (1993).

\bibitem{PBB3} M.~Gasperini,  G.~Veneziano: Phys.\ Rept.\  \textbf{373}, 1 (2003).
  
\bibitem{LEST1} C. Lovelace: Phys. Lett. B \textbf{135},  75 (1984).

\bibitem{LEST2} E. Fradkin, A. Tseytlin:  Nucl. Phys. B \textbf{ 261}, 1 (1985)

\bibitem{LEST3} C. Callan at al.:  Nucl. Phys. B \textbf{ 262}, 593 (1985).

\bibitem{G1} M.~Gasperini, M.~Giovannini, G.~Veneziano:
  Phys.\ Lett.\ B \textbf{569}, 113 (2003).
  
\bibitem{G2} M.~Gasperini, M.~Giovannini, G.~Veneziano:
  Nucl.\ Phys.\ B \textbf{ 694}, 206 (2004).

\bibitem{meis0} K. A. Meissner,  G. Veneziano: Mod. Phys. Lett.  A \textbf{ 6}, 3397 (1991).

\bibitem{meis1} K. A. Meissner  G. Veneziano: Phys. Lett. B \textbf{ 267}, 33 (1991).

\bibitem{gmv} M.~Gasperini, J.~Maharana,  G.~Veneziano:
Nucl.\ Phys.\ B \textbf{ 472}, 349 (1996).

\bibitem{heating} M.~Giovannini:  Class.\ Quant.\ Grav.\  \textbf{ 21}, 4209 (2004).

\bibitem{squeezed1} M.~Gasperini, M.~Giovannini: Phys.\ Rev.\ D \textbf{47}, 1519 (1993).

\bibitem{squeezed2} M.~Gasperini, M.~Giovannini: Phys.\ Lett.\ B \textbf{301}, 334 (1993).

\bibitem{squeezed3}  M.~Giovannini: Phys.\ Rev.\ D \textbf{61}, 087306 (2000).

\bibitem{tens1} R.~Brustein, M.~Gasperini, M.~Giovannini, G.~Veneziano, Phys.\ Lett.\ B \textbf{361}, 45 (1995).

\bibitem{maxmuk} R.~Brustein, M.~Gasperini, M.~Giovannini, V.~F.~Mukhanov, G.~Veneziano,
  Phys.\ Rev.\ D \textbf{51}, 6744 (1995).

\bibitem{sloth} K.~Enqvist  M.~S.~Sloth:
  Nucl.\ Phys.\ B \textbf{626}, 395 (2002);
  \\ M.~S.~Sloth:
  Nucl.\ Phys.\ B \textbf{656}, 239 (2003).

\bibitem{ax1}  V.~Bozza, M.~Gasperini, M.~Giovannini,  G.~Veneziano:
  Phys.\ Rev.\ D \textbf{67} (2003) 063514.\\
  V.~Bozza, M.~Gasperini, M.~Giovannini,  G.~Veneziano:  Phys.\ Lett.\ B \textbf{ 543}, 14 (2002).

\bibitem{bars} P. Astone {\em et al}.: Astron. Astrophys. {\bf 351}, 811 (1999).

\bibitem{picasso} Ph. Bernard, G. Gemme, R. Parodi, E. Picasso:
Rev. Sci. Instrum. {\bf 72}, 2428 (2001).

\bibitem{cruise} A. M. Cruise: Class. Quantum Grav. {\bf 17}, 2525 (2000); 
\\
A. M. Cruise: Mon. Not. R. Astron. Soc \textbf{204}, 485 (1983).

\bibitem{maxdan} D. Babusci and M. Giovannini: 
Int.J. Mod. Phys. D \textbf{10} 477 (2001); 
\\ D. Babusci and M. Giovannini: 
Class. Quant. Grav. \textbf{17}, 2621 (2000).

\bibitem{alex2}
P.~J.~E.~Peebles, A.~Vilenkin: Phys.\ Rev.\ D \textbf{59}, 063505 (1999).

\bibitem{maxquint} M.~Giovannini: Class.\ Quant.\ Grav.\  \textbf{16}, 2905 (1999);
\\ M.~Giovannini:
Phys.\ Rev.\ D \textbf{60}, 123511 (1999).
\\ D.~Babusci and M.~Giovannini: Phys.\ Rev.\ D \textbf{ 60}, 083511 (1999);
\\
M. Giovannini: Phys. Rev. D \textbf{ 58}, 083504 (1998).

\bibitem{nicotri} M. Gasperini, S. Nicotri: e-print [hep-th/0511039].


\bibitem{beck1} R. Beck: Astron.Nachr. \textbf{327}, 512 (2006).

\bibitem{beck2} R. Beck, A. Brenburg, D. Moss, A. Skhurov,  D. Sokoloff:
 Annu. Rev. Astron. Astrophys. \textbf{34}, 155 (1996).

\bibitem{vallee} J. Vall\'ee:  Astrophys. J. \textbf{566},  261 (2002).
 
\bibitem{ED1} E. Battaner, E. Florido: Mon. Not. R. Astron. Soc \textbf{277}, 1129 (1995).

\bibitem{ED2}  E. Battaner, E. Florido, J. Jimenez-Vincente: Astron.Astrophys. \textbf{326}, 13 (1997).

\bibitem{ED3}  E. Florido, E. Battaner: Astron.Astrophys. \textbf{327}, 1 (1997).

\bibitem{ED4}   E.~Florido, et al.:
  arXiv:astro-ph/0609384.
  
\bibitem{ED5}  E.~Battaner, E.~Florido,
  Fund.\ Cosmic Phys.\  \textbf{ 21}, 1 (2000).
 
\bibitem{f4} K.~Subramanian: Astron.Nachr. \textbf{327}, 399 (2006).
 
\bibitem{f4a} M.~Giovannini: Class.\ Quant.\ Grav.\  \textbf{23}, R1 (2006). 
 
\bibitem{f2} A.~Brandenburg,  K.~Subramanian:
  Phys.\ Rept.\  \textbf{ 417}, 1 (2005); \\
  A. Lazarian, E. Vishniac,  J. Cho: Astrophys. J. \textbf{603}, 180 
(2004); Lect. Notes Phys.  \textbf{614}, 376 (2003).

\bibitem{f3} J. Barrow,  K. Subramanian: Phys. Rev. Lett.  {\bf 81}, 
3575 (1998);\\
 J. Barrow,  K. Subramanian:  Phys. Rev.  D \textbf{ 58}, 83502 (1998);\\
C.~Tsagas, R.~Maartens: Phys.\ Rev.\ D \textbf{ 61}, 083519 (2000);\\
A.~Mack, T.~Kahniashvili,  A.~Kosowsky:  Phys.\ Rev.\ D \textbf{65}, 123004 (2002);\\
 A.~Lewis:  Phys.\ Rev.\ D \textbf{70}, 043518 (2004); \\
 T.~Kahniashvili,  B.~Ratra:  Phys.\ Rev.\ D \textbf{71}, 103006 (2005).

\bibitem{f9} G.~Chen {\it et al.}:
  Astrophys.\ J.\  \textbf{611}, 655 (2004); \\
  P.~D.~Naselsky {\it et al.}:
   Astrophys.\ J.\  \textbf{615}, 45 (2004);
  L.~Y.~Chiang, P.~Naselsky:
  Int.\ J.\ Mod.\ Phys.\ D \textbf{14}, 1251 (2005);\\
  L.~Y.~Chiang, P.~D.~Naselsky, O.~V.~Verkhodanov, M.~J.~Way:
  Astrophys.\ J.\  \textbf{590}, L65 (2003);
   \\ D.~G.~Yamazaki {\it et al.}:  Astrophys.\ J.\  \textbf{625}, L1 (2005).

\bibitem{f5} H.~V.~Peiris {\it et al.}, [WMAP Collaboration]: Astrophys.\ J.\ Suppl.\  \textbf{148}, 213 (2003).

\bibitem{wmap3} D. Spergel {\it et al.} [WMAP Collaboration]: arXiv:astro-ph/0603449.

\bibitem{f60} K.~Enqvist, H.~Kurki-Suonio  J.~Valiviita:  Phys.\ Rev.\ D \textbf{62}, 103003 (2000).
 
\bibitem{f61} H.~Kurki-Suonio, V.~Muhonen  J.~Valiviita:  Phys.\ Rev.\ D \textbf{71}, 063005 (2005). 

\bibitem{f62} K.~Moodley, M.~Bucher, J.~Dunkley, P.~G.~Ferreira  C.~Skordis:  Phys.\ Rev.\ D \textbf{70}, 103520 (2004).

\bibitem{maxprd06}  M.~Giovannini:  Phys.\ Rev.\ D \textbf{73}, 101302 (2006).

\bibitem{maxprd206} M.~Giovannini: Phys.\ Rev.\ D \textbf{74}, 063002 (2006).

\bibitem{maxcqg06}  M.~Giovannini: Class.\ Quant.\ Grav.\  \textbf{23}, 4991 (2006).

\bibitem{f3a}  J.~D.~Barrow, R.~Maartens, C.~G.~Tsagas:
  arXiv:astro-ph/0611537.
  
\bibitem{f3b}T.~Kahniashvili, B.~Ratra:
  arXiv:astro-ph/0611247.

\bibitem{f70} E. Harrison: Rev. Mod. Phys. \textbf{ 39}, 862 (1967).

\bibitem{f71}J. M. Bardeen: Phys. Rev. D \textbf{22}, 1882 (1980).

\bibitem{f72} C.-P. Ma  E. Bertschinger:  Astrophys. J. \textbf{455}, 7 (1995).

\bibitem{f73} M.~Giovannini: Phys.\ Rev.\ D \textbf{ 70}, 123507 (2004).

\bibitem{f74} M.~Giovannini: Int.\ J.\ Mod.\ Phys.\ D {\bf 14}, 363 (2005).

\bibitem{bir} M.~Giovannini: Phys.\ Rev.\ D \textbf{71}, 021301 (2005).

\bibitem{f80} J. Bardeen, P. Steinhardt,  M. Turner: Phys. Rev. D \textbf{28}, 679 (1983).

\bibitem{f81} R. Brandenberger, R. Kahn,  W. Press: Phys. Rev. D \textbf{ 28}, 1809 (1983).

\bibitem{bv1}  M.~Giovannini: Phys.\ Lett.\ B \textbf{622}, 349 (2005).

\bibitem{bv2} M.~Giovannini: Class.\ Quant.\ Grav.\  \textbf{22}, 5243 (2005).

\bibitem{wh} W. Hu  N. Sugiyama: Astrophys.\ J.\  \textbf{ 444}, 489 (1995); {\it ibid}. \textbf{ 471}, 30 (1996).

\bibitem{wmap1}  H.~V.~Peiris {\it et al.} [WMAP collaboration]:  Astrophys.\ J.\ Suppl.\  \textbf{ 148}, 213 (2003).

\bibitem{wmap2} L.~Page {\it et al.} [WMAP collaboration]: arXiv:astro-ph/0603450.

\bibitem{riess} A.~G.~Riess {\it et al.}, Astrophys.\ J.\  \textbf{607}, 665 (2005).

\bibitem{astier} P. Astier {\it et al}., astro-ph/0510447.

\bibitem{old2} P. Naselsky, I. Novikov: Astrophys. J. \textbf{413}, 14 (1993).

\bibitem{old3} H. Jorgensen, E. Kotok, P. Naselsky, I Novikov: Astron. Astrophys. 
\textbf{294}, 639 (1995).

\bibitem{TC1} P. J. E. Peebles, J. T. Yu: Astrophys. J. {\bf 162}, 815 (1970).

\bibitem{TC2}  A. G. Doroshkevich, Ya. B. Zeldovich,  R. A. Sunyaev:  Sov. Astron. \textbf{ 22}, 523 (1978).

\bibitem{TC3} M.~Zaldarriaga  D.~D.~Harari:  Phys.\ Rev.\ D \textbf{ 52} (1995) 3276.

\bibitem{CH} S.  Chandrasekar: \textit{ Radiative Transfer}, (Dover, New York, US, 1966).

\end{thebibliography}
\end{document}